\documentclass[
 reprint, 
 prb,
 amsmath,amssymb,
 aps,
 nofootinbib
]{revtex4-2}

\usepackage{graphicx}
\usepackage{dcolumn}
\usepackage{bm}
\usepackage[colorlinks, allcolors=blue]{hyperref}
\usepackage[all]{hypcap}
\usepackage{amsmath}
\usepackage{amsthm}
\usepackage{caption}
\usepackage{comment}
\usepackage{subcaption}
\usepackage{multirow}
\usepackage{algorithm}
\usepackage{algpseudocode}
\usepackage[export]{adjustbox}

\newtheorem{theorem}{Theorem}

\newtheorem{definition}{Definition}
\newtheorem{prop}{Proposition}

\newcommand{\ket}[1]{\vert #1\rangle}
\newcommand{\bra}[1]{\langle #1\vert}
\newcommand{\expect}[3]{\langle #1\vert #2 \vert #3 \rangle}
\newcommand{\braket}[2]{\langle #1\vert #2 \rangle}
\DeclareMathOperator{\Tr}{Tr}
\newcommand*{\horzbar}{\rule[.5ex]{2.5ex}{0.5pt}}

\newcommand{\pref}[2]{\hyperref[#1]{\ref{#1}(#2)}}

\DeclareMathAlphabet{\mathmybb}{U}{bbold}{m}{n}
\newcommand{\1}{\mathmybb{1}}

\newcommand{\hide}[1]{}

\begin{document}
\title{Classical simulability of Clifford+T circuits with Clifford-augmented matrix product states}

\def\urbana{
The Anthony J. Leggett Institute for Condensed Matter Theory and IQUIST and NCSA Center for Artificial Intelligence Innovation and Department of Physics, University of Illinois at Urbana-Champaign, IL 61801, USA}
\author{Zejun Liu} 
\affiliation{\urbana}
\author{Bryan K. Clark}
\affiliation{\urbana}
\date{\today}

\begin{abstract}
Determining the quantum-classical boundary between quantum circuits which can be efficiently simulated classically and those which cannot remains a fundamental question. One approach to classical simulation is to represent the output of a quantum circuit as a Clifford-augmented Matrix Product State (CAMPS) which, via a disentangling algorithm, decomposes the wave function into Clifford and MPS components and from which Pauli expectation values can be computed in time polynomial in the MPS bond-dimension.  In this work, we develop an optimization-free disentangling (OFD) algorithm for Clifford circuits either doped with $T$-gates or, 
equivalently, preceded by multi-qubit gates of the form $\alpha I+\beta P$. We give a simple and easily computed algebraic criterion which characterizes the individual quantum circuits for which OFD generates an efficient CAMPS - the bond-dimension is exponential in the null space of a GF(2) matrix induced by a tableau of the twisted Pauli strings $P$. This significantly increases the number of circuits with rigorous polynomial time classical simulations.  We also give evidence that the typical $N$ qubit random Clifford circuit doped with $N$ uniformly distributed $T$ gates of poly-logarithmic depth or greater has a CAMPS with polynomial bond-dimension.  In addition, we compare OFD against disentangling by optimization. 
We further explore the representability of CAMPS for random Clifford circuits doped with more than $N$ $T$-gates.
We also propose algorithms for sampling, probability and amplitude estimation of bitstrings, and evaluation of entanglement R\'enyi entropy from CAMPS, which, though still having exponential complexity, are more efficient than standard MPS simulations.
This work establishes a versatile framework for understanding classical simulatability of Clifford+$T$ circuits and explores the interplay between quantum entanglement and quantum magic in quantum systems. 
\end{abstract}

\maketitle
\tableofcontents
\section{Introduction}
The power of quantum computation comes from its ability to run quantum circuits which cannot be efficiently simulated by classical computers~\cite{feynman2018simulating,bouland2019complexity,hangleiter2023computational}. Nonetheless, there are certain circuits for which polynomial-time classical simulations exist~\cite{jozsa2008matchgates,Aaronson2004,gottesman1998heisenberg,napp2022efficient,vidal2003efficient,vidal2004efficient}. Further demarcating this dividing line between circuits which have efficient classical simulations and those which do not is integral to the understanding of quantum computing~\cite{pan2022solving,Pan2022SimulateCirc,villalonga2021efficient,beguvsic2024fast,tindall2024efficient}. 

\begin{figure}
\centering
\includegraphics[width=0.5\textwidth]{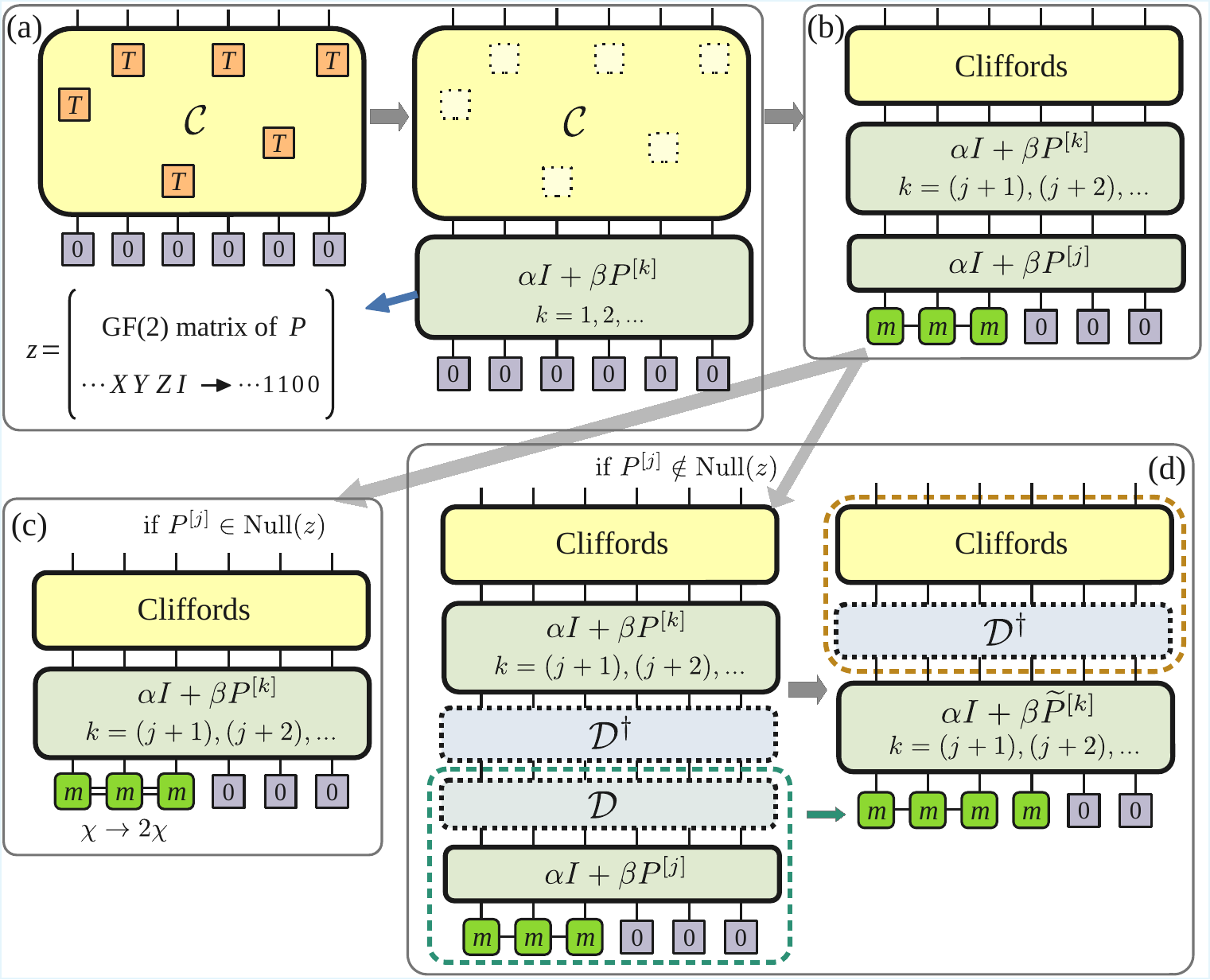}
\caption{(a) $T$-gates of a Clifford+$T$ circuit are commuted through the Clifford circuit, giving rise to a list of multi-qubit gates as $\{\alpha I+\beta P^{[k]}\}$ with twisted Pauli strings $\{P^{[k]}\}$. The GF(2) matrix $z$ (bottom left) is induced by those Pauli strings.
(b) Given a Clifford circuit plus a list of $\{\alpha I+\beta P^{[k]}\}$ acting on the initial state $\ket{mm\cdots m}\ket{0}\ket{0}\cdots\ket{0}$, for the next $\alpha I+\beta P^{[j]}$, OFD either (c) absorbs it into the MPS naively if $P^{[j]}\in\textrm{Null}(z)$, doubling the bond-dimension $\chi$ but not increasing the magic qubits, or (d) disentangles $(\alpha I+\beta P^{[j]})\ket{mm\cdots m}\ket{0}\ket{0}\cdots\ket{0}$ if $P^{[j]}\not \in\textrm{Null}(z)$, converting one free qubit into magic state and leaving the bond-dimension unchanged. Notice (c) and (d) both leave the new structure of circuit the same as in (b) as a (possibly larger) Clifford circuit plus a list of $\{\alpha I+\beta P^{[k]}\}$ acting on the initial state $\ket{mm\cdots m}\ket{0}\ket{0}\cdots\ket{0}$. The initial state has final bond-dimension exponential in the null space dimension of $z$ and a number of magic qubits equal to the rank($z$).}
\label{fig:mainfig}
\end{figure}

Both quantum entanglement and quantum magic are known to be critical for the power of quantum computing. Quantum circuits which only generate small entanglement can be simulated classically by using tensor network states with a cost that is exponential in the amount of entanglement~\cite{napp2022efficient,vidal2003efficient,vidal2004efficient,ORUS2014117}.  
States generated by Clifford circuits are stabilizer states and have no quantum magic; these circuits also can be simulated efficiently using the Gottesman–Knill theorem~\cite{Aaronson2004,gottesman1998heisenberg,nest2008classical}.   
Quantum magic can be added to these states by supplementing Clifford circuits with additional non-Clifford gates (for example, $T$ gates); such circuits become universal for quantum computing when the number of $T$ gates is polynomial in system size~\cite{nebe2001invariants,campbell2012magic,bravyi2005universal}.
As these circuits can be simulated in time exponential in the number of $T$ gates by decomposing them into a sum of Clifford, polynomial time simulations exist for 
circuits with $\mathcal{O}(\log(N))$ $T$ gates ~\cite{Bravyi2019simulationofquantum,kissinger2022simulating,beguvsic2023simulating,pashayan2022fast}.
Clifford-augmented matrix product states (CAMPS)~\cite{Lami2024b,Qian2024,Mello2024,Qian2024b} combines ideas from low-entanglement and stabilizer simulations, representing a quantum state as $\mathcal{C}\ket{\psi}$ where $\mathcal{C}$ is a Clifford circuit and $\ket{\psi}$ is an MPS (Fig.~\ref{fig:mainfig}).
In this work, we show that CAMPS can represent with polynomial complexity a large class of $t$-doped Clifford circuits with $t\leq N$ $T$ gates and consequently the expectation values of Pauli strings can be computed in polynomial time from them.
While there have been some hints of this previously from Ref.~\cite{Lami2024b} and concurrently with this work from Ref.~\cite{huang2024nonstabilizer,fux2024disentangl}, we prove that efficient simulations exist for certain circuits by giving an explicit characterization of when a Clifford+$T$ circuit is simulatable as well as describe an explicit (without optimization) algorithm which permits efficient transformation into CAMPS from the original circuit and thus the computation of the expectation values from the output state.

To characterize these circuits, we use the fact that the $T$-gate can be written as the identity operator plus another Pauli string and this structure is preserved even after we commute the $T$-gate to the beginning of the Clifford circuit. These Pauli strings can then be mapped to a GF(2) matrix. The complexity of representing the circuit with CAMPS (and then computing the expectation value of a Pauli observable for this circuit) is exponential in the dimension of the null/kernel space of this matrix. In addition, the Pauli strings which are linearly dependent on the minimal basis set of this matrix are the ones which will increase the entanglement of the MPS component of CAMPS (CMPS). 

This paper is organized as follows. After giving background on CAMPS in Sec.~\ref{sec:background}, we introduce an optimization-free disentangling algorithm (Sec.~\ref{sec:OFDone}) from which we can
explicitly characterize when a single Pauli string from a commuted $T$-gate can be disentangled.  
We then proceed to characterize when a set of twisted Pauli strings/$T$-gates can be disentangled concluding with a clean algebraic characterization (Sec.~\ref{sec:OFDmany} and~\ref{sec:GEalgor}). In Sec.~\ref{subsec:ApplyTotLessThanN} we use this characterization to consider the capability of CAMPS to represent some classes of random Clifford+$T$ circuits. Next, we consider the previous approach to disentangling CAMPS by providing improvements to accelerate it (Sec.~\ref{subsec:improvedOBD}) and comparing it to our new OFD approach (Sec.~\ref{sec:connectOFD}). In Sec.~\ref{sec:tmorethanN} we consider the application of CAMPS to random Clifford+$T$ circuits with more than $N$ $T$-gates. Finally, in Sec.~\ref{sec:othersimulation} we develop additional algorithms for computation of CAMPS beyond expectation values of Pauli observables, including sampling, probability, amplitude and R\'enyi entropy.  

\section{Background}
\label{sec:background}
We will now describe the standard approach for rewriting a $t$-doped Clifford circuit in the CAMPS representation.  Consider a Clifford circuit doped with $T$ gates where each $T$ gate can be written as $T=e^{i\pi/8} \left(\alpha I + \beta Z \right)$ with $\alpha= \cos(\pi/8)$ and $\beta=-i \sin(\pi/8)$; we can always disregard the global phase factor $e^{i\pi/8}$. Using the fact that a Pauli string $P$ can be commuted through any Clifford circuit $\mathcal{C}$ such that $P\mathcal{C}=\mathcal{C}P'$ and $P'$ is a new Pauli string
\begin{eqnarray}
P'=\mathcal{C}^\dagger P \mathcal{C} 
\end{eqnarray}
we can rewrite the output of any $t$-doped Clifford circuit with Clifford gates $\mathcal{C}$ and $t$ number of $T$ gates as 
\begin{equation}
\label{eq:camps}
\mathcal{C}\prod_{k=t}^1 \overline{T}^{[k]} \ket{0}^{\otimes N}
\end{equation}
by commuting each $T$ gate from the shallowest to the deepest layers through the Clifford circuit, where $\overline{T}^{[k]}=\alpha I+\beta \overline{Z}^{[k]}$, with a twisted Pauli string $\overline{Z}^{[k]}$ acting non-trivially on multiple qubits now due to the scrambling of Clifford gates ahead of it. This leaves us in the CAMPS representation with a Clifford circuit acting on an MPS $\ket{\psi}=\prod \overline{T}^{[k]} \ket{0}^{\otimes N}$ (CMPS).  

Although this is a valid representation, we wish to minimize the bond-dimension of the CMPS $\ket{\psi}$ and this can be done using a gauge freedom in the CAMPS representation by selecting Clifford circuit disentanglers $\mathcal{D}^{[k]}$ and writing
\begin{align}
\mathcal{C}\ket{\psi}&=  \left(\mathcal{C} \mathcal{D}^{[1]\dagger }\cdots\mathcal{D}^{[t]\dagger}\right) \left(\mathcal{D}^{[t]}\cdots \mathcal{D}^{[1]}\right) \prod_{k=t}^1\overline{T}^{[k]}\ket{0}^{\otimes N} \\
&= \left(\mathcal{C} \mathcal{D}^{[1]\dagger}\cdots\mathcal{D}^{[t]\dagger}\right) \prod_{k=t}^1\mathcal{D}^{[k]} \widetilde{T}^{[k]} \ket{0}^{\otimes N} \label{eq:T_tilde} \\
&=\mathcal{C}\mathcal{D}^{[1]\dagger}\cdots\mathcal{D}^{[t]\dagger}\ket{\psi'} 
\end{align}
where 
\begin{equation}
\widetilde{T}^{[k]}=\left(\mathcal{D}^{[k-1]}\cdots \mathcal{D}^{[1]}\right)\overline{T}^{[k]}\left(\mathcal{D}^{[1]\dagger}\cdots \mathcal{D}^{[k-1]\dagger}\right) 
\end{equation}
as the $k$-th twisted $T$-gate $\overline{T}^{[k]}$ has to commute through the first $(k-1)$ disentanglers to get into the form of Eq.~\eqref{eq:T_tilde}. As $T=\alpha I+\beta Z$, we analogously have $\widetilde{T}^{[k]}=\alpha I + \beta \widetilde{Z}^{[k]}$ where 
\begin{equation}
\label{eq:ZbartoZtilde}
\widetilde{Z}^{[k]} =\left(\mathcal{D}^{[k-1]}\cdots \mathcal{D}^{[1]}\right)\overline{Z}^{[k]}\left(\mathcal{D}^{[1]\dagger}\cdots \mathcal{D}^{[k-1]\dagger)}\right)    
\end{equation}
The goal then is to choose disentanglers $\mathcal{D}$ so that $\ket{\psi'}$
has small bond-dimensions.  
Showing how to construct these disentanglers and characterize when it is possible to keep the bond-dimension small is one of our core contributions and discussed in the next section.

\section{Disentangling Algorithm}
\subsection{Disentangling a Single Pauli String: OFD}
\label{sec:OFDone}
\begin{figure*}
\centering
\includegraphics[width=0.8\textwidth]{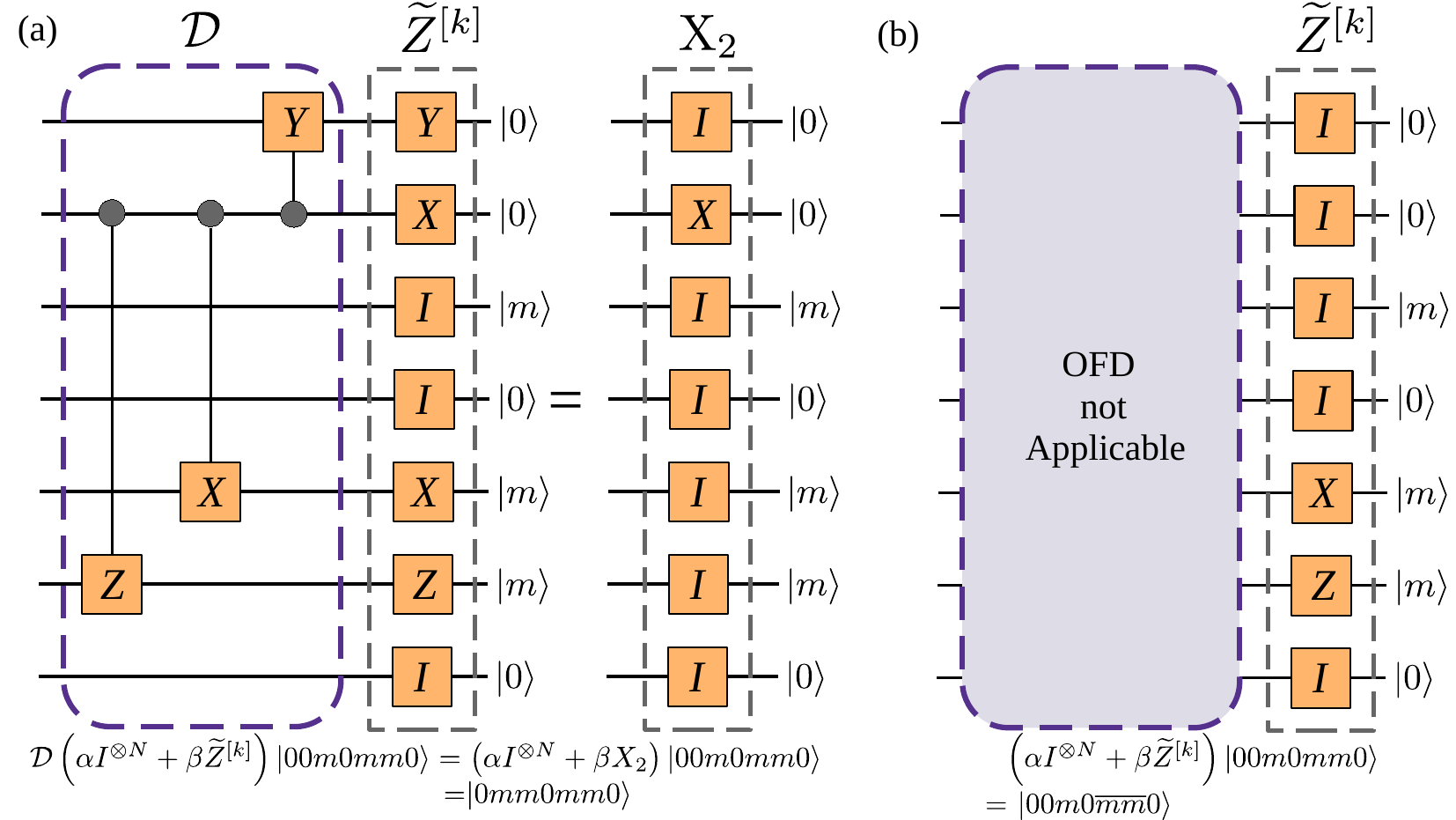}
\caption{Schematic diagram for OFD algorithm (on $N=7$ qubits). (a) for $\widetilde{Z}^{[k]}=YXIIXZI$ applying to $\ket{00m0mm0}$, choosing the second qubit as the control qubit, and constructing the appropriate disentangling circuit $\mathcal{D}$ maintain the initial state as a product state. (b) $\widetilde{Z}^{[k]}=IIIIXZI$ have multiple non-trivial Pauli terms on $\ket{m}$ but none on $\ket{0}$, OFD does not apply in this case; the resulting state has the 5th and 6th qubits entangled. }
\label{fig:opt-free-distangle}
\end{figure*}

In this section, we characterize what properties a twisted Pauli string and the CMPS it acts on need to have so that there is a disentangler which does not increase the CMPS bond-dimension (see Thm.~\ref{theo:OFD}). We also give an explicit algorithm to construct this disentangler (see the proof of Thm.~\ref{theo:OFD}, Fig.~\ref{fig:opt-free-distangle} and Algor.~\ref{alg:ofd_disentangle}).  
\begin{theorem}
\label{theo:OFD}
Call a state $\ket{\psi}$ and Pauli string $P$ \textbf{disentanglable} if $\ket{\psi}=\ket{0}_i\ket{\psi_{N\backslash i}}$ has an unentangled $\ket{0}$ state on qubit $i$\footnote{In cases where the unentangled qubits are at states such as $\ket{\pm}$ instead of $\ket{0}$, it is always possible to change them back to $\ket{0}$ using single-qubit Clifford gates from $\{H, S, X\}$.} and the $i$-th Pauli term of $P=A_i\otimes_{j\neq i}P_j$  is $A_i\in\{X_i,Y_i\}$. For any disentanglable $(P,\ket{\psi})$, there exists a disentangler 
$\mathcal{D}$ such that
\begin{equation}    
\ket{\psi'}\equiv\mathcal{D}(\alpha I + \beta P)\ket{\psi}=(\alpha I + \beta A_i) \ket{\psi} =\ket{m}_i\ket{\psi_{N\backslash i}}
\label{eq:UD_thm}
\end{equation}
where $\ket{m}\equiv\alpha \ket{0}+\beta \ket{1}$ or $\ket{m} \equiv \alpha \ket{0}+i\beta \ket{1}$ (for $A_i=X_i$ or $Y_i$, respectively).  The notation $\ket{\psi_{N\backslash i}}$ denotes the part of wave function within $\ket{\psi}$ unentangled from qubit $i$. $\alpha$ and $\beta$ in this theorem can be arbitrary scalars.
\end{theorem}

\begin{proof}
Consider a disentanglable $(P,\ket{\psi})$. Then 
\begin{equation}
    (\alpha I + \beta P)\ket{\Psi} = \alpha \ket{0}_i\ket{\psi_{N\backslash i}} + \beta'\ket{1}_i \left(\otimes_{j\neq i}P_j\right)\ket{\psi_{N\backslash i}}
    \label{eq:UD_step}
\end{equation} 
where $\beta'$ is either $\beta$ (for $A_i=X_i$) or $i\beta$ (for $A_i=Y_i$).  
We now explicitly construct the disentangler $\mathcal{D}$ with a series of control-Pauli gates. Within $\mathcal{D}$, there is a control-Pauli for each qubit $j\neq i$ and $P_j \neq I$ with control qubit $i$ and target qubit $j$ controlling the Pauli gate $P_j$:
\begin{equation}
\mathcal{D} = \prod_{\substack{j\neq i \\ P_j \neq I}}(\mathrm{CP}_j)_{i,j}   
\label{eq:U_D}
\end{equation}
Consider the r.h.s of Eq.~\eqref{eq:UD_step}.  The disentangler $\mathcal{D}$ does not affect the first term $\alpha \ket{0}_i\ket{\psi_{N\backslash i}}$ as the control gate is acting on $\ket{0}$. For the second term $\beta'\ket{1}_i \left(\otimes_{j\neq i}P_j\right)\ket{\psi_{N\backslash i}}$, the control qubits of $\mathcal{D}$ are on $\ket{1}$ and the action of each (non-identity) $P_j$ on $\ket{\psi_{N\backslash i}}$ is canceled as $(CP_j)_{i,j}P_j\ket{1}_i\ket{\psi_{N\backslash i}} = \ket{1}_i\ket{\psi_{N\backslash i}}$.  Therefore, after the application of all the control-Paulis we have $(\alpha \ket{0}_i + \beta' \ket{1}_i)\ket{\psi_{N\backslash i}}$ or $\ket{m}\ket{\psi_{N\backslash i}}$.
\end{proof}
For simplicity, we will refer to the qubits at state $\ket{0}$ as \textit{free qubits} and the qubits at state $\ket{m}$ as \textit{magic qubits}. Notice that $\ket{\psi'}$ has the same bond-dimension as $\ket{\psi}$ and the disentangler $\mathcal{D}$ consumes one free qubit at site $i$ leaving the rest of the state unaffected. Importantly, any other free qubits remain unentangled and still at state $\ket{0}$.  

Our proof of Theorem~\ref{theo:OFD} provides an algorithm (see Algorithm~\ref{alg:ofd_disentangle}), the \textbf{optimization-free disentanglement algorithm (OFD)}, to generate an explicit disentangler $\mathcal{D}$. The OFD disentangler consists of control-Pauli gates with the control qubit of all these gates on a single qubit (a $\ket{0}$ being acted on by $X/Y$) and the Pauli gate of each gate is chosen as the Pauli term from the Pauli string being disentangled.
(see Fig.~\ref{fig:opt-free-distangle}).
This explicit approach contrasts with the standard approach for disentangling a twisted $T$-gate applied to CMPS, which uses a heuristic iterative optimization that minimizes the CMPS entanglement~\cite{Lami2024b,Qian2024,Qian2024b,Mello2024,huang2024nonstabilizer,fux2024disentangl}; we call this standard approach the optimization-based disentangler (OBD). In Sec.~\ref{sec:connectOFD} we more explicitly discuss the connection between OFD and OBD. We find that OBD often will reproduce by heuristic optimization the OFD algorithm but also give explicit counter-examples where this is not the case, i.e., OBD fails while OFD succeeds (Append.~\ref{sec:failOBD}).
An alternative approach for building explicit disentanglers is the ICCR approach described in Ref.~\cite{paviglianiti2024}. Though the disentanglers share some similarities including the use of cascades of control-Pauli gates, ICCR differs from OFD in requiring an additional ancilla and is not applicable to all multi-qubit gates of the form $\alpha I+\beta P$.

\begin{algorithm}[H]
\caption{Optimization-free Disentangling Algorithm (OFD)}\label{alg:ofd_disentangle}
\begin{flushleft}
\hspace*{\algorithmicindent} \textbf{Input}: $N$-qubit CMPS $\ket{\psi}$, $N$-qubit Pauli string $P$.
\end{flushleft}
\begin{algorithmic}[1]
\State Initialize: Clifford disentangler ${\mathcal{D}}=I^{\otimes N}$. 
\For{$n=1,2,\cdots,N$}
\If {$\ket{\psi [n]}=\ket{0}$ and $P[n]\in\{X,Y\}$}
\For{$m=1,2,\cdots,n-1,n+1,\cdots,N$}
\If{$P[m]\in\{X,Y,Z\}$}
\State{${\mathcal{D}}\gets {\mathcal{D}}\cdot(\mathrm{CP}[m])_{n,m}$;}
\EndIf
\EndFor
\State{\textbf{break for loop};}
\EndIf
\EndFor
\end{algorithmic}
\begin{flushleft}
\hspace*{\algorithmicindent} \textbf{Return}: Disentangler $\mathcal{D}$.
\end{flushleft}
\end{algorithm}  

\subsection{Disentangling Many Pauli Strings}
\label{sec:OFDmany}
In this section, we will consider applying the twisted $T$ gates $\{ \widetilde{T}^{[k]}\}$ sequentially to the CMPS and disentangling the resulting CMPS after each application. If $\widetilde{Z}^{[k]}$ and the current CMPS is disentanglable, then the CMPS bond-dimension stays constant and a free qubit is used as a control qubit and consumed. Otherwise, the CMPS bond-dimension at most doubles and no free qubits are consumed, following from the theorem below.
\begin{theorem}
If $(P,\ket{\psi})$ is not disentanglable, then $\ket{\psi'}\equiv(\alpha I+\beta P)\ket{\psi}$ has (i) an MPS representation with at most double of the bond dimension from $\ket{\psi}$ and (ii) the same number of free qubits as $\ket{\psi}$. 
\end{theorem}
\begin{proof}
$(P,\ket{\psi})$ being not disentanglable, implies that all the $\ket{0}$ qubits from $\ket{\psi}$ are acted on by the $I$ or $Z$ Pauli terms from $P$. Then we can rewrite them as
\begin{eqnarray}
\ket{\psi}&&=\ket{0}_{J_1}\ket{0}_{J_2}\ket{\psi_{N\backslash J}}\\P&&=\otimes_{j\in J_1}I_j \otimes_{j\in J_2}Z_j\otimes_{k\notin J}P_k \\
&&\equiv I_{J_1}\otimes Z_{J_2}\otimes P_{N\backslash J}        
\end{eqnarray}
where $J=J_1\cup J_2$ is the set of qubits at state $\ket{0}$, and $J_1$ and $J_2$ are the sets of qubits with $I$ and $Z$ acted on from $P$, respectively. Next, we have
\begin{equation}
\begin{split}    
\ket{\psi'}&=(\alpha I+\beta I_{J_1}\otimes Z_{J_2}\otimes P_{N\backslash J})\ket{0}_J\ket{\psi_{N\backslash J}}\\
&=\ket{0}_J(\alpha\ket{\psi_{N\backslash J}}+\beta P_{N\backslash J}\ket{\psi_{N\backslash J}})
\end{split}
\label{eq:add_mps}
\end{equation}
The property (ii) is obvious from Eq.~\eqref{eq:add_mps}; and property (i) follows because both $\ket{\psi_{N\backslash J}}$ and $P_{N\backslash J}\ket{\psi_{N\backslash J}}$ have the same entanglement rank as $\ket{\psi}$ and therefore their sum has at most double of the entanglement rank.
\end{proof}

Therefore, the CMPS bond dimension will be exponential in the number of twisted Pauli strings which cannot be disentangled. Because there are $N$ initial unentangled $\ket{0}$'s to consume, it is possible (in the best case) that we could convert the circuit into CAMPS and compute the expectation value of up to $N+\log(N)$ $T$-gates doped in a Clifford circuit (the additional $\log(N)$ coming from the number of entangling $T$-gates we can simulate in polynomial time using entangled CMPS). 

Although we now have an explicit prescription for which sets of twisted Pauli strings $\{\widetilde{Z}^{[k]} \}$ acting on $\ket{0}^{\otimes N}$ can be written as compact CAMPS (e.g. there are enough disentanglable twisted Pauli strings), this characterization is somewhat implicit.  For example, both the twisted Pauli string $\widetilde{Z}^{[4]}$ and the CMPS it acts on depend on all the previous disentanglers $\mathcal{D}^{[1]},\cdots, \mathcal{D}^{[3]}$, in essence requiring us to run OFD for the first three Pauli strings before determining whether the fourth Pauli string is disentanglable.  It would be preferable if there were a characterization which depended only on the twisted Pauli strings $\{\overline{Z}^{[k]} \}$ without running OFD to get $\{\widetilde{Z}^{[k]} \}$. Such a characterization will be built in the next subsection.

\begin{algorithm}[H]
\caption{OFD on $t$-doped Clifford circuits (OFDS)}\label{alg:disentangle}
\begin{flushleft}
\hspace*{\algorithmicindent} \textbf{Input}: A sequence of $N$-qubit Pauli strings $\{\overline{Z}^{[k]}\}_{k=1}^t$ as generated in Eq.~\eqref{eq:camps}.
\end{flushleft}
\begin{algorithmic}[1]
\State Initialize: Disentangling Clifford circuit ${\mathcal{D}}=I^{\otimes N}$, $N$-qubit CMPS $\ket{\psi}=\ket{0}^{\otimes N}$.
\For{$k=1,2,\cdots,t$}
\State $\mathcal{D}^{[k]}=\text{OFD}(\ket{\psi},\overline{Z}^{[k]})$.
\State $\ket{\psi}\gets \mathcal{D}^{[k]}(\alpha I+\beta \overline{Z}^{[k]})\ket{\psi}$.
\For{$j=k+1,k+2,\cdots,t$}
\State $\overline{Z}^{[j]}\gets\mathcal{D}^{[k]} \overline{Z}^{[j]}\mathcal{D}^{[k]\dagger}$.
\EndFor
\State $\mathcal{D}\gets\mathcal{D}\cdot\mathcal{D}^{[k]\dagger}$.
\EndFor
\end{algorithmic}
\begin{flushleft}
\hspace*{\algorithmicindent} \textbf{Return}: Disentangler $\mathcal{D}$, CMPS $\ket{\psi}$.
\end{flushleft}
\end{algorithm}  

\subsection{Characterizing Disentanglable Pauli Strings}
\label{sec:GEalgor}

In this section, we show that the null space of the GF(2) matrix induced by the Pauli strings $\overline{Z}$ under the map $\{I,Z\}\rightarrow 0$ and $\{X,Y\}\rightarrow 1$ characterizes which Pauli strings increase the bond-dimension of the CAMPS thereby bounding the cost to represent Clifford+$T$ circuits with CAMPS (and compute Pauli expectation values) to be exponential in the dimensional of this null space.  Therefore, any Clifford+$T$ circuit for which the null space dimension is less than $\log(N)$ can be represented by a polynomial-size CAMPS from which expectation values can be efficiently evaluated. The key understanding needed here is how the twisted Pauli strings $\overline{Z}$ and $\widetilde{Z}$ (for which we know the disentangable condition, see Thm.~\ref{theo:OFD}) are related.

To accomplish this, we start with the following theorem which tells us how a Pauli string $P$ is transformed when commuting through a disentangler $\mathcal{D}$.  
\begin{theorem}
Consider a disentangler $\mathcal{D}$ generated from OFD for twisted Pauli string $\widetilde{Z}$ with control qubit $r$.  Let $P$ be an arbitrary Pauli string and $P'=\mathcal{D}^{\dagger} P \mathcal{D}$. Let $p$, $p'$ and $\widetilde{z}$ be the binary representations of $P$, $P'$ and $\widetilde{Z}$, respectively, and $1_r$ be the bit string with an 1 in the $r$-th place and 0 otherwise. Then
\begin{equation}
p' = \begin{cases}
p \oplus  \widetilde{z} \oplus 1_r, & \text{if } p_r=1 \\
p, & \text{if } p_r=0
\end{cases}    
\end{equation}
\label{theo:Commute-BitRep}
\end{theorem}
\begin{proof}
\begin{table}[h]
\caption{\label{tab:pauli-controlP}%
Transformation of two-qubit Pauli strings under the conjugation with control-Pauli gates. 
}
\begin{ruledtabular}
\begin{tabular}{lccr}
$P$ &
Control-$X$&
Control-$Y$&
Control-$Z$\\
\colrule
$II$ & $II$ & $II$ & $II$\\
$IX$ & $IX$ & $ZX$ & $ZX$\\
$IY$ & $ZY$ & $IY$ & $ZY$\\
$IZ$ & $ZZ$ & $ZZ$ & $IZ$\\
$XI$ & $XX$ & $XY$ & $XZ$\\
$XX$ & $XI$ & $YZ$ & $YY$\\
$XY$ & $YZ$ & $XI$ & $YX$\\
$XZ$ & $YY$ & $YX$ & $XI$\\
$YI$ & $YX$ & $YY$ & $YZ$\\
$YX$ & $YI$ & $XZ$ & $XY$\\
$YY$ & $XZ$ & $YI$ & $XX$\\
$YZ$ & $XY$ & $XX$ & $YI$\\
$ZI$ & $ZI$ & $ZI$ & $ZI$\\
$ZX$ & $ZX$ & $IX$ & $IX$\\
$ZY$ & $IY$ & $ZY$ & $IY$\\
$ZZ$ & $IZ$ & $IZ$ & $ZZ$
\end{tabular}
\end{ruledtabular}
\end{table}
\begin{table}[h]
\caption{\label{tab:bin-controlP}%
Transformation of bitstring representation for two-qubit Pauli strings under the conjugation with control-Pauli gates.
}
\begin{ruledtabular}
\begin{tabular}{lcr}
$p$ &
Control-$0$\footnote{control-0: control-Z gate.}&
Control-$1$\footnote{control-1: control-X and control-Y gates.}\\
\colrule
$00$ & $00$ & $00$\\
$01$ & $01$ & $01$\\
$10$ & $10$ & $11$\\
$11$ & $11$ & $10$
\end{tabular}
\end{ruledtabular}
\end{table}

First, let us consider what the disentangler $\mathcal{D}$ with control qubit $r$ looks like. For all non-identity Pauli term $\widetilde{Z}_{j\neq r}$ within $\widetilde{Z}$,  we have a $(C\widetilde{Z}_j)_{rj}$ gate in $\mathcal{D}$.  As shown in Tab.~\ref{tab:pauli-controlP} and Tab.~\ref{tab:bin-controlP}, if $p_r=0$, conjugating $P$ with any type of control-Pauli gates will not affect the bitstring for the resulting $P'$ - i.e. $0x \rightarrow 0x$. Therefore, $p'=p$.  If $p_r=1$, then the CX and CY gates flip target bits $p_j$ in $p$; notice that the target qubits of CX and CY gates are where $\widetilde{z}_{j\neq r}=1$. This is then equivalent to $p'=p \oplus \widetilde{z}$ for all but the control bit $r$. The control bit never flips in this conjugating transformation but gets flipped in $p'=p \oplus \widetilde{z}$, to compensate for this difference, we flip it back with the additional $1_r$, leading to $p'=p \oplus \widetilde{z} \oplus 1_r$.
\end{proof}

Next, we will show that the OFD is closely related to Gaussian elimination and the way in which earlier disentanglers affect later twisted Pauli strings (e.g. the transformation from $\overline{P}^{[k]}\rightarrow \widetilde{P}^{[k]}$ as it commutes through $D^{[j]}$ where $j<k$) is analogous to the step of Gaussian elimination where the row below the current pivot is zeroed out). 

Let us define $\overline{z}$ as the binary matrix with each row initially as the bitstring corresponding to $\overline{Z}^{[k]}$:
\begin{equation}
z=
\begin{pmatrix}
\horzbar \overline{z}^{[1]} \horzbar \\
\horzbar \overline{z}^{[2]} \horzbar \\
\horzbar \dots \horzbar \\
\horzbar \overline{z}^{[t]} \horzbar
\end{pmatrix}
\end{equation}

Below we describe simultaneously three algorithms which perform essentially the same operations: standard Gaussian elimination (GE) to reduce the matrix $z$ to reduced row echelon form; a small variant of Gaussian elimination (VGE); and the transformations induced by the implementation of OFD on $\{\widetilde{Z}^{[k]}\}$ (OFDS).  
In particular, we will see that VGE and OFDS generate the same binary matrix at each step and perform the \textit{same} row-operations as GE does. 
Here the \textit{same} row-operations are defined as swapping the same columns and doing the ``adding'' and replacement of the same rows, where ``adding'' will be defined differently for GE (step iiia) and VGE/OFDS (step iiib).  
When simulating OFDS, we will preserve the invariant (by qubit swapping) that when working on twisted Pauli string $\widetilde{Z}^{[k]}$, qubits $j<k$ are the control qubits consumed in previous $(k-1)$ steps.  
\\
\noindent \textbf{Algorithm(s):} Assume we are working on the $k$-th row of $z$ (GE/VGE) or equivalently Pauli string $\widetilde{Z}^{[k]}$ (for OFDS) that have been transformed from $\overline{Z}^{[k]}$ via commuting through all the disentanglers for previous Pauli strings (Eq.~\eqref{eq:ZbartoZtilde}, note that $\widetilde{Z}^{[1]}=\overline{Z}^{[1]}$).\\ \\
\noindent (i) If it exists, GE/VGE chooses a pivot column $j\geq k$ with $z^{[k]}_{j}=1$. Equivalently, OFDS chooses the same $j$ as the control qubit for $\widetilde{Z}^{[k]}$. 
This is always consistent because, based on Thm.~\ref{theo:Commute-BitRep}, our algorithms preserve the property that the available columns $j(\geq k)$ with $z^{[k]}_{j}=1$ in GE/VGE correspond to the free qubits in OFDS with $\widetilde{Z}_j^{[k]} \in \{X,Y\}$.   
If there is no such column $j$ (and hence no control qubit available), move to step \textit{(nd)}.\\

\noindent (ii) Swap column $j$ and $k$ in GE/VGE;  this corresponds in OFDS to swapping qubit $j$ and $k$ which we will permit without loss of generality. The control qubit is now at position $k$ with $z^{[k]}_{k}$. This preserves the property that the control qubits are in $j\leq k$. At this step, the disentangler $\mathcal{D}^{[k]}$ consists of a SWAP gate only.\\

\noindent (iiia) \textit{GE only:} For all rows $j>k$ where $z^{[k]}_{j}=1$, take $z_{j}=z^{[j]} \oplus z^{[k]}$.  Define this step as the ``adding'' and replacing row-operation for GE.\\

\noindent (iiib) \textit{VGE or OFDS}: In VGE, for all rows $j>k$ where $z^{[j]}_k=1$, take $z^{[j]}=z^{[j]}\oplus z^{[k]} \oplus 1_k $. Define this step as the``adding'' and replacing row-operation for VGE.  Steps \textit{(iiia)} and \textit{(iiib)} are the only difference between GE and VGE, but only lead to different lower triangular (below diagonal) part of the matrix $z$ after the operations. 
This is obvious from the following facts: (a) when GE/VGE do the same row-operation to row $j$, only the bits below the diagonal $z^{[j]}_k$ ($j>k$) are different; (b) the row operations to perform only depend on the bits in the upper triangular (including diagonal) part of $z$ (these bits correspond to the free qubits when dealing with each $\widetilde{Z}^{[k]}$ in OFDS). A row operation that makes the implementations of these two algorithms on the same initial $z$ diverge will never happen, since this would require the upper triangular part to be different, which in turn would have required different row operations prior to this point.
In OFDS, at this step we can construct the disentangler $\mathcal{D}^{[k]}$ consisting of control-Pauli gates $\{(\mathrm{C}\widetilde{Z}^{[k]}_j)_{k,j}\}$ for those non-trivial Pauli terms in $\widetilde{Z}^{[k]}$. Now we commute all the rest $\overline{Z}^{[j]}$ ($j>k$) through $\mathcal{D}^{[k]\dagger}$ (Eq.~\eqref{eq:ZbartoZtilde}). By Thm.~\ref{theo:Commute-BitRep}, the effect of this commutation is that for all bit-strings $z^{[j]}$ ($j>k$) with $z^{[j]}_k=1$, we have that  $z^{[j]} = z^{[j]} \oplus z^{[k]} \oplus 1_k $. Note that this is exactly the transformation of VGE.  

We now continue with step \textit{(i)} with row $k \rightarrow k+1$.\\

\noindent (n.d.) If $z^{[k]}_j=0$ for all  $j\geq k$ then this corresponds to a vector in the null space of the matrix $z$ (for GE/VGE) or a Pauli string $\widetilde{Z}^{[k]}$ which is not disentanglable (for OFDS).   Typically, in standard Gaussian elimination, this row would be swapped to the bottom of the matrix; here instead (for GE/VGE and implicitly for OFDS as the twisted Pauli string is applied to the CMPS sequently) we leave the row in its current position but relabel it from row $k$ to row $nd_i$ (for the $i$-th not disentanglable string) and reorder the remaining rows starting from row $k$. This relabeling simplifies the exposition of these algorithms by keeping the control qubit at position $k$ for $k$-th disentanglable Pauli string.  \\ \\

Each of our steps preserves the properties that the binary matrix $z$ is the same between VGE and OFDS and the row-operations (Pauli transformation) are performed simultaneously to the same rows (Pauli strings) for all three methods (GE/VGE/OFDS). Therefore the rank of the resulting GF(2) matrix $z$ from GE, which is the number of rows with $z^{[k]}_k=1$ both in GE and VGE, corresponds to the number of twisted Pauli strings in $\{\overline{Z}^{[k]}\}$ that can be disentangled by OFDS. The rows for which  $z^{[k]}_k \neq 1$ in GE and VGE are the twisted Pauli strings not disentanglable by OFDS and the null space of the Gaussian elimination.

In other words, by simply performing GE on $z$ (generated from $\overline{Z})$, we learn which Pauli strings can be disentangled by OFDS and therefore that the complexity of representing Clifford+$T$ circuits with CAMPS is exponential in the dimension of the null space. 

From this identification, we can also conclude some additional interesting properties about OFDS.   The null space dimension of a GF(2) matrix is a property of the rows of the matrix and does not depend on the pivot chosen in doing Gaussian elimination nor the ordering of the rows. The null space itself does not depend on the pivot.  Therefore, the disentanglable strings and the CMPS bond-dimension does not depend on the control qubit chosen in OFDS (here we assume that the disentanglable strings have the maximal effect on the bond-dimension). The maximal CMPS bond-dimension does not depend on the Pauli string order, although when constructing the CMPS, we need to disentangle the Pauli strings with the given order (as one cannot generically reorder $\{\alpha I+\beta P^{[k]}\}$ except in cases where they commute such as when the $T$-gates are outside each other's lightcone).  

Note that the above algorithm bounds the maximal bond-dimension of running OFDS.  There are some situations where the bond-dimension could be less.   First, if the row $k$ is a not disentanglable Pauli string, and there is only one non-idenity Pauli term on site $j<k$, or multiple Pauli $Z$'s on site $j\geq k$, this string does not increase the bond-dimension either; note that one cannot tell this just from the binary string. Secondly, some undisentanglable strings might not double the maximal bond-dimension, such as those Pauli strings entangling qubits in separate regions. It is also possible that the consumed qubits turn from $\ket{m} \rightarrow \ket{0}$ by OFDS when two $T$-gates combine into a Clifford gate in the original circuit.  

\subsection{Application to Random $(t\leq N)$-doped Clifford Circuits}\label{subsec:ApplyTotLessThanN}
We have previously given an explicit characterization of disentangablity of multi-qubit twisted Pauli strings and shown the cost of representing Clifford+$T$ circuits with CAMPS (and then computing the expectation of Pauli observables) is exponential in the number of entangling Pauli strings (or $T$ gates). 

In this section we focus on different classes of random Clifford+$T$ circuits to give evidence for the expected number of entangling twisted Pauli strings generated from such circuits.  
Let us consider circuits composed of layers of random two-qubit Clifford gates in a brick-wall pattern. The twisted Pauli strings $\overline{Z}^{[k]}$ for $T$-gates at depths $d$   will have (some) non-identity Pauli terms on a window of qubits with width $w=4d$ caused by the induced light-cone; this width significantly effects the relevant rank.

We start by considering $T$-gates at large depth ($d\geq N/4$) whose light-cone spans the entire system and consider a simplified model, which can be addressed analytically, where the distribution of each $\overline{Z}^{[k]}$ is uniform over all $4^N$ Pauli strings. While formally such $T$ gates may induce some correlated distribution over their strings, the numerics (see below) suggest the effects of this minimal.  
Of course, one could append a deep random Clifford circuit to each $T$-gate to exactly generate this distribution of twisted Pauli-strings~\cite{fux2024disentangl}. 
The number of disentanglable strings is then the rank of a random matrix over $\mathrm{GF}(2)$; the average rank of such a matrix is $N-\gamma$ (with $\gamma\approx 0.85$) with a very low probability of having a rank less than $N-4$ for all $N \geq 4$  (see Fig.~\ref{fig:rankprob})~\cite{kolchin1999random}. Alternatively, the expected number of twisted Pauli strings needed to consume all the free qubit in CAMPS is no more than $N+2$ (see Append.~\ref{sec:derivetfree}). In the limit of uniform Pauli strings, the complexity of representing $(t\lesssim N)$-doped Clifford circuits with CAMPS (e.g. deep enough $T$ gates) is therefore polynomial in $N$~\footnote{The computational complexity is dominated by the manipulation of Clifford stabilizer and CMPS.}.  
Note that recently Ref.~\cite{fux2024disentangl} also used the probability of successfully encoding $T$ gates into CAMPS to estimate the expected number of $T$ gates before disentangling fails, finding a similar result. 
   
In cases where the window-size is smaller, it is harder to analytically reason about the number of entangling gates. Instead, we numerically generate the twisted Pauli strings to determine their rank. This is done by commuting $T$ gates through random Clifford gates; alternatively, we considered random Pauli strings in the induced window size of $4d$ with nearly identical results (see Append.~\ref{sec:moreNumericOntlessN}) 

\begin{figure}
\centering
\includegraphics[width=0.50\textwidth]{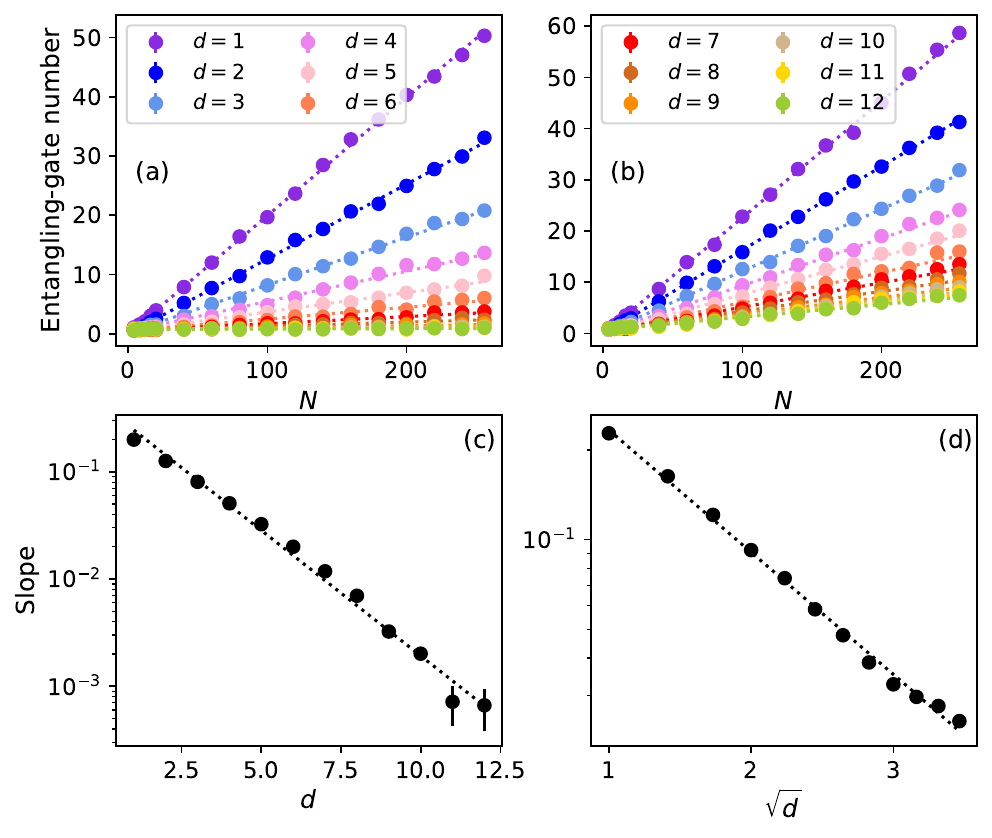}
\caption{Average number of entangling gates versus qubit number $N$ for OFDS at various $T$-gate depths $d$: (a) Clifford+$T$ circuits with $N$ $T$-gates being put at depth $d$; (b) Clifford+$T$ circuits with $N/2$ $T$-gates being randomly put at depth $d$ and another $N/2$ $T$-gates being randomly put at depth $d+1$. (c)/(d) The slopes of the fitting lines from (a)/(b) versus $d$ (or $\sqrt{d}$). }
\label{fig:entanglingGateModel}
\end{figure}

Of primary importance here is not only a bound on the window-size (generically larger window sizes will have larger rank) but also their location.   We consider two different $T$-gate locations with varying uniformity: either $N$ $T$-gates at depth $d$ or $N/2$ $T$-gates on random (possibly overlapping) sites at both depths $d$ and $d+1$. In both cases, we observe from Fig.~\ref{fig:entanglingGateModel} that the entangling-gate number grows linearly with system size for a fixed $d$ with a slope which decays exponentially with $d$  and $\sqrt{d}$ respectively.  The scaling from the numerical results would indicate that all depths greater than $d_{\text{min}}(c)$ have a constant number ($c$) of entangling gates independent of $N$ with 
\begin{equation}
d_{\text{min}}(c) = \frac{b}{a}+\frac{1}{a}\log_{10}\frac{N}{c}    
\end{equation}
or 
\begin{equation}
d_{\text{min}}(c) = \left(\frac{b}{a}+\frac{1}{a}\log_{10}\frac{N}{c} \right)^2   
\end{equation}
for the circuit ensembles in Fig.~\pref{fig:entanglingGateModel}{a} and Fig.~\pref{fig:entanglingGateModel}{b}, respectively. Here $a$ and $b$ are the linear fitting coefficients for Fig.~\pref{fig:entanglingGateModel}{c} or Fig.~\pref{fig:entanglingGateModel}{d}, i.e., $\log_{10}(\text{Slope})=-a\cdot d+b$.

In the opposing limit where uniformity of $T$-gates is strongly violated, we can see that the rank will be large.  Consider the case where the windows from all $N$ $T$-gates actually span only a sub-linear number of $k$ qubits; then the GF(2) matrix only has non-zeros on those $k$ columns and its rank is bounded by $k$.
It is worth pointing out that this ``uniformity'' requirement is on the location of the Pauli-string windows and so $T$-gates that are all to the left-half of the system but at varying depths still are likely to be sufficiently uniform in window-location given the spread of the light cone. 

Here we have focused on $T$-gates which are deep in the circuit.  Interestingly if all the $T$-gates are shallow (e.g. less than $\log(N)$ depth) then it is also true that CAMPS can represent them efficiently as the first $\log(N)$ layers can all be absorbed into a polynomial size MPS.

\section{Optimization-based Disentangling Algorithm}\label{sec:OBD}
\subsection{Improved Optimization-based Disentangling Algorithm}\label{subsec:improvedOBD}
OFD is effective when there are free qubits left, which is often the case for random $(t\leq N)$-doped Clifford circuits. However, once all the free qubits are consumed as control qubits, OFD is no longer applicable. From this section, we consider the Clifford circuits doped with $(t\geq N)$ $T$-gates, and the application of an alternative disentangling algorithm. To start with, we describe the optimization-based disentangling algorithm (OBD) and present a modification to this algorithm which results in a substantial speed-up -- by several orders of magnitude over the original approach. 
OBD has been utilized in the Clifford disentangling procedures from Ref.~\cite{Lami2024b,Qian2024,Qian2024b,Mello2024,huang2024nonstabilizer,fux2024disentangl}. The main idea is to iteratively examine each pair of neighboring qubits and select the optimal gate from the two-qubit Clifford group to minimize the entanglement entropy over the cut between the pair of qubits. Our core modification will be a way to more efficiently compute the objective function for this optimization.

After moving the canonical center of CMPS to the cut between qubit $n$ and $n+1$, the original OBD algorithm uses the entanglement entropy over this cut as the objective function; this entanglement entropy is computed by obtaining singular values through the singular value decomposition (SVD), with a time complexity of $\mathcal{O}(\chi^3)$ and a memory complexity of $\mathcal{O}(\chi^2)$ (where $\chi$ is the MPS bond dimension), for each two-qubit Clifford gate $U$. 

We show instead that it is more efficient to compute the exponential of the second R\'enyi entropy using the following tensor contraction  
\begin{equation}
\mathcal{L}(U, n)=e^{-S_2(U, n)}=
\includegraphics[width=0.25\textwidth, valign=c]{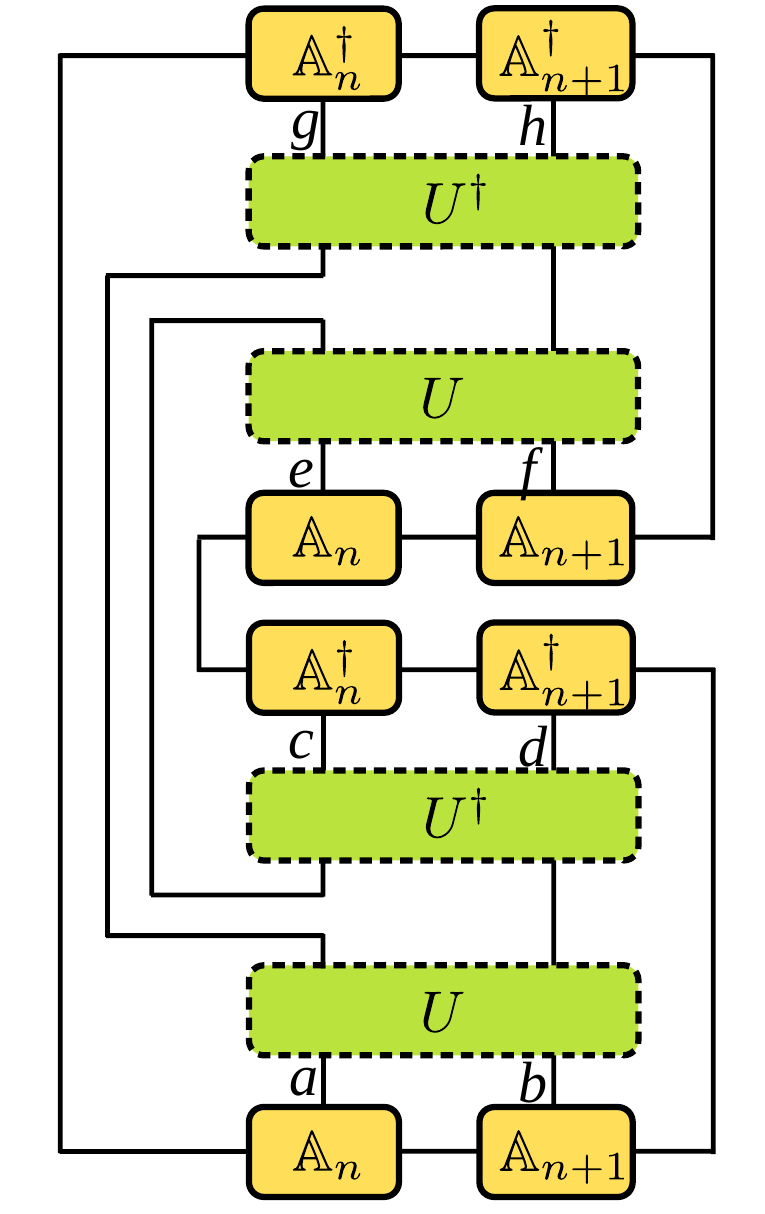}
\label{eq:loss}
\end{equation}
where $\mathbb{A}_n$ represents the tensor for qubit $n$ in the CMPS, and $S_2(U,n)$ denotes the second R\'enyi entropy at the cut between qubit $n$ and $n+1$ after the application of $U$ to qubit $n$ and $n+1$. 

One can then compute $S_2$ for all 2-qubit Clifford gates in time $\mathcal{O}(\chi^3)+\mathcal{O}(|\text{Cl}_2|)$ ($|\text{Cl}_2|=11,520$ is the size of the two-qubit Clifford group~\cite{gidney2021stim})\footnote{Here we write the complexity as a sum of two $\mathcal{O}$'s to emphasize the additive cost from the computation for the Clifford gates.}.
This can be done by first contracting the tensors $\mathbb{A}_n$ and $\mathbb{A}_{n+1}$, leaving physical bonds (labelled as letters $\{a,b,c,d,e,f,g,h\}$ in Eq.~\eqref{eq:loss}) uncontracted. This preprocessing step requires $\mathcal{O}(\chi^3)$ time and $\mathcal{O}(\chi^2)$ memory. The resulting tensor, with $2^8=256$ elements, is then dispatched to a tensor-contraction computation for each two-qubit Clifford gate $U$. Notably, this step has a time and memory complexity of $\mathcal{O}(1)$, independent of the MPS bond dimension $\chi$. 

Overall, for sequential computation, the original method has a time complexity involving a multiplicative product $|\text{Cl}_2|\chi^3$ while in the improved version these terms contribute to the complexity additively as $\chi^3+|\text{Cl}_2|$.  If parallel computation is employed for each two-qubit Clifford gate $U$, the memory cost goes from including a multiplicative contribution $|\text{Cl}_2|\chi^2$ to an additive contribution $\chi^2+|\text{Cl}_2|$ 

Finally, the OBD is completed by sweeping over the CMPS from qubit 1 to $N$, identifying the optimal Clifford gate for each qubit pair. These sweeps continue until convergence, resulting in a disentangling Clifford circuit $\mathcal{D}$ in a stair-step pattern~\cite{Lami2024b,Qian2024,Qian2024b,Mello2024,huang2024nonstabilizer,fux2024disentangl}.

\subsection{Connection between OFD and OBD}
\label{sec:connectOFD}

\begin{figure*}
\centering
\includegraphics[width=\textwidth]{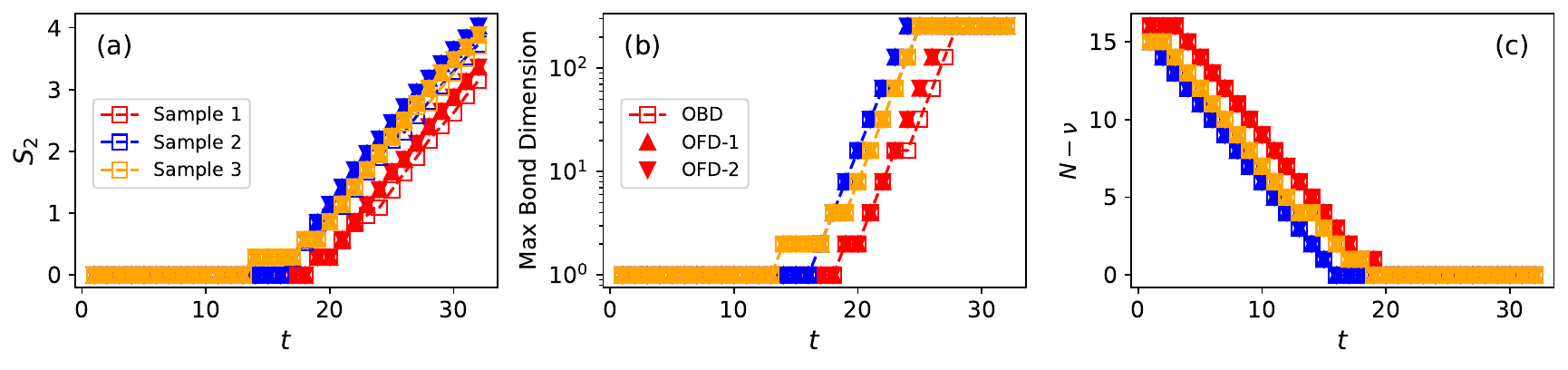}
\caption{Results of applying different types of disentangling algorithms (square, upward and downward triangular markers) to three samples of random Clifford+$T$ circuits (red, blue, and yellow colors) with $N=16$ qubits and $N_T=1$ $T$-gate at every layer: (a) the second R\'enyi entropy of the CMPS; (b) the maximal bond dimension of CMPS; (c) the nullity $\nu$ of the CMPS (see Append.~\ref{sec:stablearn}). Square markers: OBD; upward triangular markers: OFD with control qubit chosen from leftmost; downward triangular markers: OFD with control qubit chosen from rightmost.}
\label{fig:optfree}
\end{figure*}

In Fig.~\ref{fig:optfree}, we compare the performance of both OBD and OFD on the same random Clifford+$T$ circuits. For these examples, when $t\leq N$, OBD and OFD perform identically. Unsurprisingly, as OFD stops disentangling the qubits when all the free qubits are consumed (e.g. when $t>N$) we observe that OBD sometimes performs better as it is still able to disentangle CAMPS occasionally. In previous literature~\cite{Lami2024b,huang2024nonstabilizer,fux2024disentangl} it is also shown that OBD often disentangles many of the first $N$ twisted Pauli strings generated from $T$-gates. This motivates understanding the relation between OBD and the OFD introduced in this work.

Here we argue that the similar performance happens because OBD's heuristic optimization will often (but not always) essentially find the OFD disentangler (albeit with swap gates used to emulate long-range gates and a possibility of having multiple control-bits).  It is easiest to see this similarity in cases where the twisted Pauli strings are all disentanglable.  In cases where only some strings are disentanglable, the CMPS can build up an entanglement structure which prevents OBD's optimization from finding the OFD disentangler for disentanglable strings;  see Append.~\ref{sec:failOBD} for a concrete example of where this happens. 

\begin{figure}
\centering
\includegraphics[width=0.3\textwidth]{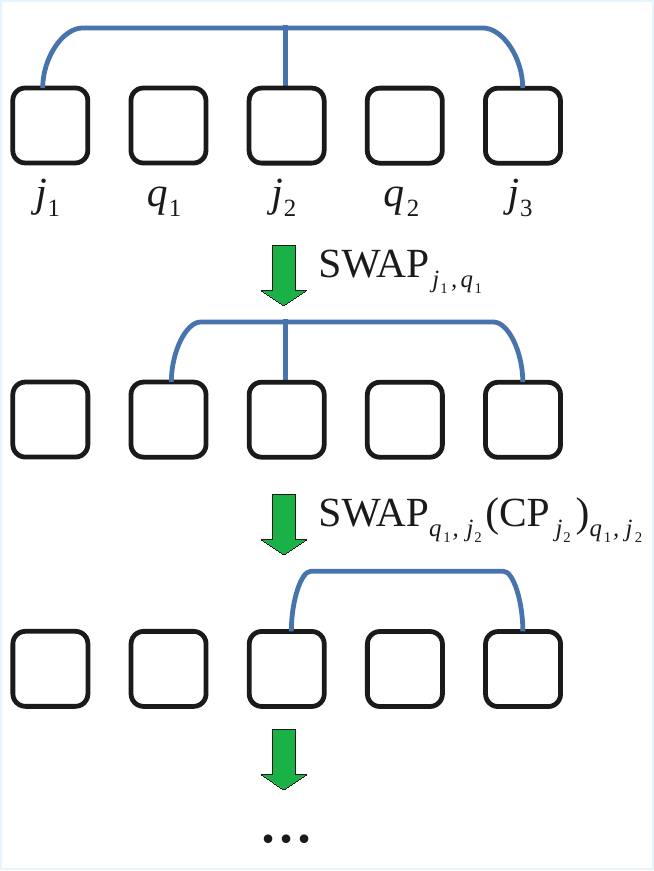}
\caption{Illustration of OBD procedure removing entanglement within CMPS bond by bond. $\{j_1,j_2,j_3\}=J$ is the qubit set entangled by the twisted Pauli string, indicated by the connection of wires, $\{q_1,q_2\}=N\backslash J$ are the remaining qubits. Qubit $j_1$ is assumed to be at state $\ket{0}$ initially so that it can be a control qubit.}
\label{fig:OBD_procedure}
\end{figure}

Here we argue why OBD generally finds essentially the OFD disentangler focusing on the case where all strings are disentanglable.  Assume the twisted Pauli strings $\{\overline{Z}^{[k]}\}$ come from $T$-gates (so that the coefficients $\alpha$ and $\beta$ are fixed) and are all disentanglable according to Thm.~\ref{theo:OFD}. When it comes to the $k$-th twisted Pauli string $\overline{Z}^{[k]}$, the CMPS is still a product state $\ket{\psi}=\otimes_{i=1}^N \ket{\psi_i}$, with $\ket{\psi_i}=\ket{0}\text{ or }\ket{m}$. Let $J$ be the subset of qubits which are acted on non-trivially by the Pauli terms from $\widetilde{Z}^{[k]}$. These qubits are entangled in the resulting CMPS $\ket{\psi'}=(\alpha I+\beta \widetilde{Z}^{[k]})\ket{\psi}$. As shown in Fig.~\ref{fig:OBD_procedure}, when OBD looks at a qubit pair where one of them belongs to $J$, and the other belongs to $N\backslash J$, OBD chooses the SWAP gate to remove the entanglement at the cut between two qubits, effectively moving the qubits in $J$ closer (one such qubit at the edge is always movable in this way); when OBD looks at a qubit pair where both belong to $J$, and one of them can be a control qubit, then a control-Pauli gate (followed by a SWAP gate, if necessary) is applied to remove the entanglement at this cut.   If neither of them can be used as a control qubit (i.e., two $\ket{m}$ qubits), then OBD generically does nothing (although in some cases OBD can remove some entanglement). If both belong to $N\backslash J$, OBD does nothing.  By a sequence of these moves, OBD then will be able to typically disentangle this string.  

Notice that this disentangler consists of control-Pauli's and SWAPs but can differ from the OFD disentangler both in the trivial way where SWAPs are needed to mediate the long-range control-Pauli gates from OFD, and also non-trivially by having multiple control-qubits in OBD while OFD uses only one. These differences in disentangler will induce different twisted Pauli strings $\widetilde{Z}^{[k]}$ when the remaining $\overline{Z}^{[k]}$ commute through them. 
It is then important to ask whether the different choices of disentanglers from OBD or OFD affect the disentanglability of remaining twisted Pauli strings. 
The rank of the GF(2) matrix of the Pauli substrings on the remaining free qubits along with the current control qubit, determines the number of disentanglable Pauli strings left.
Notice that, regardless of the choices of OBD or OFD, the disentangler will target the flipped free qubits with control-$X/Y$ gates, except the one which will become a magic qubit after disentangling. For these gates, including the SWAPs, the transformation on the GF(2) submatrix for the free qubits is reversible, implying that the rank of the GF(2) submatrix is unchanged.   

\section{Application to Random $(t>N)$-doped Clifford Circuits}
\label{sec:tmorethanN}
In this section, we show numerical results for various class of random Clifford+$T$ circuits with $T$-gate number $t>N$. An important quantity we will mainly display is the entanglement entropy of the CMPS (e.g., the second R\'enyi entropy $S_2$), which determines the complexity of simulation with CAMPS. 

\subsection{Random $t$-doped Clifford Circuits with $L_T=1$}

\begin{figure*}
\centering
\includegraphics[width=\textwidth]{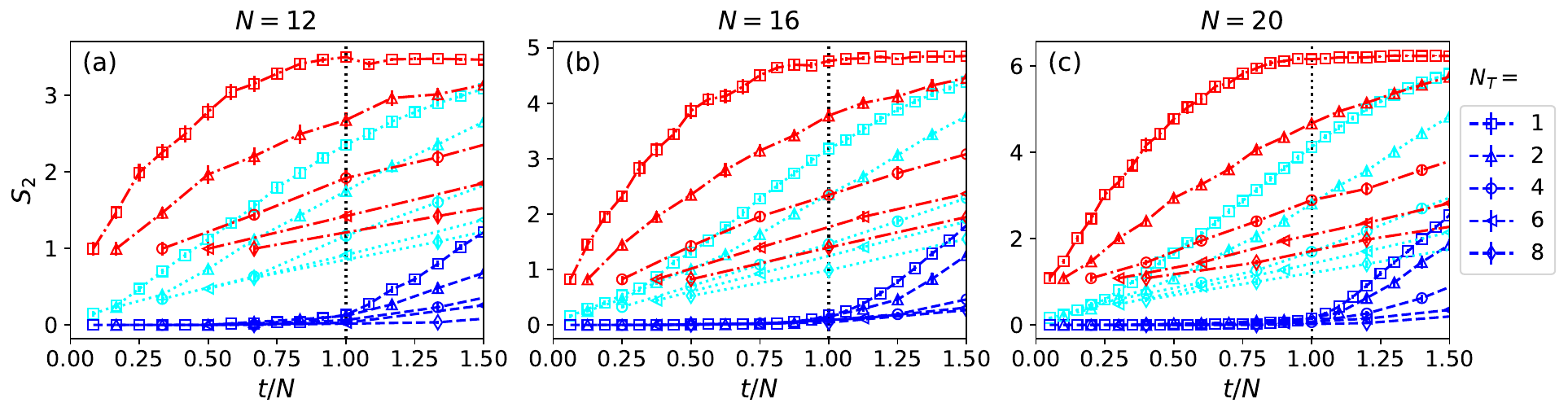}
\caption{The CMPS $S_2$ from the simulation of random $t$-doped Clifford circuits using MPS (red), CAMPS with disentangling algorithm (blue) and CAMPS without Clifford disentangling (cyan) for system size $N\in\{12,16,20\}$ and $T$-gate number per layer $N_T\in\{1,2,4,6,8\}$. Each data is averaged over 16 circuit realizations and error bars are present.}
\label{fig:s2all}
\end{figure*}

We start with considering random Clifford+$T$ circuits with $N_T=1$ $T$-gate at every layer ($L_T=1$).
In Fig.~\ref{fig:s2all}, for CAMPS with disentangling algorithm, the CMPS $S_2$ remains almost 0 until $t\simeq N$. In contrast, MPS and CAMPS without disentangling have their $S_2$ increased from the beginning. These show that adding a Clifford circuit on top of MPS helps reducing the simulation cost significantly, and a disentangling algorithm is crucial to minimize the entanglement within CMPS.
Moreover, we observe that MPS (without Clifford augmentation) reaches the maximal entanglement at circuit depth $d\simeq N$ as expected. However, using CAMPS postpones the saturation significantly, even without the Clifford disentangling as the CAMPS incorporates the entanglement from the Clifford part with stabilizer representation. The main effect of disentangling seems to primarily be that the first $\approx N$ $T$-gates do not increase the entanglement but the rate of entanglement growth afterwards seems to be similar.

\begin{figure}
\centering
\includegraphics[width=0.5\textwidth]{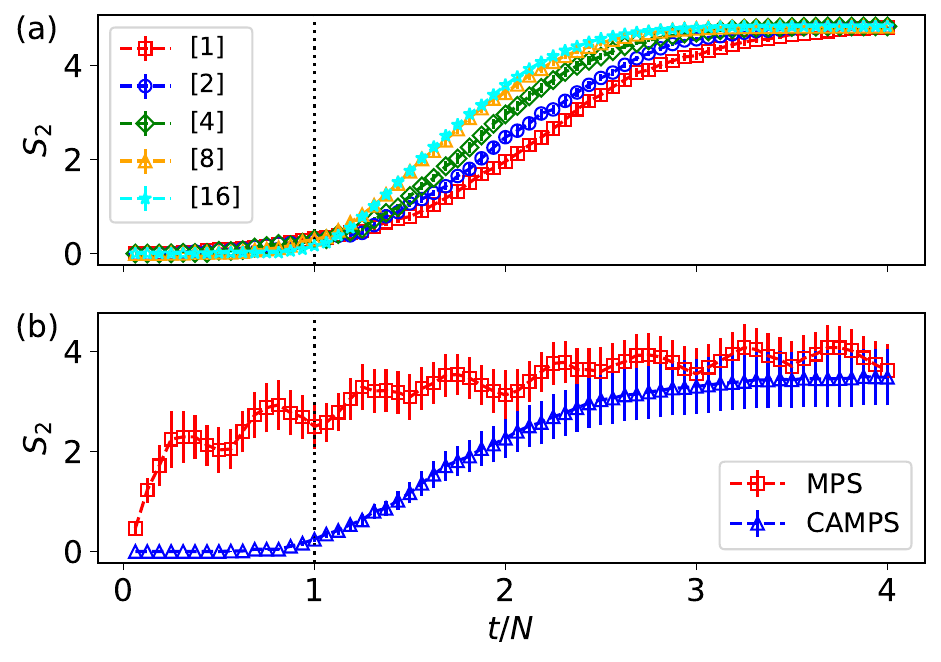}
\caption{The CMPS $S_2$ from the simulation of random $t$-doped Clifford circuits with specific restrictions. (a) The positions of $T$-gates are distributed among the first $n_r$ qubits. Results are shown for $n_r\in\{1,2,4,8,16\}$ with $N=16$, $N_T=1$. (b) All the two-qubit gates are identical for the whole circuit, with MPS simulation results present for comparison. Each data is averaged over 16 circuit realizations and error bars are present.}
\label{fig:tonfirstfew}
\end{figure}

We now put restrictions on the random circuits. For example, in Fig.~\ref{fig:tonfirstfew}, we consider two specific scenarios: (a) distributing the $T$-gates only among the first $n_r$ qubits, and (b) using the same two-qubit Clifford gate throughout the entire circuit. In both of these cases, we see essentially qualitatively the same behaviors as in the above random Clifford case.  

\begin{figure}
\centering
\includegraphics[width=0.5\textwidth]{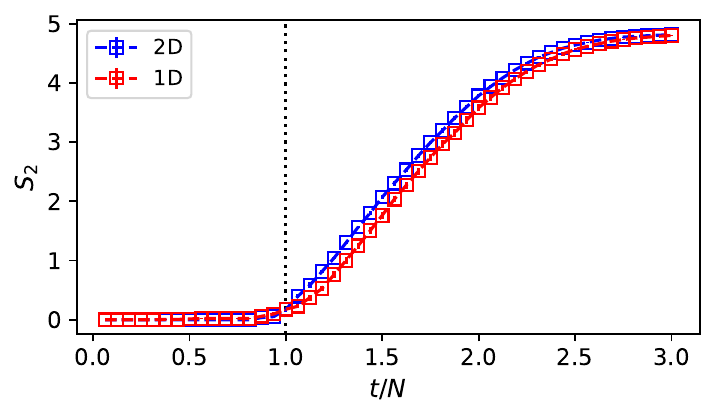}
\caption{The CMPS $S_2$ from the simulation for 2D random $t$-doped Clifford circuits ($N=4\times 4$, $N_T=1$, averaged over 16 circuit samples). Results of 1D circuits (from Fig.~\ref{fig:s2all}) are included for comparison.}
\label{fig:2dentropy}
\end{figure}

In Fig.~\ref{fig:2dentropy}, we show that the entanglement entropy within CAMPS for random 2D circuits is close to that in 1D case, suggesting that CAMPS remains as an efficient representation for higher-dimensional circuits. Indeed, the key steps in CAMPS simulation --- such as absorbing Clifford gates into the stabilizer tableau or commuting the non-Clifford gates through the Clifford circuits (see Append.~\ref{sec:tableau}) --- maintain the same computational complexity regardless of locality of the Clifford gates and the twisted Pauli strings for OFD to disentangle.

\begin{figure}
\centering
\includegraphics[width=0.5\textwidth]{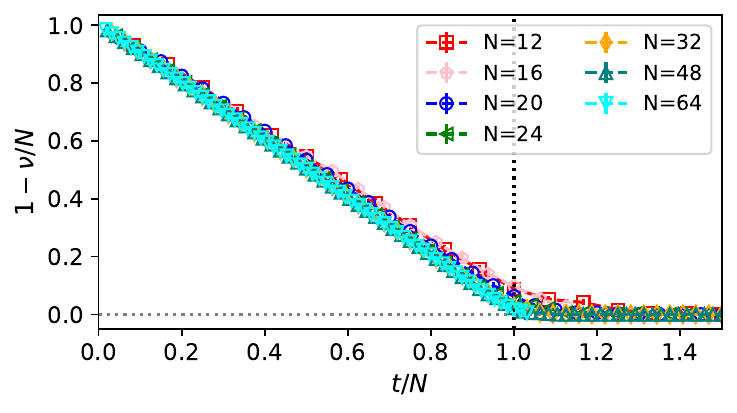}
\caption{Stabilizer nullity $\nu$ of CAMPS as a function of the number of $T$-gates $t$ ($N_T=1$) for various $N$. Each nullity data is averaged over 16 circuit realizations with error bar present.}
\label{fig:nullity}
\end{figure}

\begin{figure}
\centering
\includegraphics[width=0.5\textwidth]{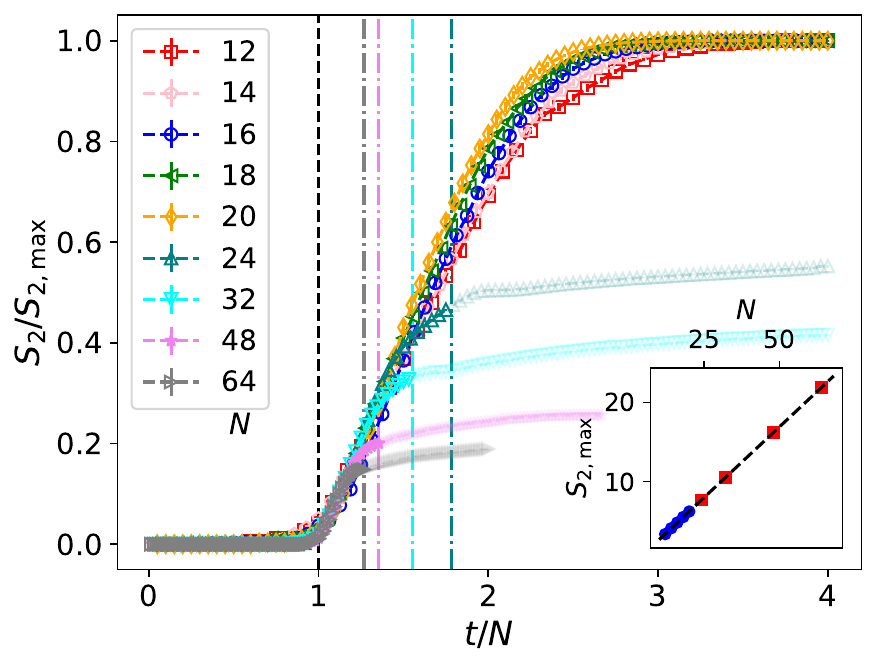}
\caption{The CMPS $S_2$ of CAMPS over $S_{2,\mathrm{max}}$ as a function of $t/N$ with $N_T=1$. Each data point is averaged over 16 circuit realizations with error bars included.
Black dashed vertical line: position of $t/N=1$; dashdot vertical lines: positions of $t/N$ at which $\widetilde{\mathcal{F}}$ for each $N$ reduces to $e^{-1}$. Inset: linear fit to $S_{2,\mathrm{max}}$ as a function of $N$: blue circles: the exact values for $N\leq 20$; dashed line: linear fitting to the exact values; red squares: estimation of $S_{2,\text{max}}$ from the fitting line for $N>20$.}
\label{fig:s2max_tn}
\end{figure}

At last, we extract some universal behaviors from CAMPS simulations. Fig.~\ref{fig:nullity} shows that the stabilizer nullity (i.e., the number of magic qubits) of CAMPS in average obeys a simple function form across a wide range of system sizes:
\begin{equation}
1-\frac{\nu}{N}=
\begin{cases}
1-\frac{t}{N} & t\leq N, \\
0 & t>N.
\end{cases}
\end{equation}
indicating that in average one $T$-gate turns one qubit into a magic qubit.
For individual circuit sample, as shown in Fig.~\pref{fig:optfree}{c}, the nullity changes by at most one for each $T$-gate, as expected from OFD.

In Fig.~\ref{fig:s2max_tn}, we plot $S_2$ as a function of the $T$-gate number $t$ for system sizes up to 64.
We observe that CAMPS reaches the maximal entanglement at $t\simeq 2.5N$ with the CMPS entanglement dynamics collapsing across system sizes.
We find that $S_{2,\mathrm{max}}$ scales linearly as $S_{2,\mathrm{max}}\simeq0.356N-0.878$ by fitting the maximum $S$ for $N \leq 20$ where we can simulate the $t$-doped Clifford circuits using CAMPS without bond truncation;  we extrapolate this to determine $S_{2,\mathrm{max}}$ for $N>20$ where truncation prevents us from getting the exact result at large number of $t$ gates (see inset of Fig.~\ref{fig:s2max_tn}).  Interestingly, the $S_2$ of the truncated CMPS diverges from the expected behavior when the norm of the CMPS~\cite{Zhou2020,Ayral2023} $\widetilde{F}=|\braket{\psi}{\psi}|^2 \approx e^{-1}$.
Once bond truncation is introduced, the benefits of disentangling algorithms diminish. This is because the bond truncation discards small singular values in the CMPS; additional entangling gates from disentangling introduce new small singular values, leading to alternating cycles of bond truncation and disentangling that degrade the fidelity of the CAMPS by continuously discarding these tails. Therefore, we cease disentangling once bond truncation begins.

\begin{figure}
\centering
\includegraphics[width=0.5\textwidth]{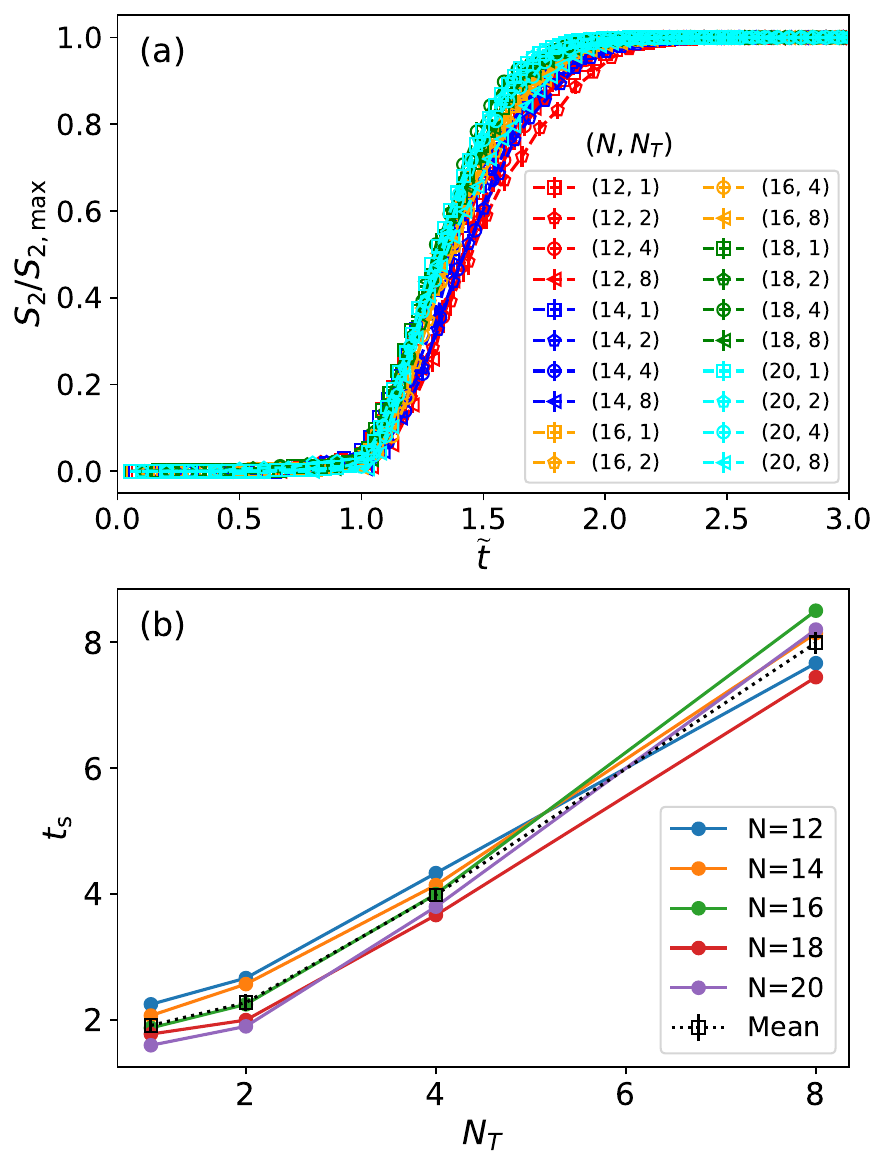}
\caption{(a) The CMPS $S_2/S_{2,\mathrm{max}}$ as a function of the rescaled $t$, for $(N,N_T)\in\{12,14,16,18,20\}\times\{1,2,4,8\}$. (b) Scaling factor for $t$ as a function of $N_T$. Each data is averaged over 16 circuit realizations, with error bars included.}
\label{fig:S2vsNt}
\end{figure}

We now consider putting more than $N_T=1$ $T$ gates per layer for various system sizes. In Fig.~\pref{fig:S2vsNt}{a}, we find that the entropy data collapses by rescaling $t$ as 
\begin{equation}
\widetilde{t}=    
\begin{cases}
t-N
& t\leq N, \\
\frac{1}{t_s}(t-N) 
& t>N.
\end{cases}
\end{equation}
where $t_s$ is the average number of $T$-gates (not including the first `free' $N$ T-gates) needed to increase the MPS entropy from 0 to the maximum scaled by $1/N$. 
We also find from Fig.~\pref{fig:S2vsNt}{b} that $t_s \approx N_T$ with a slight deviation for $N_T=1$, suggesting that, for $L_T=1$ circuits, a unit increase in entanglement happens after every $N_T$ gates or equivalently a layer of two-qubit Clifford+$T$ gates.
\subsection{Random $t$-doped Clifford Circuits for various $(L_T,N_T)$}

\begin{figure}
\centering
\includegraphics[width=0.48\textwidth]{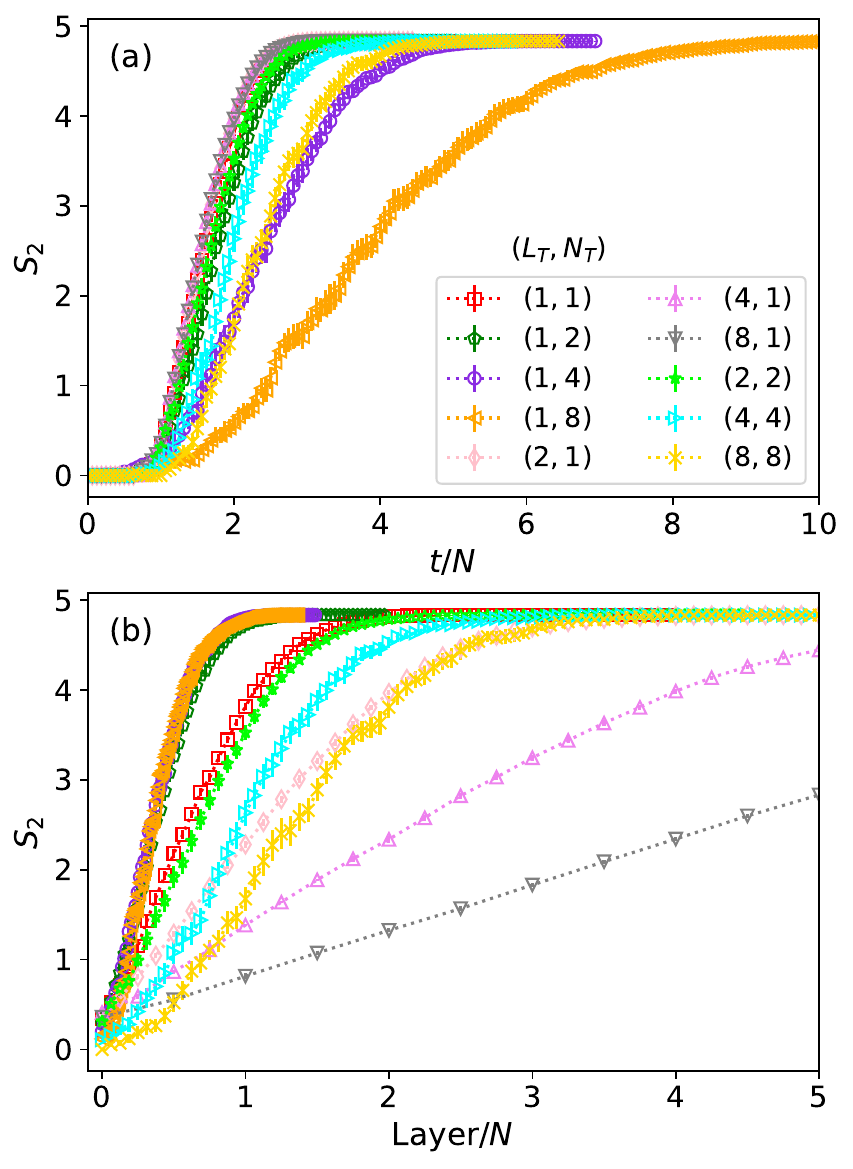}
\caption{The CMPS $S_2$ as a function of (a) $t/N$ and (b) the circuit layer over $N$ after $N=16$ $T$-gates have been applied, for various values of $(L_T,N_T)$, i.e., Layer=$L_T\cdot(t-N)/N_T$. The data are averaged over 16 circuit realizations, with error bars included.}
\label{fig:S2vsLtNt}
\end{figure}

In this section, we fix the qubit number $N=16$, and tune the values of $N_T$ and $L_T$. We observe that
\begin{enumerate}
\item 
In Fig.~\pref{fig:S2vsLtNt}{b}, when $L_T=1$, $S_2$ has the same rate of growth per layer for $N_T>1$, while $N_T=1$ displays obvious deviation.
\item 
In Fig.~\pref{fig:S2vsLtNt}{a}, when $N_T=1$, $S_2$ has the same rate of growth per $T$-gate, regardless of the number of Clifford layers $L_T$.
\item 
In Fig.~\pref{fig:S2vsLtNt}{a}, when $N_T>1$, $L_T$ affects $S_2$ increment per $T$-gate in a non-trivial way, for example, the rate of $S_2$ growth per $T$-gate for $(L_T, N_T)=(2,2)$ is close to that of $(L_T, N_T)=(1,1)$, and $(L_T, N_T)=(4,4)$ is close to $(L_T, N_T)=(1,2)$. 
\end{enumerate}

\begin{figure*}
\centering
\includegraphics[width=\textwidth]{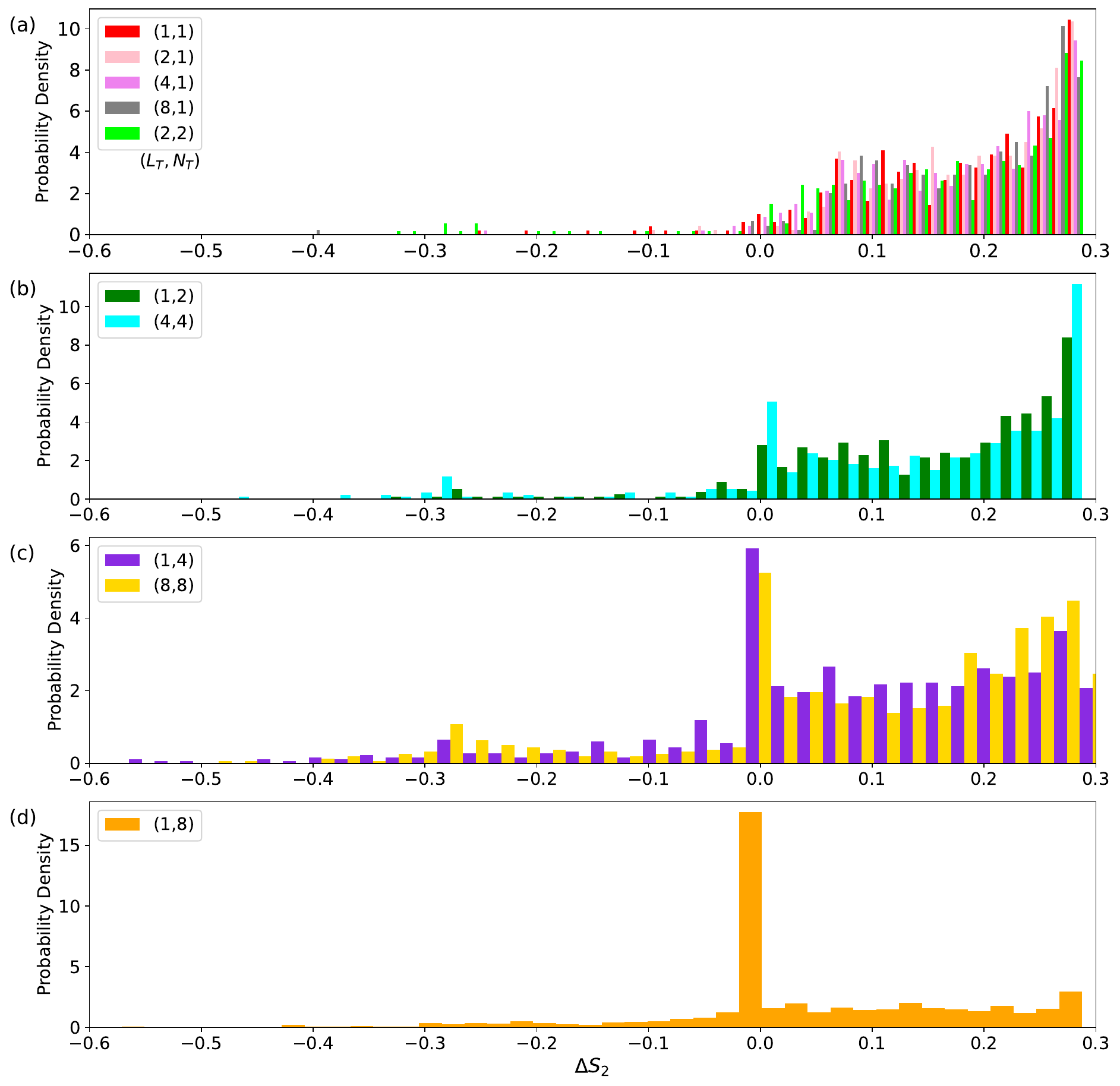}
\caption{In the same setting to Fig.~\ref{fig:S2vsLtNt}, the probability densities of the CMPS entanglement entropy change $\Delta S_2$ after a $T$-gate is absorbed into the CMPS and disentangled for various groups of $(L_T,N_T)$.}
\label{fig:S2density}
\end{figure*}

We also show in Fig.~\ref{fig:S2density} the distribution of $\Delta S_2$ -- the amount $S_2$ changes from each $T$-gate -- before $S_2$ reaches the maximum. 
We observe where the group of $(L_T,N_T)$ collapse that there is similar probability distribution of $\Delta S_2$. We also observe that as $N_T$ increases (i.e., more $T$-gates are placed in one layer), more $T$-gates contributes zero increased entropy to the MPS component. It remains as an open question regarding why $L_T$ and $N_T$ together affect the CMPS entanglement entropy in such an intriguing way and we leave it for future work.

\section{Simulation Tasks}
\label{sec:othersimulation}
We have demonstrated that CAMPS provides a compact representation for the quantum states from $t$-doped Clifford circuits with $t\lesssim N$. However, to accomplish the classical simulation, it is necessary to be able to obtain useful information from the CAMPS states. Given the access to CAMPS, it is straightforward to evaluate the expectation values of Pauli strings (see Appendix~\ref{sec:pauliexpect} for the numerical results). In addition, we have used a polynomially efficient algorithm (see Appendix~\ref{sec:stablearn}) to calculate the stabilizer nullity.
In this section, we present the algorithms for the following tasks: (1) the probability calculation or the sampling of bitstrings, computation of (2) the amplitude of bitstrings and (3) the entanglement R\'enyi entropy of the entire CAMPS states.

\subsection{Probability and Sampling of Bitstring}
\label{sec:sampleandprob}
We leverage the disentangling algorithms to efficiently sample bitstrings from CAMPS or compute the probability of any given bitstring: 
\begin{equation}
\text{Prob}(\bm{s})=|\expect{\bm{s}}{\mathcal{C}}{\psi}|^2    
\end{equation}
In this context,  measurements are implemented as projection operators:
\begin{equation}
\text{P}^{[0]} = \ket{0}\bra{0}=\frac12I+\frac12Z, \quad \text{P}^{[1]} = \ket{1}\bra{1}=\frac12I-\frac12Z.    
\end{equation}
For a CAMPS $\mathcal{C}\ket{\psi}$, we commute these projection operators through the Clifford circuit part and absorb them with CMPS, e.g., when measuring the $k$-th qubit:
\begin{equation}
\text{P}^{[0]}_k \mathcal{C}\ket{\psi}=\mathcal{C}\left[\left(\frac{1}{2}I+\frac{1}{2}\widetilde{Z}_k\right)\ket{\psi}\right]\equiv\mathcal{C}\ket{\phi}   
\end{equation}
where the disentangling algorithm will be applied to $\ket{\phi}=\left(\frac{1}{2}I+\frac{1}{2}\widetilde{Z}_k\right)\ket{\psi}$ to reduce the bond dimension. The marginal probability on the $k$-th qubit is then given by
\begin{equation}
\pi(s_k=0)=\bra{\psi}\mathcal{C}^{\dagger}\text{P}^{[0]}_k \mathcal{C}\ket{\psi}=\braket{\phi}{\phi}  
\end{equation}
Before we move on to measure the next qubit, we shall update the CAMPS with the new state $\text{P}^{[0]}_k\mathcal{C}\ket{\psi}$. 
Repeating this process until all the qubits are measured, it will yield the probability for a bitstring $\bm{s}=(s_1,s_2,\dots,s_N)$ as
\begin{equation}
\text{Prob}(\bm{s})=\pi(s_0)\pi(s_1|s_0)\dots\pi(s_N|\bm{s}_{1:N-1})    
\end{equation}

Additionally, we can sample the $0/1$ state at each site $k$ using the marginal probability on $k$-th qubit. These two algorithms are outlined in Algorithm~\ref{alg:bitprob} and Algorithm~\ref{alg:bitsample}, respectively.

\begin{algorithm}[H]
\caption{Bitstring Probability from Clifford-MPS}\label{alg:bitprob}
\begin{flushleft}
\hspace*{\algorithmicindent} \textbf{Input}: Clifford-MPS $\mathcal{C}\ket{\psi}$, binary bitstring $\bm{s}\in\{0,1\}^N$
\end{flushleft}
\begin{algorithmic}[1]
\State Initialize: $\text{Prob}(\bm{s})=1$. 
\For{$n=1,2,\dots,N$}
\State $\widetilde{Z}_n\gets {\mathcal{C}}^{\dagger}Z_n{\mathcal{C}}$.
\State $\ket{\psi} \gets \left(\frac{1}{2}I+\frac{1}{2}(-1)^{s_n}\widetilde{Z}_n\right)\ket{\psi}$.
\State Perform disentangling on $\ket{\psi}$.
\State $\pi(s_n|s_{1:n-1})\gets\braket{\psi}{\psi}$.
\State $\text{Prob}(\bm{s})\gets\text{Prob}(\bm{s})\cdot\pi(s_n|s_{1:n-1})$
\EndFor
\end{algorithmic}
\begin{flushleft}
\hspace*{\algorithmicindent} \textbf{Output}: probability for bitstring $\bm{s}$: $\text{Prob}(\bm{s})$.
\end{flushleft}
\end{algorithm}

\begin{algorithm}[H]
\caption{Bitstring Sampling from Clifford-MPS}\label{alg:bitsample}
\begin{flushleft}
\hspace*{\algorithmicindent} \textbf{Input}: Clifford-MPS ${\mathcal{C}}\ket{\psi}$. 
\end{flushleft}
\begin{algorithmic}[1]
\State Initialize: bistring $\bm{s}=()$, $\text{Prob}(\bm{s})=1$. 
\For{$n=1,2,\dots,N$}
\State $\widetilde{Z}_n\gets {\mathcal{C}}^{\dagger}Z_n{\mathcal{C}}$.
\State $\ket{\phi} \gets \left(\frac{1}{2}I+\frac{1}{2}\widetilde{Z}_n\right)\ket{\psi}$.
\State Marginal probability $\pi(s_n=0|s_{1:n-1})\gets\braket{\phi}{\phi}$.
\State Sample $s_n\in\{0,1\}$ according to probability $\{\pi(s_n=0|s_{1:n-1}), \pi(s_n=1|s_{1:n-1})=1-\pi(s_n=0|s_{1:n-1})\}$
\State $\ket{\psi} \gets \left(\frac{1}{2}I+\frac{1}{2}(-1)^{s_n}\widetilde{Z}_n\right)\ket{\psi}$.
\State Perform Clifford disentangling on $\ket{\psi}$.
\State $\bm{s}\gets(s_1,s_2,\dots,s_n)$.
\State $\text{Prob}(\bm{s})\gets\text{Prob}(\bm{s})\cdot\pi(s_n|s_{1:n-1})$.
\EndFor
\end{algorithmic}
\begin{flushleft}
\hspace*{\algorithmicindent} \textbf{Output}: bitstring $\bm{s}$ and the corresponding probability $\text{Prob}(\bm{s})$.
\end{flushleft}
\end{algorithm}

\begin{figure}
\centering
\includegraphics[width=0.48\textwidth]{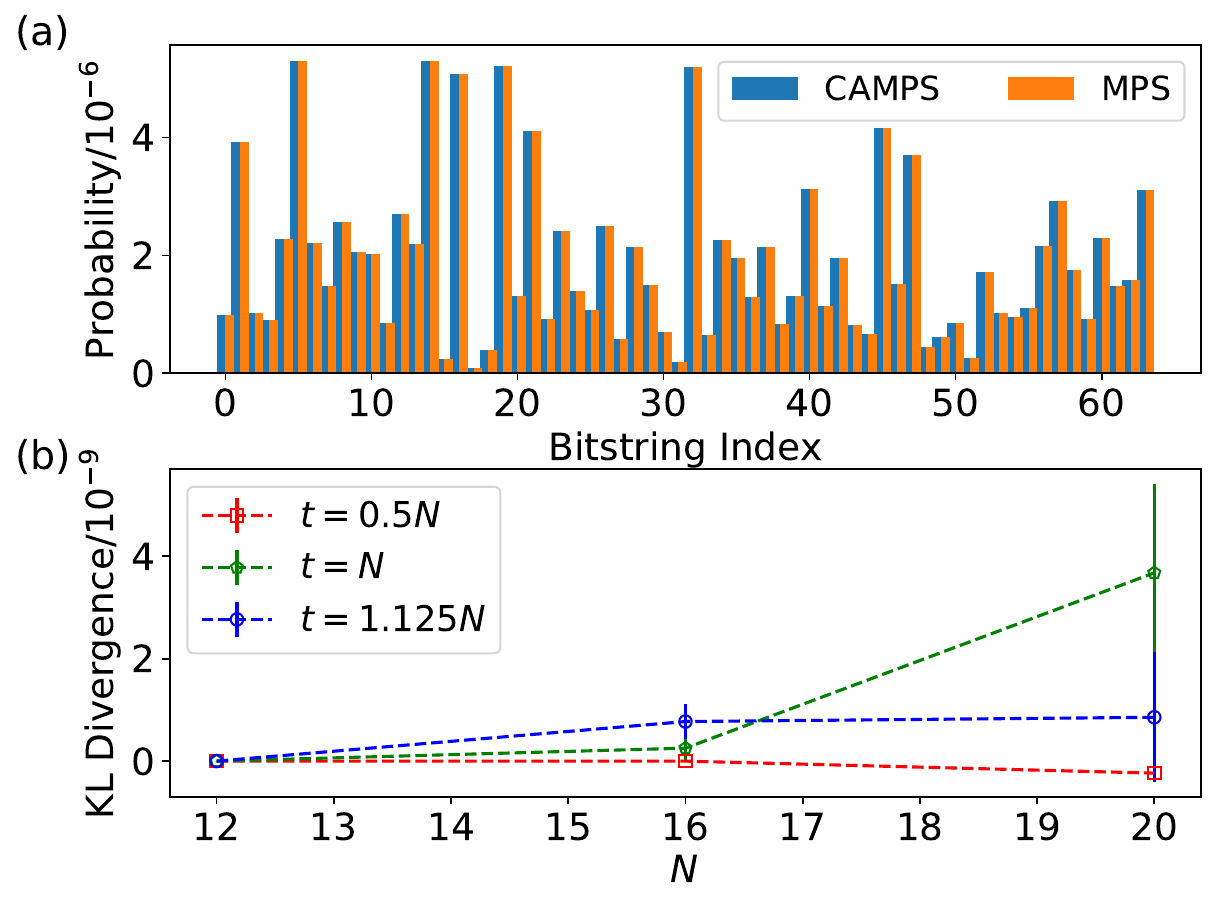}
\caption{Probability measurement from CAMPS: (a) Comparison with the exact probability from MPS simulation ($N=20, N_T=1, t=22$); (b) KL divergence of the bitstring probabilities between CAMPS and the exact MPS. Each data from (a) and (b) is averaged over 9 circuit realizations, with error bars included.}
\label{fig:prob}
\end{figure}

\begin{figure*}
\centering
\includegraphics[width=\textwidth]{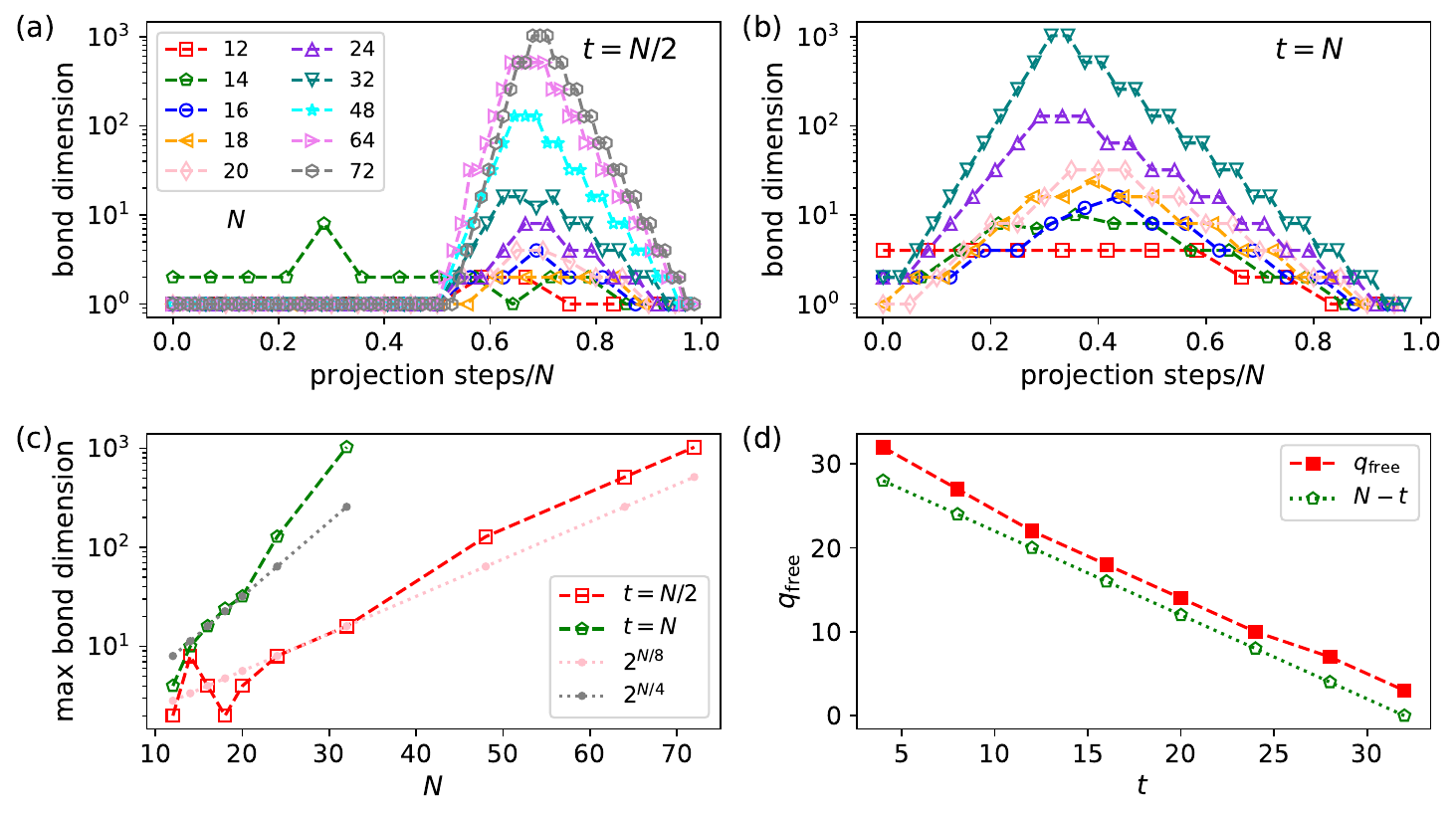}
\caption{Evolution of the MPS bond dimension during the probability measurement using CAMPS: (a) $t=0.5N$; (b) $t=N$; (c) the maximal bond dimension as a function of system size $N$. Dotted lines $2^{N/8}$ and $2^{N/4}$ are drawn to contrast with the exponential growth from $t=0.5N$ and $t=N$, respectively. (d) The number of measured qubits for which the MPS bond dimension starts increasing during the measurement ($q_\mathrm{free}$) as a function of $t$ ($N=32$, $N_T=1$); dotted line $N-t$ is drawn for comparison. }
\label{fig:prob2}
\end{figure*}

Fig.~\pref{fig:prob}{a} shows the bitstring probabilities calculated from CAMPS, are identical to the exact probabilities from MPS simulation on the same $t$-doped Clifford circuit. Moreover, Fig.~\pref{fig:prob}{b} shows the nearly 0 (up to machine precision) KL divergence of the bitstring probabilities between the CAMPS and MPS simulations. 

In Fig.~\ref{fig:prob2}, we examine the required CMPS bond dimension during the measurement: when $t=N/2$ (see Fig.~\pref{fig:prob2}{a}), the CMPS bond dimension stays the same, but starts increasing after $N/2$ qubits are measured, and eventually decreases to 1 again when all qubits are measured. Similarly, for $t=N$, the CMPS bond dimension increases first and then decreases to 1. In Fig.~\pref{fig:prob2}{c}, we show the maximal CMPS bond dimension during the measurement grows exponentially with system size. 

From these numerical simulation, we observe that the measurement on each qubit projects out one qubit from CMPS which, in the subsequent steps, typically will not be affected by $\widetilde{Z}$ from the projection operators; however, those $\widetilde{Z}$ will entangle the rest of the unprojected qubits. For $t<N$, if a non-trivial Pauli term from $\widetilde{Z}_k$ hits a free qubit, a disentangling circuit can be constructed using OFD. However, once all the free qubits have been projected out by the measurement, the entanglement starts increasing among the remaining magic qubits. Generically, as indicated by Fig.~\pref{fig:prob2}{d}, the marginal measurement on $q\approx N-t$ qubits is efficient without increasing the CMPS bond dimension. For more general cases ($q>N-t$), one can measure the marginal probability on $q$ qubits, with a computational cost exponential in $q-\max\{0,N-t\}$, and the measurement over all the qubits is exponentially costly in the system size.

\subsection{Wavefunction Amplitude}
\label{sec:amplitude}
Leveraging the ease of Pauli measurements on CAMPS states, we can also obtain the (unnormalized) wavefunction amplitude for any bitstring. Access to the amplitude is useful~\cite{mello2025clifford}, for instance, when performing imaginary-time projection from variational states to approximate the exact ground state properties~\cite{Ceperley1980,martin2016interacting}. The approach we present here is inspired by the scheme from quantum certification~\cite{huang2024}, but is formalized in a deterministic manner, owing to the special structure of CAMPS.

We begin by setting the (unnormalized) wavefunction amplitude for $\ket{\bm{s}^{[0]}}=\ket{0}^{\otimes N}$ to 1. To compute the amplitude for any bitstring $\ket{\bm{s}}$, such as $\ket{\bm{s}}=\ket{0\dots001\dots11}$, we define a path that connects $\ket{\bm{s}^{[0]}}$ and $\ket{\bm{s}}$ by turning on the 1’s in the following bitstring sequence, one by one:
\begin{equation}
\bm{s}^{[0]}\rightarrow \bm{s}^{[1]} \rightarrow \bm{s}^{[2]} \cdots \rightarrow \bm{s}^{[w(\bm{s})-1]} \rightarrow \bm{s}^{[w(\bm{s})]}=\bm{s}   
\end{equation}
where $\bm{s}^{[j]}$ and $\bm{s}^{[j+1]}$ differ by a single bit, and $w(\bm{s})=|\bm{s}|$ is the Hamming weight (i.e., the number of 1's) of $\bm{s}$. 

Since we have defined $\expect{\bm{s}^{[0]}}{\mathcal{C}}{\psi}=1$, the calculation of $\expect{\bm{s}}{U_\mathcal{C}}{\psi}$ is completed by evaluating each ratio in the following expression:
\begin{equation}
\label{eq:amplitude}
\expect{\bm{s}}{\mathcal{C}}{\psi}=\expect{\bm{s}^{[0]}}{\mathcal{C}}{\psi}\frac{\expect{\bm{s}^{[1]}}{\mathcal{C}}{\psi}}{\expect{\bm{s}^{[0]}}{\mathcal{C}}{\psi}}\cdots\frac{\expect{\bm{s}}{\mathcal{C}}{\psi}}{\expect{\bm{s}^{[w(\bm{s})-1]}}{\mathcal{C}}{\psi}}    
\end{equation}
Note that $\bm{s}^{[j]}$ and $\bm{s}^{[j+1]}$ differ by one bit, say the $k_j$-th bit. We can rewrite these bitstrings as $\bm{s}^{[j]}=\bm{s}^{[j]}_{1:k_j-1}0\bm{s}^{[j]}_{k_j+1:N}=\bm{s}^{[j]}_{N\backslash{k_j}}0$, $\bm{s}^{[j+1]}=\bm{s}^{[j]}_{1:k_j-1}1\bm{s}^{[j]}_{k_j+1:N}=\bm{s}^{[j]}_{N\backslash{k_j}}1$. For simplicity, we define
\begin{equation}
z^{[0]}=\expect{\bm{s}^{[j]}_{N\backslash{k_j}}0}{\mathcal{C}}{\psi},\quad z^{[1]}=\expect{\bm{s}^{[j]}_{N\backslash{k_j}}1}{\mathcal{C}}{\psi}.
\end{equation}
We can evaluate the probability of any bitstring using the method described in Sec.~\ref{sec:sampleandprob}. Thus, we can obtain $p_0=|z^{[0]}|^2$, $p_1=|z^{[1]}|^2$, $p_2=|z^{[0]}+z^{[1]}|^2/2$, $p_3=|z^{[0]}-z^{[1]}|^2/2$, $p_4=|z^{[0]}+iz^{[1]}|^2/2$, and $p_5=|z^{[0]}-iz^{[1]}|^2/2$. The latter four terms are computed by appending Hadamard and phase gates to $\mathcal{C}$, changing the eigenstates of Pauli-$Z$ operator to those of Pauli-$X/Y$. Finally, the ratio factor in Eq.~\eqref{eq:amplitude} is given by
\begin{equation}
\frac{z^{[1]}}{z^{[0]}}=\frac{p_2-ip_4-\frac12(1-i)(p_0+p_1)}{p_0}.    
\end{equation}

\begin{figure}
\centering
\includegraphics[width=0.5\textwidth]{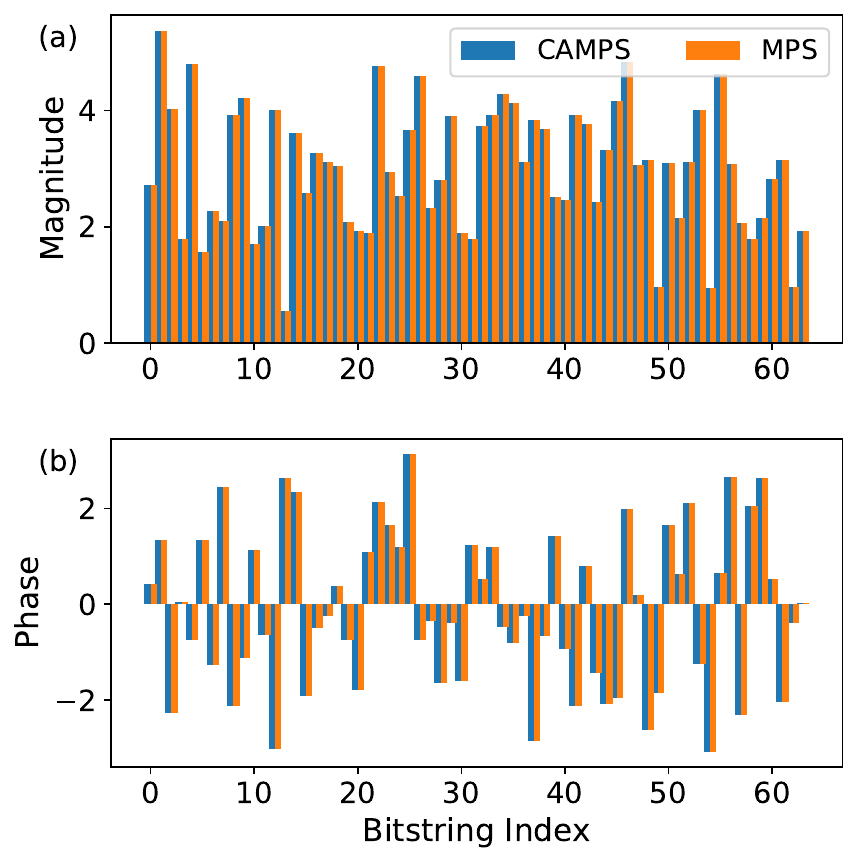}
\caption{(a) The magnitude and (b) the phase of the unnormalized wave function amplitude from CAMPS ($N=12, N_T=1, t=2N$). Exact results obtained from MPS simulation are included for comparison.}
\label{fig:wfamp}
\end{figure}

In Fig.~\ref{fig:wfamp}, we present the amplitudes from a CAMPS representing the state from a $t$-doped Clifford circuit ($N=12, N_T=1, t=2N$), and note that it reproduces the same results as the exact ones from the MPS simulation.

\subsection{Entanglement R\'enyi Entropy}
We present an algorithm for estimating the entanglement R\'enyi entropy (ERE) of the full CAMPS states when the CMPS represents a product state. Specifically, this algorithm generalizes the standard entropy-calculation method for stabilizer states~\cite{fattal2004} to cases where the product states include magic states, and it outperforms traditional entropy calculation methods that rely on using MPS to simulate entire states, offering a more efficient, albeit still exponential, approach for calculating the entanglement entropy of $(t \leq N)$-doped Clifford circuits, in particular when studying the interplay between quantum entanglement and magic~\cite{viscardi2025interplay}. 

We denote the Clifford-Product state with density matrix $\rho$ as:
\begin{equation}
\rho = {\mathcal{C}}\ket{\phi}\bra{\phi}{\mathcal{C}}^\dagger   
\end{equation}
and partition the $N$-qubit system into regions $A$ and $\bar{A}$ containing $N_A$ and $N_{\bar{A}}$ qubits, respectively ($N_A+N_{\bar{A}}=N$). The density matrix for the product state can be decomposed into a product of single-qubit density matrices, i.e.,
\begin{equation}
\Tilde{\rho}=\ket{\phi}\bra{\phi}=\prod_{j=1}^N \Tilde{\rho}_j   
\end{equation}

The second entanglement R\'enyi entropy is defined as
\begin{equation}
S_2=-\ln\Tr_A\left(\rho_A^2\right) 
\label{eq:def_renyi}
\end{equation}
where $\rho_A=\Tr_{\bar{A}}\rho$. Representing the density matrices in terms of Pauli strings, we write:
\begin{equation}
\rho=\frac{1}{2^N}\sum_{\bm{\sigma}\in \mathcal{P}_N}a_{\bm{\sigma}}\bm{\sigma}   
\end{equation}
where $a_{\bm{\sigma}}=\Tr\left(\bm{\sigma}\rho\right)$, The reduced density matrix for region $A$ then becomes:
\begin{equation}
\rho_A=\frac{1}{2^{N_A}}\sum_{\bm{\sigma}\in \mathcal{P}_A}a_{\bm{\sigma}}\bm{\sigma}    
\label{eq:reducedensityA}
\end{equation}
where $\mathcal{P}_A$ denotes the Pauli strings restricted to region $A$. Thus, calculating the entanglement R\'enyi entropy Eq.~\eqref{eq:def_renyi} reduces to evaluating the coefficients $a_{\bm{\sigma}}$ for Pauli strings $\bm{\sigma}$ that have identity operators outside region $A$:
\begin{equation}
S_2=-\ln\left(\frac{1}{2^{N_A}}\sum_{\bm{\sigma}\in \mathcal{P}_A}a_{\bm{\sigma}}^2\right)  
\label{eq:s2_reducedA}
\end{equation}
Although the number of terms grows exponentially with the size of region $A$, the computational cost can be reduced by identifying the Pauli strings with nonzero coefficients, though the exponential complexity is not guaranteed to always be removed~\cite{gu2024}. The standard entropy-calculation method for stabilizer states~\cite{fattal2004}, along with the algorithm proposed here, can be effectively interpreted as realizations of this strategy. 

To find the non-vanishing Pauli strings for $\rho$, we begin by identifying the non-vanishing generators from $\Tilde{\rho}$. Since $\Tilde{\rho}$ is a product state, this process is efficient; we only need to examine single-qubit Pauli operators for each qubit in $\Tilde{\rho}$. The non-vanishing generators thus originate from the $3N$ single-qubit Pauli operators, defined as
\begin{equation}
\{\Tilde{G}_j|\Tr\left(\Tilde{G}_j\Tilde{\rho}\right)\neq0, j=1,2,\dots,\Tilde{M} \}    
\end{equation}
where the number of these generators $\Tilde{M}\leq 3N$. 

Now, let us delve into the efficiency of this step. For a stabilizer state, the action of Pauli strings maps the state back to itself with a coefficient of $\pm 1$ or 0. This gives rise to the favorable property for stabilizer states $\ket{\psi}$:
\begin{equation}
\Tilde{a}_{\bm{\sigma}_1\bm{\sigma}_2}=\expect{\psi}{\bm{\sigma}_1\bm{\sigma}_2}{\psi}=\expect{\psi}{\bm{\sigma}_1}{\psi}\expect{\psi}{\bm{\sigma}_2}{\psi}=\Tilde{a}_{\bm{\sigma}_1}\Tilde{a}_{\bm{\sigma}_2}    
\label{eq:a_productlaw}
\end{equation}
For a non-product $\Tilde{\rho}$, where there exist qubits entangled in clusters, Eq.~\eqref{eq:a_productlaw} does not hold in general.
However, the product states, or states with small clusters, partially mitigate this problem: if $\bm{\sigma}_1$ and $\bm{\sigma}_2$ belongs to two unentangled regions respectively, we still have $\Tilde{a}_{\bm{\sigma}_1\bm{\sigma}_2}=\Tilde{a}_{\bm{\sigma}_1}\Tilde{a}_{\bm{\sigma}_2}$,
Consequently, it suffices to check Pauli strings within each cluster, with complexity growing only exponentially with cluster size, i.e., $\sup \Tilde{M}=\sum_{R} 3^{|R|}$, where $|R|$ is the qubit number in each cluster $R$ ($\sum_{R}|R|=N$). For product states, each cluster size $|R|$ is always 1, yielding $\sup \Tilde{M}=3N$. 

For clusters of size 1 (i.e., single qubit unentangled from the rest), we need to check all the three Pauli operators on that qubit, because for magic qubits created by the $T$ gates, two out of the three Pauli operators can have nonzero coefficients, while for stabilizer states, only one non-vanishing Pauli operator exists for each qubit.

After identifying the generator set $\{\Tilde{G}_j\}_{j=1}^{\Tilde{M}}$ with non-zero coefficients from $\Tilde{\rho}$, we map these generators to the Pauli strings for $\rho$ and select the subset needed to generate the Pauli strings in Eq.~\eqref{eq:reducedensityA}. The mapping is achieved by
\begin{equation}
G_j = \mathcal{C}\widetilde{G}_j \mathcal{C}^\dagger    
\end{equation}
To isolate the subset of generators that correspond to $\bm{\sigma}\in\mathcal{P}_A$, we use Gaussian elimination to remove, as many as possible, the non-identity operators in region $\bar{A}$, yielding:
\begin{equation}
\{Q_j\subseteq\langle \{G_j\}_{j=1}^{\Tilde{M}}\rangle|Q_j\in\mathcal{P}_A\otimes I^{\otimes N_{\bar{A}}}, j=1,2,\dots,M \}    
\end{equation}
where $M\leq\Tilde{M}$ represents the number of generators identified through Gaussian elimination. During this process, we track the decomposition of each $Q$'s in terms of $G$'s:
\begin{equation}
Q_j=\prod_{k=1}^{\Tilde{M}}G_{k}^{K^j_k}    
\end{equation}
where $K^j_k\in\{0,1\}$ indicates the presence of $G_k$ in the composition of $Q_j$. Consequently, we find
\begin{equation}
\begin{split}
a_{Q_j}=&\Tr\left(Q_j\rho\right)=\Tr\left(\prod_{k=1}^{\Tilde{M}}G_{k}^{K^j_k}\rho\right)\\
=&\Tr\left(\prod_{k=1}^{\Tilde{M}}G_{k}^{K^j_k}{\mathcal{C}}\Tilde{\rho} {\mathcal{C}}^{\dagger}\right)=\Tr\left({\mathcal{C}}^{\dagger}\prod_{k=1}^{\Tilde{M}}G_{k}^{K^j_k}{\mathcal{C}}\Tilde{\rho}\right)\\
=&\Tr\left(\prod_{k=1}^{\Tilde{M}}\widetilde{G}_{k}^{K^j_k}\Tilde{\rho}\right)
\end{split}
\label{eq:QdecompseK}
\end{equation}
This expression enables the calculation of the coefficients $a_{Q_j}$ based on the transformations applied to the generators $\{\widetilde{G}\}$.

As we have noted, even for Clifford-Product states, Eq.~\eqref{eq:a_productlaw} does not hold if, for instance, $\bm{\sigma}_1$ and $\bm{\sigma}_2$ are Pauli operators on the same magic qubit. Thus, the set $\{Q_j\}_{j=1}^M$ obtained from Gaussian elimination may still produce Pauli strings with zero coefficients. In contrast, for stabilizer states, Eq.~\eqref{eq:a_productlaw} ensures that $\{Q_j\}_{j=1}^M$ will only generate Pauli strings with non-zero coefficients, which are already known to be $\pm 1$; 
thus for stabilizer states, there are $2^M$ Pauli strings in region $A$ with non-vanishing coefficients of $\pm 1$, yielding $S_2=(N_A-M)\ln2$ according to Eq.~\eqref{eq:s2_reducedA}. 

This process of identifying the non-zero Pauli generators essentially mirrors the standard method for calculating entropy in stabilizer states~\cite{fattal2004}. We now introduce two approaches to finalize the computation of entanglement entropy with the set of non-zero Pauli generators. 

The first approach involves enumerating all Pauli strings generated by $\{Q_j\}_{j=1}^M$ and calculating the coefficients for each. While straightforward, this method has exponential complexity, $\mathcal{O}(2^M)$, making it practical only for modest values of $M$ regardless of the system size. 

The second approach aims to represent these coefficients using a new MPS, from which Eq.~\eqref{eq:s2_reducedA} can be derived as the norm of this new MPS. 
This approach, while still potentially exponential, allows a series of systematically improvable approximations associated with truncating the MPS bond-dimension. 
We begin by rewriting the reduced density matrix $\rho_A$ in terms of $\{Q_j\}_{j=1}^M$:
\begin{equation}
\rho_A=\frac{1}{2^{N_A}}\left(I+a_{Q_1}Q_1\right)\circ\left(I+a_{Q_2}Q_2\right)\circ\cdots\left(I+a_{Q_M}Q_M\right)   
\label{eq:rhoAfromQs}
\end{equation}
where $\circ$ is defined as:
\begin{equation}
\begin{split}
(a_{\bm{\sigma}_1}\bm{\sigma}_1)\circ(a_{\bm{\sigma}_2}\bm{\sigma}_2)&=a_{\bm{\sigma}_1\bm{\sigma}_2}\bm{\sigma}_1\bm{\sigma}_2 \\
(a_{\bm{\sigma}_1}\bm{\sigma}_1+a_{\bm{\sigma}_2}\bm{\sigma}_2)\circ(a_{\bm{\sigma}_3}\bm{\sigma}_3)&=a_{\bm{\sigma}_1\bm{\sigma}_3}\bm{\sigma}_1\bm{\sigma}_3+a_{\bm{\sigma}_2\bm{\sigma}_3}\bm{\sigma}_2\bm{\sigma}_3\\
(I)\circ(a_{\bm{\sigma}}\bm{\sigma})&=a_{\bm{\sigma}}\bm{\sigma}\\
\circ_{k=1}^K \left(a_{\bm{\sigma}_k}\bm{\sigma}_k\right)=\left(a_{\bm{\sigma}_1}\bm{\sigma}_1\right)&\circ\left(a_{\bm{\sigma}_2}\bm{\sigma}_2\right)\circ\cdots\circ\left(a_{\bm{\sigma}_K}\bm{\sigma}_K\right)
\end{split}    
\end{equation}
Note that even if $a_{Q_j}=0$ for one factor in Eq.~\eqref{eq:rhoAfromQs}, combining it with others may still yield non-zero Pauli strings. According to Eq.~\eqref{eq:QdecompseK}, we find
\begin{equation}
a_{Q_j}Q_j=\circ_{k=1}^{\Tilde{M}}\left(\Tilde{a}_{\Tilde{G}_k}\widetilde{G}_k\right)^{K_k^j}
\end{equation}
Now, we introduce a new type of quantum state as
\begin{equation}   
\ket{\rho_A}=\frac{1}{2^{N_A/2}}\prod_{j=1}^{M}\left(1+\prod_{k=1}^{\Tilde{M}}\left(c_k^\dagger\right)^{K_k^j}\right)\ket{0}    
\label{eq:rho_Afromcreations}
\end{equation}
where the creation operators $\{c_k^{\dagger}\}_{k=1}^{\Tilde{M}}$ aligns with $\{\Tilde{a}_{\Tilde{G}_k}\Tilde{G}_k\}_{k=1}^{\Tilde{M}}$ via: 
\begin{equation}
\Tilde{a}_{\widetilde{G}_k}\widetilde{G}_k \Longleftrightarrow c^\dagger_k\ket{0}=\Tilde{a}_{\widetilde{G}_k}\ket{1}      
\label{eq:prop_1}
\end{equation}
\begin{equation}
\circ\left(\Tilde{a}_{\widetilde{G}_k}\widetilde{G}_k\right)^2=I^N \Longleftrightarrow c^\dagger_k\ket{1}=\frac{1}{\Tilde{a}_{\widetilde{G}_k}}\ket{0}  
\label{eq:prop_2}
\end{equation}
\begin{equation}
\left(\Tilde{a}_{\widetilde{G}_{k_1}}\widetilde{G}_{k_1}\right)\circ\left(\Tilde{a}_{\widetilde{G}_{k_2}}\widetilde{G}_{k_2}\right)=0\Longleftrightarrow c^\dagger_{k_1}c^\dagger_{k_2}\ket{00}=0
\label{eq:prop_3}
\end{equation}
The first two properties define a matrix product operator for $c_k^\dagger$:
\begin{equation}
\mathbb{O}_k=\begin{pmatrix}
0 & \frac{1}{\Tilde{a}_{\widetilde{G}_k}} \\
\Tilde{a}_{\widetilde{G}_k} & 0
\end{pmatrix}    
\end{equation}
Since the signs of the coefficients do not impact R\'enyi entropy, we define the commutation relation as $c^\dagger_kc^\dagger_j=c^\dagger_jc^\dagger_k$, $\forall j,k$. Expanding $\ket{\rho_A}$ in basis states, such as $\ket{0100\dots0}$, gives the $a_{\bm{\sigma}}$ coefficients from Eq.~\eqref{eq:reducedensityA}. The square norm, 
\begin{equation}
\braket{\rho_A}{\rho_A}=\sum_{\bm{\sigma}}a_{\bm{\sigma}}^2    
\end{equation}
yields the second R\'enyi entropy as per  Eq.~\eqref{eq:s2_reducedA}.

Representing $\ket{\rho_A}$ as an MPS follows the standard MPO-MPS contraction process~\cite{ORUS2014117}. Specifically, for the term $\prod_{k=1}^{\Tilde{M}}\left(c_k^\dagger\right)^{K_k^j}$ from Eq.~\eqref{eq:rho_Afromcreations}, we construct the MPO as $\otimes_{k=1}^{\Tilde{M}}\mathbb{O}_k^{K_k^j}$. At each step, we obtain a new MPS as $\ket{\rho_A^m}=\left(1+\prod_{k=1}^{\Tilde{M}}\left(c_k^\dagger\right)^{K_k^j}\right)\ket{\rho_A^{m-1}}$ (with $\ket{\rho_A^0}=\ket{0}$). We then implement Eq.~\eqref{eq:prop_3} by updating the tensors $\mathbb{A}_{k_1}$ and $\mathbb{A}_{k_2}$ within $\ket{\rho_A^m}$ with $\mathbb{A}_{k_1}'$ and $\mathbb{A}_{k_2}'$ such that $\mathbb{A}_{k_1}'^{1}\mathbb{A}_{k_2}'^{1}=0$ and $\mathbb{A}_{k_1}'^{s_1}\mathbb{A}_{k_2}'^{s_2}=\mathbb{A}_{k_1}^{s_1}\mathbb{A}_{k_2}^{s_2}$ otherwise. 

Since the MPO involves long-range creation of quantum states, the resulting MPS can require a large bond dimension during this process. We can optimize the sequence of the creation operators from Eq.~\eqref{eq:rho_Afromcreations}, such that the bond dimension in the intermediate steps does not increase severely. For instance, the property Eq.~\eqref{eq:prop_3} is likely to decrease the bond dimension, so we prefer to implement them as early as possible. To this end, we use Gaussian elimination to move the Pauli strings satisfying Eq.~\eqref{eq:prop_3} to the first few MPO-MPS contraction steps. Besides, the qubit positions corresponding to the creation of $c_k^\dagger$ or $\widetilde{G}_{k}$ are adjustable, so that the long-range correlation incurred in MPS can be further reduced. We expect this direction of optimization can be further explored in future work. Noticeably, if we use this second approach to calculate the entropy for stabilizer states, we have $\mathbb{O}_k=X$, then, even if the number of $Q$'s can be large, $\ket{\rho_A}$ forms a stabilizer state, allowing efficient representation through the Clifford-MPS and Clifford disentangling algorithm. We refer to the two methods introduced here as the Pauli Coefficient Enumeration (PCE) and Pauli Coefficient Matrix Product State (PCMPS) methods.

\begin{figure*}
\centering
\includegraphics[width=\textwidth]{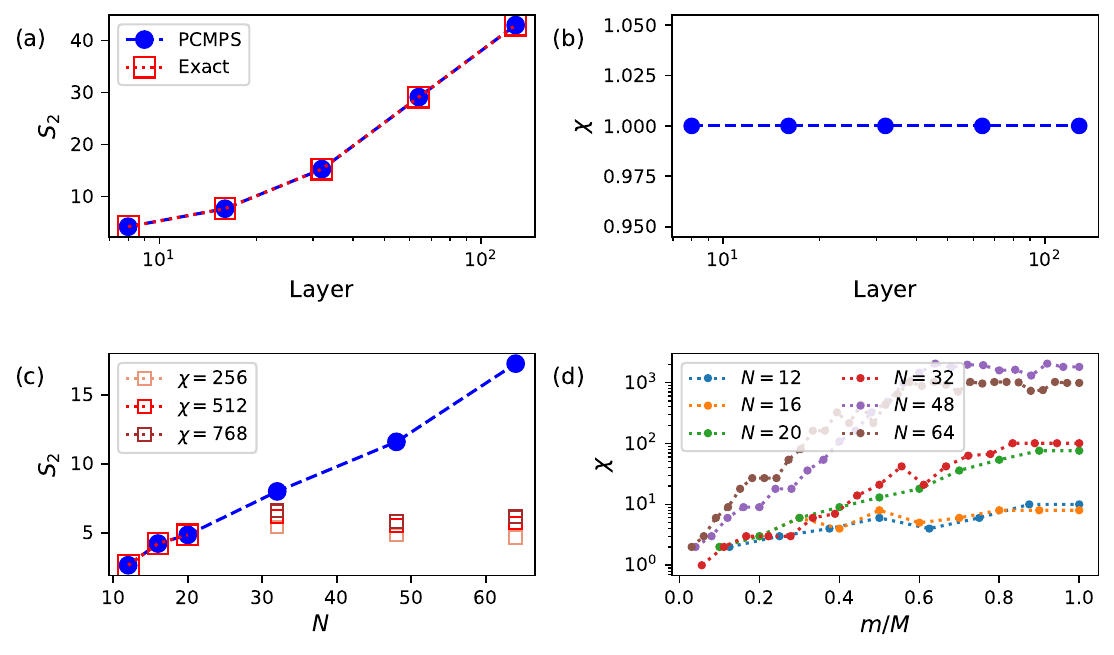}
\caption{(a) Second R\'enyi entropy of CAMPS on a Clifford circuit ($N=128$, brick-wall circuit architecture), with exact results from the standard Clifford entropy algorithm included for comparison, and (b) the bond dimension of PCMPS as a function of circuit layer; (c) Second R\'enyi entropy of CAMPS on $t$-doped Clifford circuits ($N_T=1, t=N/2$) as a function of qubit number, with the results from MPS simulation included for comparison (for $N\geq 32$, $\chi_{\text{max}}$ for MPS is set to as $\{256, 512, 768\}$), and (d) the bond dimension of PCMPS during the simulation process.  }
\label{fig:ere}
\end{figure*}

In Fig.~\ref{fig:ere}, we show the R\'enyi entropy calculated using PCMPS method. First, in Fig.~\pref{fig:ere}{a}, we apply PCMPS to a Clifford circuit ($N=128$) and demonstrate that it reproduces the exact results. Moreover, the bond dimension of PCMPS remains as 1, as shown in Fig.~\pref{fig:ere}{b}, indicating the application of PCMPS to Clifford circuits is polynomially efficient. Next, we apply it to $t$-doped Clifford circuits up to 64 qubits: as shown in Fig.~\pref{fig:ere}{c}, it reproduces the exact results in small systems, and outperforms the results from MPS simulation with comparable bond dimensions. Fig.~\pref{fig:ere}{d} shows the bond dimension from PCMPS during the calculation of entropy: for $N=48$ and $N=64$, bond truncation is involved.

\section{Conclusions}
In this work, we delineate a class of quantum circuits whose Pauli expectation values can be computed efficiently on classical computers. Interestingly, these circuits are characterized by a simple algebraic property - that the GF(2) matrix $z$ induced by the circuit is nearly full rank.  The matrix $z$ is simply the tableau of twisted Pauli strings (mapping $\{X,Y\} \rightarrow 1, \{I,Z\} \rightarrow 0$) once the $T$-gates have been commuted through the Clifford circuit towards the initial $\ket{0}$ states.  
The null space of $z$ contains the twisted Pauli strings which increase the CMPS bond-dimension (each one by at most of factor of 2).  In addition to this characterization, we also give an explicit algorithm, OFD, which removes CMPS entanglement from those linearly independent twisted Pauli strings,  ensuring a constructive way to generate a CAMPS whose CMPS bond-dimension is at most exponential in the null space dimension of $z$. 

We have additionally considered the null space dimension of particular classes of random Clifford+$T$ circuits.
Random GF(2) matrices, as one would expect to get from $T$-gates deep in a random circuit have nearly full rank and therefore small CMPS bond-dimension.  In fact, we give numerical evidence that as long as the $T$-gates induce uniformly distributed light cones of polylogarithmic size (in $N$) (e.g. polylog-depth $T$-gates), the null space dimension is at most logarithmic leading to polynomial CMPS bond-dimension.  It is an interesting open question to determine the rank of various different distributions of $T$-gates.  In addition, while OFD successfully disentangles many circuits, it is unclear if one can generate even better disentanglers for wider distributions of circuits.   A particularly interesting case is random Clifford circuits where there are an extensive number of $T$ gates both below and above logarithmic depth.

We also compared our OFD algorithm with previous OBD algorithm, and establish that for some scenarios, OFD can find the right disentanglers while OBD cannot. In addition, the simplicity of OFD also makes it a better choice for Clifford circuits doped with $t<N$ $T$-gates.

Furthermore, for $t>N$, we have numerically considered the growth of CMPS entanglement entropy with respect to the $T$-gate number, and observed the collapse  of the CMPS entanglement across various choices of $T$-gate number per layer and system size, suggesting some universal features of the CMPS entanglement dynamics. 

Finally, we introduce algorithms for the calculation of (1) the probability (and the sampling) of bitstrings from the CAMPS, (2) the wave function amplitude, and (3) the entanglement R\'enyi entropy of CAMPS. We show these approaches have polynomial cost on Clifford circuits and, while still exponential on $t$-doped Clifford circuits, outperform the traditional tensor-network simulation method.

While this work establishes a computational framework for the simulation of $t$-doped Clifford circuits, we expect it also contributes to the growing interest on the interplay between quantum entanglement and quantum magic and their roles in the boundary between classical simulability and quantum advantage, the connections between different classical algorithms for simulating Clifford+$T$ circuits, as well as development of new algorithms for the simulation of quantum systems~\cite{huang2024nonstabilizer,viscardi2025interplay,fux2024disentangl,gu2024,Bejan2024,zhang2024classicalsimul,wei2025long,Nakhl2025Stabilizer}.

\begin{acknowledgments}    
We would like to thank X. Feng for insightful discussions. This work made use of the Illinois Campus Cluster, a computing resource that is operated by the Illinois Campus Cluster Program (ICCP) in conjunction with the National Center for Supercomputing Applications (NCSA) and which is supported by funds from the University of Illinois at Urbana-Champaign. We acknowledge support from the NSF Quantum Leap Challenge Institute for Hybrid Quantum Architectures and Networks (NSF Award No. 2016136).
\end{acknowledgments}

\bibliography{main}

\appendix
\section{Notations}
\label{sec:clifmore}
In this section we introduce the notations used throughout this work. We denote the identity and Pauli operators as 
\begin{equation}
\begin{split}    
I = 
\begin{pmatrix}
1 & 0 \\ 1 & 0    
\end{pmatrix},&\quad
X = 
\begin{pmatrix}
0 & 1 \\ 1 & 0    
\end{pmatrix}, \\
Y = 
\begin{pmatrix}
0 & -i \\ i & 0    
\end{pmatrix},&\quad
Z = 
\begin{pmatrix}
1 & 0 \\ 0 & -1    
\end{pmatrix}.
\end{split}
\end{equation}
For a quantum system composed of $N$ qubits, we represent Pauli strings as the tensor product of Pauli and identity operators across the system, denoted by $\bm{\sigma}$, i.e., $\bm{\sigma}=\otimes_{j=1}^N \sigma_j=\sigma_1\sigma_2\cdots\sigma_N$, where $\sigma_j\in\{I,X,Y,Z\}$, A substring, for example, from qubit 1 to qubit $n$, is denoted as $\bm{\sigma}_{1:n}\equiv\sigma_1\sigma_2\cdots\sigma_n$. The set of all $N$-qubit Pauli strings is called the Pauli group, denoted as $\mathcal{P}_N$. 

Clifford circuits are composed of Clifford gates, a class of unitary operators that can be generated by the Hadamard, phase, and CNOT (control-X) gates:
\begin{equation}
H = \frac{1}{\sqrt{2}}
\begin{pmatrix}
1 & 1 \\ 1 & -1    
\end{pmatrix},\quad
S =
\begin{pmatrix}
1 & 0 \\ 0 & i    
\end{pmatrix},
\end{equation}
\begin{equation}
\text{CX} = 
\begin{pmatrix}
1 & 0 & 0 & 0 \\
0 & 1 & 0 & 0 \\
0 & 0 & 0 & 1 \\
0 & 0 & 1 & 0
\end{pmatrix}.
\end{equation}
An important property of the Clifford group is that the element $U_\mathcal{C}$ of it leaves the Pauli group invariant under conjugation: for all $P\in\mathcal{P}_N$, $\mathcal{C} P \mathcal{C}^\dagger\in\mathcal{P}_N$.

In addition, we list the two-qubit Clifford gates useful in this work. The basis states in the matrix format $O_{j,k}$ are on the order of $\ket{s_js_k}=\ket{00}$, $\ket{01}$, $\ket{10}$, $\ket{11}$. 
\begin{equation}
\text{SWAP}= 
\begin{pmatrix}
1 & 0 & 0 & 0 \\
0 & 0 & 1 & 0 \\
0 & 1 & 0 & 0 \\
0 & 0 & 0 & 1
\end{pmatrix}
\end{equation}

\begin{equation}
\text{XC}= 
\begin{pmatrix}
1 & 0 & 0 & 0 \\
0 & 0 & 0 & 1 \\
0 & 0 & 1 & 0 \\
0 & 1 & 0 & 0
\end{pmatrix}
\end{equation}

\begin{equation}
\text{CZ}= 
\begin{pmatrix}
1 & 0 & 0 & 0 \\
0 & 1 & 0 & 0 \\
0 & 0 & 1 & 0 \\
0 & 0 & 0 & -1
\end{pmatrix}, \quad
\text{ZC}= 
\begin{pmatrix}
1 & 0 & 0 & 0 \\
0 & 1 & 0 & 0 \\
0 & 0 & 1 & 0 \\
0 & 0 & 0 & -1
\end{pmatrix}
\end{equation}

\begin{equation}
\text{CY}= 
\begin{pmatrix}
1 & 0 & 0 & 0 \\
0 & 1 & 0 & 0 \\
0 & 0 & 0 & -i \\
0 & 0 & i & 0
\end{pmatrix}, \quad
\text{YC}= 
\begin{pmatrix}
1 & 0 & 0 & 0 \\
0 & 0 & 0 & -i \\
0 & 0 & 1 & 0 \\
0 & i & 0 & 0
\end{pmatrix}
\end{equation}
Note that here we have introduced the notations such as XC for the control-Pauli gates whose control qubit is the second qubit and target qubit is the first qubit.

The non-Clifford gate considered in this work is the $T$-gate:
\begin{equation}
T=
\begin{pmatrix}
1 & 0 \\ 0 & e^{i\pi/4}    
\end{pmatrix}.
\end{equation}
Notably, adding $T$-gates to Clifford circuits enables universal quantum computation. 

Throughout this paper, we denote vectors with bold lowercase letters (e.g., $\bm{a}$, $\bm{g}$), and matrices with bold uppercase letters (e.g., $\bm{B}$, $\bm{C}$), with elements represented by regular letters with indices (e.g., $a_j$, $B_{jk}$). Meanwhile, for a Pauli string $P$, we can express it as $P=i^{\sum_n x_nz_n}\prod_{n=1}^NX_n^{x_n}Z_n^{z_n}$, where $X_n$ ($Z_n$) are Pauli $Z$ ($X$) operators on the $n$-th qubit, and the elements of vector $\bm{x}$ ($\bm{z}$) $\in\{0,1\}^N$ indicate the presence of $X_n$ ($Z_n$).  

\section{Stabilizer Tableau for Clifford Circuits}
\label{sec:tableau}
The most common approach to representing a Clifford circuit involves tracking the set of stabilizer generators that leave the output state of the circuit invariant~\cite{Aaronson2004,gottesman1998heisenberg}. Specifically, for an $N$-qubit Clifford circuit ${\mathcal{C}}$ acting on the computational basis state $\ket{0}^{\otimes N}$, the resulting state $\ket{\psi}={\mathcal{C}}\ket{0}^{\otimes N}$ is stabilized by any operator $C$ generated by the stabilizer generators $\{g_j\}_{j=1}^N$, meaning $C\ket{\psi}=\ket{\psi}$. These generators can be expressed as $g_j={\mathcal{C}}Z_j{\mathcal{C}}^\dagger$. However, for our purposes, it is not sufficient to track only ${\mathcal{C}}Z_j{\mathcal{C}}^\dagger$, since the Clifford circuit acts on arbitrary initial states, such as MPS, which are generally not stabilized by a Pauli $Z$ operator, Thus, additional information about the Clifford circuit -- specifically, the transformations on the Pauli $X$ operator -- is necessary to enable efficient simulation. To address this need, we introduce another definition of the stabilizer tableau used throughout this work for representing the Clifford circuit. 

\begin{definition}
\label{def:tableau}
(Stabilizer tableau for Clifford circuits) Let $U_{\mathcal{C}}$ be an $N$-qubit Clifford circuit. The corresponding tableau is defined by the action of $U_{\mathcal{C}}$ on the Pauli operators at each site $n$:
\begin{equation}
\begin{split}
U_{\mathcal{C}}^\dagger X_n U_{\mathcal{C}}=i^{g_n}\prod_{k=1}^N X_k^{D_{nk}}Z_k^{F_{nk}}  \\
U_{\mathcal{C}}^\dagger Z_n U_{\mathcal{C}}=i^{a_n}\prod_{k=1}^N X_k^{B_{nk}}Z_k^{C_{nk}} 
\end{split}
\end{equation}
where $g_n, a_n\in\{0,1,2,3\}$ and $D_{nk},F_{nk},B_{nk},C_{nk}\in\{0,1\}$. The tableau is thus constructed as a set of vectors and matrices: $\{\bm{g}$, $\bm{a}$, $\bm{D}$, $\bm{F}$, $\bm{B}$, $\bm{C}\}$.
\end{definition}
Similar to the phase-sensitive stabilizer tableau introduced in Ref.~\cite{Bravyi2019simulationofquantum}, Definition~\ref{def:tableau} focuses solely on the Clifford operators themselves (allowing for applicability to arbitrary initial states) and also tracks phase factors. This approach (Definition~\ref{def:tableau}), however, formalizes the representation in a simpler manner, without the need to classify Clifford gates into different types and treat them separately as done in Ref.~\cite{Bravyi2019simulationofquantum}.  

With Definition~\ref{def:tableau}, we can commute any Pauli string $P$ through the Clifford circuit, obtaining the resulting twisted Pauli string as $P'=\mathcal{C}^\dagger P \mathcal{C}$ using the following algorithm:
\begin{algorithm}[H]
\caption{Pauli string commuting algorithm}\label{alg:CommutePauli}
\begin{flushleft}
\hspace*{\algorithmicindent} \textbf{Input}: Pauli string $P=i^{\sum_n x_nz_n}\prod_{n=1}^NX_n^{x_n}Z_n^{z_n}$, stabilizer tableau for the Clifford circuit $\bm{g}$, $\bm{a}$, $\bm{D}$, $\bm{F}$, $\bm{B}$, $\bm{C}$.
\end{flushleft}
\begin{algorithmic}[1]
\State Initialize: $\bm{X}=\bm{0}_{2N\times N}$, $\bm{Z}=\bm{0}_{2N\times N}$, $s=\sum_{n=1}^N x_nz_n$. 
\For{$n=1,2,\cdots,N$}
\If{$x_n=1$}
\State $\bm{X}[2n]\gets \bm{D}[n]$ \Comment{$\bm{X}[2n]=(2n)$-th row of $\bm{X}$}
\State $\bm{Z}[2n]\gets \bm{F}[n]$
\State $s \gets s+g_n$.
\EndIf
\If{$z_n=1$}
\State $\bm{X}[2n+1]\gets \bm{B}[n]$
\State $\bm{Z}[2n+1]\gets \bm{C}[n]$
\State $s \gets s+a_n$.
\EndIf
\EndFor
\State $x_n'\gets\sum_{k=1}^{2N} X_{kn}$; $z_n'\gets\sum_{k=1}^{2N} Z_{kn}$.
\State $\bm{Z}^c=$ cumulative sum of $\bm{Z}$ along the rows.
\State $s\gets s+2\sum_{n=1}^N\sum_{k=1}^{2N-1}Z^c_{kn}X_{k+1,n}$.
\end{algorithmic}
\begin{flushleft}
\hspace*{\algorithmicindent} \textbf{Output}: The Pauli string after commuting through the Clifford circuit $P'=i^s\prod_{n=1}^NX_n^{x'_n}Z_n^{z'_n}$.
\end{flushleft}
\end{algorithm}

Here we provide justification for Algorithm~\ref{alg:CommutePauli}. By writing Pauli string as $P=i^{\sum_n x_nz_n}\prod_{n=1}^NX_n^{x_n}Z_n^{z_n}$, where $x_n,z_n\in\{0,1\}$, and using Definition~\ref{def:tableau}, we have
\begin{equation}
\begin{split}    
P'=U_\mathcal{C}^\dagger P U_\mathcal{C}=i^{\bm{x}\cdot\bm{z}}\prod_{n=1}^N(\mathcal{C}^\dagger X_n \mathcal{C})^{x_n}(\mathcal{C}^\dagger Z_n \mathcal{C})^{z_n}\\
=i^{\bm{x}\cdot\bm{z}+\bm{x}\cdot\bm{g}+\bm{z}\cdot\bm{a}}\prod_{n=1}^N\left(\prod_kX_k^{D_{nk}}Z_k^{F_{nk}}\right)^{x_n}\left(\prod_kX_k^{B_{nk}}Z_k^{C_{nk}}\right)^{z_n}
\end{split}
\end{equation}
Our goal is to rearrange the above expression as $P'=i^s\prod_{k}X_k^{x'_k}Z_k^{z'_k}$. To this end, we need to commute the $X$ operators through the $Z$ operators such that the $X/Z$'s acting on the same site can be combined together. This process will give an extra sign factor as $(-1)^{s'}$ with
\begin{equation}
\begin{split}    
s'=&\sum_{n=1}^N\sum_{k=2}^N\left(\sum_{k'=1}^{k-1}F_{nk'}x_n+C_{nk'}z_n\right)\left(B_{nk}z_n+D_{nk}x_n\right)\\
&+\sum_{n=1}^N\sum_{k=1}^N F_{nk}x_n B_{nk}z_n
\end{split}
\end{equation}
It is easy to verify that this extra sign factor is taken into account by Line 16 in Algorithm~\ref{alg:CommutePauli}. 

While a stabilizer tableau can be efficiently constructed when appending simple Clifford gates (e.g., CNOT, H, or S gates) to a larger Clifford circuit with a known stabilizer tableau~\cite{Aaronson2004,gottesman1998heisenberg,Bravyi2019simulationofquantum}, here we provide the formal update method for stabilizer tableau when two arbitrary Clifford circuits are concatenated. This is summarized in the following proposition:
\begin{prop}
\label{prop:U2U1concate}
(Concatenation of Clifford circuits) Let $U_1$ and $U_2$ be two $N$-qubit Clifford circuits with stabilizer tableaux $\{\bm{g}^{[1]}, \bm{a}^{[1]}, \bm{D}^{[1]}, \bm{F}^{[1]}, \bm{B}^{[1]}, \bm{C}^{[1]}\}$ and $\{\bm{g}^{[2]}, \bm{a}^{[2]}, \bm{D}^{[2]}, \bm{F}^{[2]}, \bm{B}^{[2]}, \bm{C}^{[2]}\}$, respectively. Then, the stabilizer tableau for the concatenated circuit $U_\mathcal{C}=U_2U_1$ is given by $\{\bm{g}, \bm{a}, \bm{D}, \bm{F}, \bm{B}, \bm{C}\}$, as follows:    
\begin{equation}
\label{eq:DF}
\bm{D}=\bm{D}^{[2]}\bm{D}^{[1]}+\bm{F}^{[2]}\bm{B}^{[1]},    
\quad
\bm{F}=\bm{D}^{[2]}\bm{F}^{[1]}+\bm{F}^{[2]}\bm{C}^{[1]},    
\end{equation}
\begin{equation}
\label{eq:BC}
\bm{B}=\bm{B}^{[2]}\bm{D}^{[1]}+\bm{C}^{[2]}\bm{B}^{[1]},    
\quad
\bm{C}=\bm{B}^{[2]}\bm{F}^{[1]}+\bm{C}^{[2]}\bm{C}^{[1]},    
\end{equation}
\begin{equation}
\label{eq:gn}
\begin{split}
g_n=g^{[2]}_n+\sum_{k=1}^N (D^{[2]}_{nk}g^{[1]}_k+F^{[2]}_{nk}a^{[1]}_k)+2\sum_{k,l=1}^N D^{[2]}_{nk}F^{[2]}_{nk}F^{[1]}_{kl}B^{[1]}_{kl}\\
+\sum_{l=1}^N\sum_{k=2}^N(\sum_{k'=1}^{k-1}D^{[2]}_{nk'}F^{[1]}_{k'l}+F^{[2]}_{nk'}C^{[1]}_{k'l})(D^{[2]}_{nk}D^{[1]}_{kl}+F^{[2]}_{nk}B^{[1]}_{kl}),    
\end{split}
\end{equation}
\begin{equation}
\label{eq:an}
\begin{split}
a_n=a^{[2]}_n+\sum_{k=1}^N (B^{[2]}_{nk}g^{[1]}_k+C^{[2]}_{nk}a^{[1]}_k)+2\sum_{k,l=1}^N B^{[2]}_{nk}C^{[2]}_{nk}F^{[1]}_{kl}B^{[1]}_{kl}\\
+\sum_{l=1}^N\sum_{k=2}^N(\sum_{k'=1}^{k-1}B^{[2]}_{nk'}F^{[1]}_{k'l}+C^{[2]}_{nk'}C^{[1]}_{k'l})(B^{[2]}_{nk}D^{[1]}_{kl}+C^{[2]}_{nk}B^{[1]}_{kl}).    
\end{split}
\end{equation}
\end{prop}

Eq.~\eqref{eq:DF} and~\eqref{eq:BC} involve straightforward matrix multiplications and are thus computationally manageable. While the phase factor calculations Eq.~\eqref{eq:gn} and~\eqref{eq:an} may appear more intricate, they can be implemented efficiently by counting how many times Pauli $X$ operators with the Pauli $Z$ operators and consolidating the Pauli $X$/$Z$ for each qubit. This approach requires only tracking the number of anti-commuting $Z$ operators preceding each Pauli $X$ operator, a strategy that applies identically in Algorithm~\ref{alg:CommutePauli}.

Here we prove Proposition~\ref{prop:U2U1concate} by deriving Eq.~\eqref{eq:DF} and~\eqref{eq:gn}, whereas the other two formulas can be derived similarly. 
\begin{equation}
\begin{split}
&U_\mathcal{C}^\dagger X_n U_\mathcal{C}=U_1^\dagger(U_2^\dagger X_nU_2)U_1\\
=&i^{g^{[2]}_n}\prod_{k=1}^N(U_1^\dagger X_k U_1)^{D^{[2]}_{nk}}(U_1^\dagger Z_k U_1)^{F^{[2]}_{nk}}\\
=&i^{g^{[2]}_n}\prod_{k=1}^N\left( i^{g^{[1]}_k}\prod_{l=1}^N X_l^{D^{[1]}_{kl}} Z_l^{F^{[1]}_{kl}} \right)^{D^{[2]}_{nk}} \left( i^{a^{[1]}_k}\prod_{l=1}^N X_l^{B^{[1]}_{kl}} Z_l^{C^{[1]}_{kl}} \right)^{F^{[2]}_{nk}}
\end{split}
\end{equation}
From the above expression, it is easy to see that 
\begin{equation}
D_{nl}=\sum_{k=1}^N D^{[2]}_{nk}D^{[1]}_{kl}+F^{[2]}_{nk}B^{[1]}_{kl}    
\end{equation}
\begin{equation}
F_{nl}=\sum_{k=1}^N D^{[2]}_{nk}F^{[1]}_{kl}+F^{[2]}_{nk}C^{[1]}_{kl}    
\end{equation}
which are exactly the results from Eq.~\eqref{eq:DF}. As of the exponent for the phase factor, i.e., $g_n$, we do the same procedure as for Algorithm~\ref{alg:CommutePauli}, that is, recording the additional sign factors from the process of commuting the $X$ operators through the $Z$ operators. This process eventually produces Eq.~\eqref{eq:gn}. For algorithm implementation, the phase factor can be computed in a similar way to Algorithm.~\ref{alg:CommutePauli}.  

\section{Example where OBD fails finding the disentangler}
\label{sec:failOBD}
In this section, we give an explicit example of twisted Pauli strings $\{\overline{Z}^{[k]}\}$ that OBD cannot find the correct disentangler while OFD can. 

\begin{equation}
\overline{Z}=
\begin{pmatrix}
I & X & I & I & I \\
I & I & X & I & I \\
I & I & I & X & I \\
I & X & X & X & I \\
I & X & X & I & I \\
X & I & I & I & X
\end{pmatrix}
\end{equation}
In this example, the first three Pauli strings are simply single-qubit Pauli operators, and do not require disentanglers for them, making the initial state $\ket{00000}$ into $\ket{0mmm0}$ after the applications. The next two Pauli strings $Z^{[4]}$ and $Z^{[5]}$ are undisentanglable for OFD, according to Theorem~\ref{theo:OFD}. They are undisentanglable for OBD either, by examining all the two-qubit Clifford gates. The resulting CMPS has the second qubit entangled with other qubits by twice: $\ket{0\overline{\overline{mm}m}0}$, hence the entanglement entropy at the cut between the second and third qubits is larger. When the final twisted Pauli string $Z^{[6]}$ is applied, OBD cannot swap the first qubit or the fifth qubit towards each other because of this entanglement barrier between the second and the third qubits. However, OFD is able to find the long-range control-Pauli gate $\mathrm{CX}_{1,5}$ to disentangle $Z^{[6]}$. 

\section{Rank of random GF(2) matrix}
\label{sec:derivetfree}
For a random $N\times N$ GF(2) matrix whose entries are 0/1 with equal likelihoods, the probability for this matrix to be of rank $N-s$ (here $s$ will be the number of entangling $T$ gates among the first $N$ $T$-gates) is~\cite{kolchin1999random} 
\begin{equation}
\begin{split}
\Pr(N,s)=2^{-s^2}& \prod_{k=s+1}^{N}\left(1-\frac{1}{2^k}\right) \\
&\times\left(\sum_{0\leq k_1\leq \cdots\leq k_s\leq N-s}2^{-k_1-\cdots-k_s}\right). 
\end{split}
\end{equation}
When $N\to\infty$, it simplifies to 
\begin{equation}
\Pr(s)=2^{-s^2}\prod_{k=s+1}^{\infty}\left(1-\frac{1}{2^k}\right)\prod_{k=1}^{s}\left(1-\frac{1}{2^k}\right)^{-1}. 
\end{equation}

\begin{figure}
\centering
\includegraphics[width=0.5\textwidth]{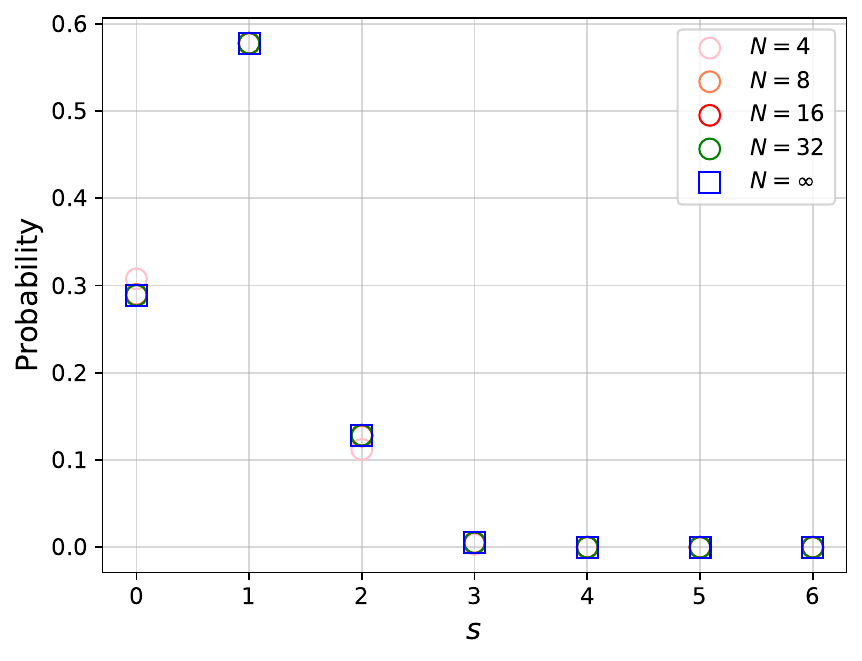}
\caption{Probability for an $N\times N$ random GF(2) matrix to be of rank $(N-s)$.}
\label{fig:rankprob}
\end{figure}

As shown in Fig.~\ref{fig:rankprob}, the probability for the random GF(2) matrix to be of rank $N-s$ quickly goes to 0 as $s$ exceeds 2. We can get the average rank by only considering the first $s_{\text{cut}}$ number of them, given the fact that the total contribution of remaining $s$ is less than $e^{-s_{\text{cut}}^2}$. This gives rise to the average rank as $N-\gamma$ with $\gamma\approx0.85$ independent of $N$.

On the other hand, we can consider the number of $T$-gates needed to consume all the free qubits in CAMPS. In such a situation, the distribution of the twisted Pauli strings is invariant even after commuting through the disentanglers as any Clifford operator is an one-to-one mapping between Pauli strings. In fact, even in the case where $w<N$, one can show that for any qubit (or window of qubits) which are uniformly distributed over Pauli strings, they map to a uniform distribution on that same window while potentially swapping some Pauli terms outside that window; this helps ensure that uniformity of the distribution prior to the disentanglers is preserved after commuting through them. 
The probability for turning one qubit into magic, if there are already $N_m$ qubits in magic states, goes as 
\begin{equation}
p(N_m)=1-\left(\frac{1}{2}\right)^{N-N_m},\quad N_m=0,1,2,\dots,N-1
\end{equation}
which is the probability for $\widetilde{Z}^{[k]}$ having at least one non-trivial Pauli term on the non-magic qubits, such that OFD is applicable and one of these qubits become magic after disentangling. The expected number of Pauli strings to find a usable control qubit is $1/p(N_m)$ and therefore, the overall average number of Pauli strings needed to consume all the $N$ free qubits is given by
\begin{equation}
t_{\text{free}}=\sum_{N_m=0}^{N-1} \frac{1}{1-\left(\frac{1}{2}\right)^{N-N_m}}   
\label{eq:tfree}
\end{equation}
from which, we have (see below for the derivation)
\begin{equation}
N+1-\frac{1}{2^N}<t_{\text{free}}<N+2-\frac{1}{2^{N-1}}, \quad \forall N
\label{eq:treebound}
\end{equation}
In other words, (in expectation) all but a small constant number of $T$-gates amongst the first $N$ gates are free. This expectation is an upper bound; in fact, we ignored the probability for $\widetilde{Z}^{[k]}$ being a trivial Pauli string, i.e., with $I/Z$ on the non-magic qubits in $\ket{0}$, and $I$ on the magic qubits, which would not require any disentangler;  nonetheless, this will have only a small effect on this upper bound as this latter probability is small ($\left(1/2\right)^{N-N_m}\left(1/4\right)^{N_m}$) compared to $p(N_m)$.

Here we provide the derivation for Eq.~\eqref{eq:treebound}. First, we rewrite Eq.~\eqref{eq:tfree} as
\begin{equation}
t_{\text{free}}=N+\sum_{N_m=1}^N\frac{1}{2^{N_m}-1}. 
\end{equation}
On the one hand, we have 
\begin{equation}
\frac{1}{2^{N_m}-1}>\frac{1}{2^{N_m}}    
\end{equation}
which gives
\begin{equation}
t_{\text{free}}>N+\sum_{N_m=1}^N\frac{1}{2^{N_m}}=N+1-\frac{1}{2^N}    
\end{equation}
On the other hand, we have
\begin{equation}
\frac{1}{2^{N_m}-1}=\frac{\frac{1}{2^{N_m}}}{1-\frac{1}{2^{N_m}}}<\frac{2}{2^{N_m}}    
\end{equation}
which gives
\begin{equation}
t_{\text{free}}<N+\sum_{N_m=1}^N\frac{2}{2^{N_m}}=N+2-\frac{1}{2^{N-1}}.    
\end{equation}

\section{More numerical results on random $(t\leq N)$-doped Clifford circuits}
\label{sec:moreNumericOntlessN}
\begin{figure}
\centering
\includegraphics[width=0.50\textwidth]{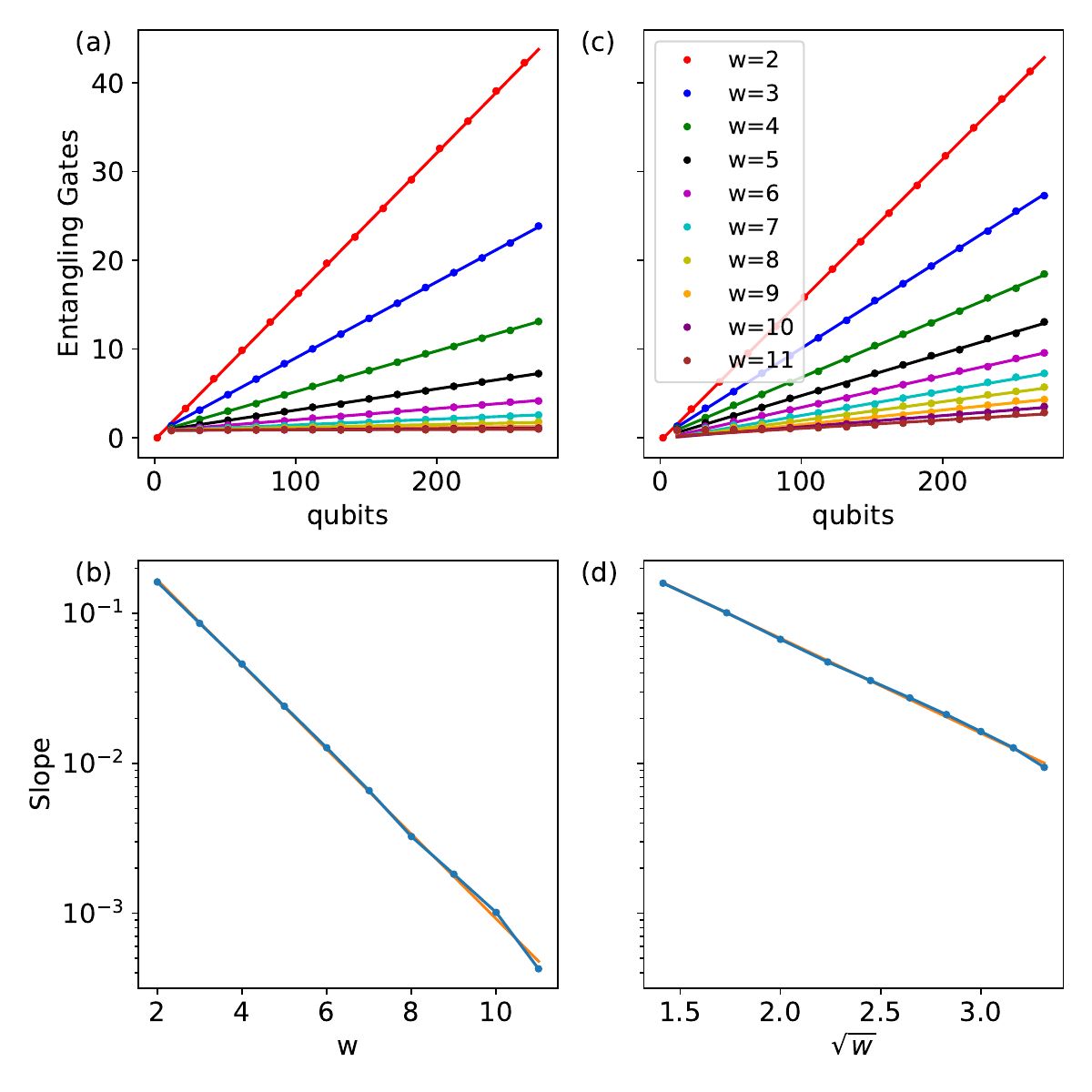}
\caption{Average number of entangling gates versus qubit number $N$ from uniform-distribution model for twisted Pauli strings $\{\overline{Z}^{[k]}\}_{k=1}^N$ of various window-size $w$: (a) a set of $N$ windows of size $w$ and (c) a sets of $N/2$ windows of size $w$ and $N/2$ windows of size
$w+2$. Error bars are below the size of dots. The lines are best-fit to all but the first five data points. Windows within each set are centered on unique qubits. (b,d) Slopes of (a) and (c) versus $w$ and $\sqrt{w}$ respectively.}
\label{fig:entanglingGateModel_uniform}
\end{figure}

In Fig.~\ref{fig:entanglingGateModel_uniform}, we show the numerical results from the uniform model for twisted Pauli strings $\{\overline{Z}^{[k]}\}_{k=1}^N$, where we generate each Pauli string of a given window size $w$ randomly from $\{I,X,Y,Z\}$ and centered at different random qubits. They display the same scaling behaviors as the those from Fig.~\ref{fig:entanglingGateModel} in the main text (note that $w=4d$), which are produced directly from random Clifford+$T$ circuits.

\begin{figure}
\centering
\includegraphics[width=0.5\textwidth]{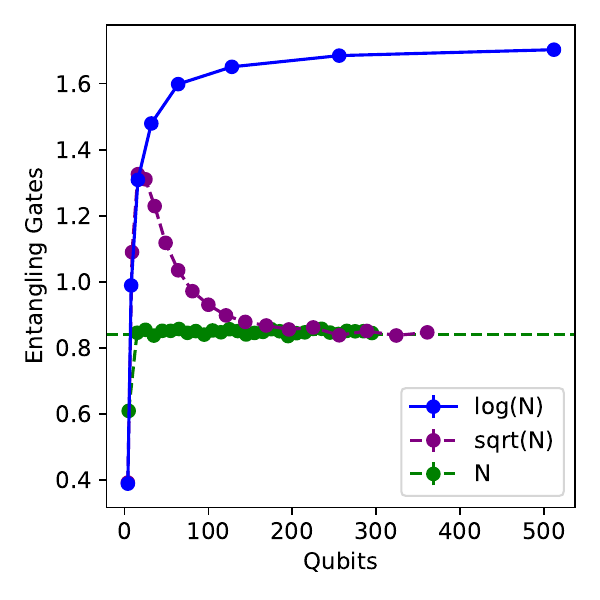}
\caption{Expected number of entangling gates from the uniform-distribution model, as a function of qubits $N$ for $N$ different windows of size $w$ where $w$ scales as in the legend and where each window is centered on one of the $N$ qubits.}
\label{fig:entanglingGates}
\end{figure}

In Fig.~\ref{fig:entanglingGates}, within the uniform-distribution model, we show that the number of entangling gates scales logarithmically or sub-logarithmically with $N$ when $w\propto\log(N)$, while for other kinds of $w$ scaling, this number converges to constant as $N$ grows. Windows of size larger than $\log(N)$ can only be easier to disentangle as one can always use the control qubits as if one has access to a smaller window. We also see that the distribution of displayed data in Fig.~\ref{fig:entanglingGates} is tightly centered around the average, suggesting that the typical circuit matches the average circuit. Note that a logarithmic number of entangling $T$-gates can always be dealt with in polynomial time as the bond-dimension of CAMPS scales at most exponentially in the number of such gates.

\begin{figure}
\centering
\includegraphics[width=0.5\textwidth]{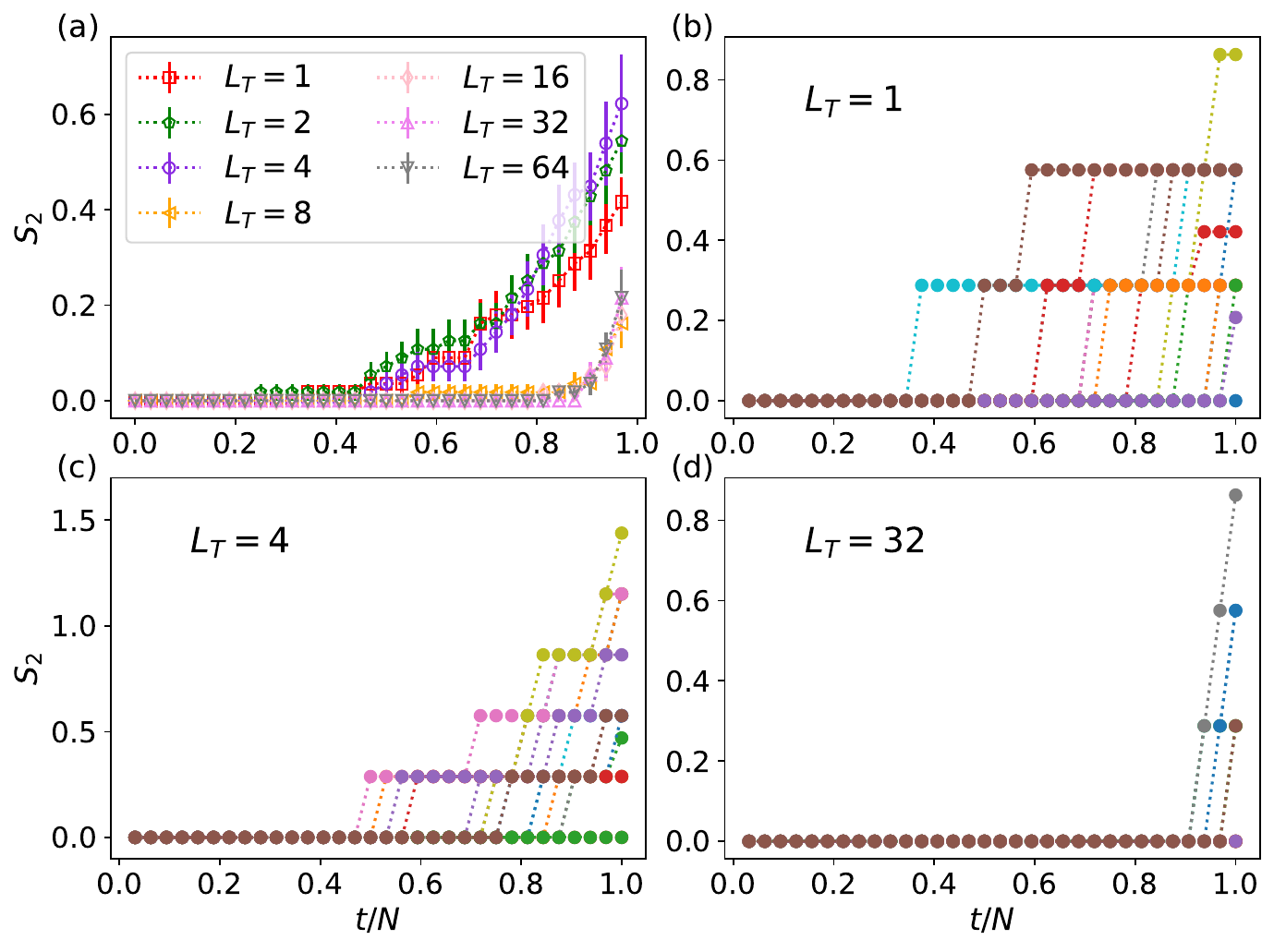}
\caption{The CMPS second R\'enyi entropy from the simulation of 1D $t$-doped Clifford circuits with CAMPS ($N_T=N=32$). (a) $S_2$ averaged over 16 random circuits for various $L_T$; (b)-(d) $S_2$ from individual circuit realizations. The $T$-gates from the same layer are absorbed into CAMPS in random order.}
\label{fig:t1layer}
\end{figure}

\begin{figure}
\centering
\includegraphics[width=0.5\textwidth]{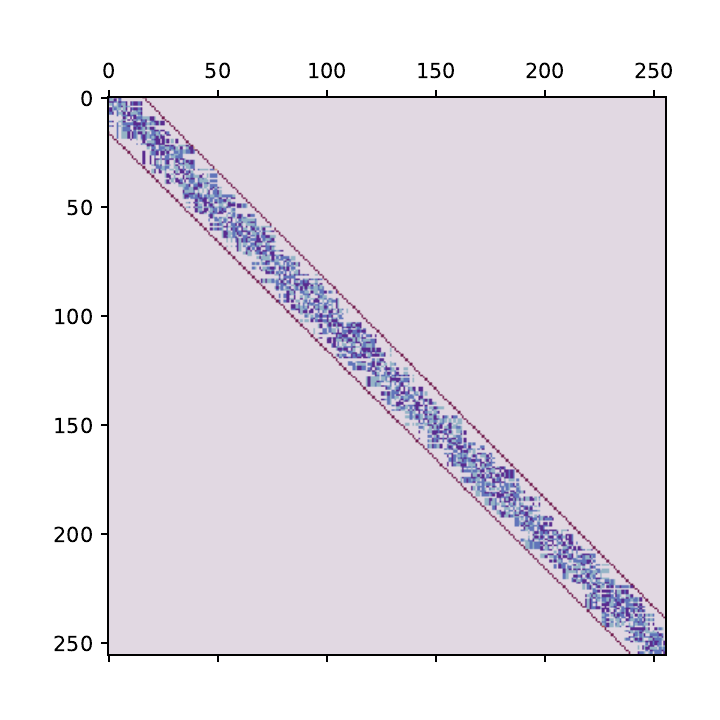}
\caption{Prototypical example of a distribution of Pauli-strings coming from the 256 Pauli terms on 256 qubits with $N=256$ $T$-gates uniformly located at depth $\log 256=8$ (so that $w=16$).  Light purple color corresponds to identity terms with other darker colors corresponding to $X,Y,Z$.  The red lines show the maximum window size that the Pauli-terms can be located on. 
}
\label{fig:distribution}
\end{figure}

In Fig.~\ref{fig:t1layer}, we present the CMPS entanglement entropy for representing the $t$-doped Clifford circuits with CAMPS using OFD. The circuits have all the first $N$ $T$-gates on the $L_T$-th layer. From Fig.~\pref{fig:t1layer}{a}, we observe that $S_2$ starts increasing when $t\approx 0.9N$ for $w=2L_T>\log(N)$, which, along with the $S_2$ displayed in Fig.~\pref{fig:t1layer}{b-d} for individual circuit samples, indicate only a small number of entangling $T$-gates exist for $w>\log(N)$.
However, for $T$-gates of depth smaller than $\log(N)$ (i.e constant depth) there are too many entangling gates;  therefore, naively, one might expect if one has a large number of $T$-gates at constant depth, this would be problematic. 

In the limit where uniformity of $T$-gates is strongly violated, we can see that our simulation techniques break down.  Consider the case where the windows from all $N$ $T$-gates actually span only a sub-linear number of $k$ qubits.  In such a case, this is equivalent to running on a circuit of $k$ qubits and we will not achieve more than $k$ free $T$-gates as there are only $k$ $\ket{0}$ to consume.  It is worth pointing out that this "uniformity" requirement is on the location of the Pauli-string windows and so $T$-gates that are all to the left-half of the system but at varying depths still are likely to be sufficiently uniform in window-location given the spread of the light cone. 

In our simplified model, we've been using an assumption that the Pauli strings are uniformly distributed.  In practice, this model is not fully realistic.  For example, in Fig.~\ref{fig:distribution} we can see that in the window size induced by the light-cone where Pauli strings can be supported, there is significantly less non-identity Pauli terms at the edge of the window then in the center for a prototypical example on a random Clifford circuit.  It is an interesting open question whether there are more realistic models of Pauli string distributions that are applicable to different distributions of circuits. 

\section{Improved Stabilizer Group Learning Algorithm}
\label{sec:stablearn}
In this section we present an improved algorithm to compute the stabilizer nullity efficiently, one of the measures for quantum magic. Nullity has been studied as a crucial quantity for the boundary of classical simulability~\cite{gu2024}, and here will provide further insights on the classical simulation of $t$-doped Clifford circuits with Clifford-MPS states. Stabilizer nullity is defined as $\nu=N-|G|$, where $G$ is a generator set (of size $|G|$) that produces Pauli strings $\bm{\sigma}$ (stabilizers) satisfying $|\expect{\psi}{\bm{\sigma}}{\psi}|=1$. 

To compute the stabilizer nullity, or equivalently, determine the number of stabilizer generators 
$|G|$ for Clifford-MPS states, we employ an algorithm improved upon the one from Ref.~\cite{Lami2024}. And as we will show soon, the improvement reduces the original exponential cost to a linear cost. The algorithm leverages a theorem (for convenience, we call it as \textit{threshold condition} here) from Ref.\cite{Lami2024}:

\begin{theorem}\label{theo:threshold}
(Threshold condition Ref.~\cite{Lami2024}) Given an $N$-qubit matrix product state $\ket{\psi}=\sum_{\bm{s}}\mathbb{A}^{s_1}_1\mathbb{A}^{s_2}_2\cdots\mathbb{A}^{s_N}_N\ket{\bm{s}}$, where $\bm{s}=(s_1,s_2,\cdots,s_N)$, and $\mathbb{A}^{s_j}_j$ are $\chi_{j-1}\times\chi_j$ matrices ($\chi_0=\chi_N=1$), if an $N$-qubit Pauli string $\bm{\sigma}\equiv\sigma_1\sigma_2\cdots\sigma_N$ stabilizes $\ket{\psi}$, i.e., $\bm{\sigma}\ket{\psi}=\pm\ket{\psi}$, then the marginal probability of any first $n$-qubit substring $\bm{\sigma}_{1:n}\equiv\sigma_1\sigma_2\cdots\sigma_n$ from $\bm{\sigma}$, which is defined by $\pi(\bm{\sigma}_{1:n})=\sum_{\bm{\sigma}'_{n+1:N}}\Pi(\bm{\sigma}_{1:n}\bm{\sigma}'_{n+1:N})$ and $\Pi(\bm{\sigma}_{1:n}\bm{\sigma}'_{n+1:N})=\frac{1}{2^N}|\expect{\psi}{\bm{\sigma}_{1:n}\bm{\sigma}'_{n+1:N}}{\psi}|^2$, must satisfy
\begin{equation}
\pi(\bm{\sigma}_{1:n})\geq\frac{1}{2^n\chi_n}    
\end{equation}
For right-normalized $\ket{\psi}$, i.e., $\sum_{s_j}\mathbb{A}^{s_j}_j(\mathbb{A}^{s_j}_j)^{\dagger}=\1$, $\pi(\bm{\sigma}_{1:n})$ can be efficiently calculated as 
\begin{equation}
\begin{split}    
\pi(\bm{\sigma}_{1:n})=\Tr(\mathbb{L}(\bm{\sigma}_{1:n})^\dagger\mathbb{L}(\bm{\sigma}_{1:n})) \text{,  with} \\
\mathbb{L}(\bm{\sigma}_{1:n})=\frac{1}{\sqrt{2}}\sum_{s',s}\sigma_n^{s's}(\mathbb{A}^{s'}_n)^{\dagger}\mathbb{L}(\bm{\sigma}_{1:n-1})\mathbb{A}^{s}_n, \\
\mathbb{L}(\bm{\sigma}_{1:0})=1.
\end{split}
\end{equation}
\end{theorem}

The original algorithm’s primary insight is that only Pauli strings meeting the threshold condition during autoregressive sampling from qubit 1 to $N$ are retained. Our improvement reduces redundancy by tracking only the maximal generator set (via methods like Gaussian elimination) for Pauli substrings that satisfy the threshold condition throughout the autoregressive sampling. The complete algorithm is detailed in Algorithm~\ref{alg:stabMPS}. 

\begin{algorithm}[H]
\caption{Stabilizer group learning algorithm for MPS}\label{alg:stabMPS}
\begin{flushleft}
\hspace*{\algorithmicindent} \textbf{Input}: an $N$-qubit right-normalized MPS $\ket{\psi}$.
\end{flushleft}
\begin{algorithmic}[1]
\State Initialize: Generator set $G=\{\}$. 
\For{$n=1,2,\cdots,N$}
\State Initialize: Pauli substring set $\Sigma=\{\}$.
\State Compute $\pi(\bm{\sigma}_{1:n-1}\sigma_n)$ for $\sigma_n\in\{I,X,Y,Z\}$ and $\bm{\sigma}_{1:n-1}\in G$.
\State Add $\bm{\sigma}_{1:n-1}\sigma_n$ to $\Sigma$ \textbf{if} $\pi(\bm{\sigma}_{1:n-1}\sigma_n)\geq 1/(2^n\chi_n)$. 
\State Sort $\Sigma$ in descending order according to  $\pi(\bm{\sigma}_{1:n})$.
\State $G=\{\}$.
\For{($j=1,2,\cdots,|\Sigma|$)}
\State Add $\bm{\sigma}^{j}$ to $G$ \textbf{if} $\bm{\sigma}^{j}\in\Sigma$ cannot be generated by $G$.   
\EndFor
\EndFor
\end{algorithmic}
\begin{flushleft}
\hspace*{\algorithmicindent} \textbf{Output}: stabilizer group generators $G=\{\bm{\sigma}^{j}\}_{j=1}^{|G|}$ for $\ket{\psi}$.
\end{flushleft}
\end{algorithm}

\begin{figure}
\centering
\includegraphics[width=0.5\textwidth]{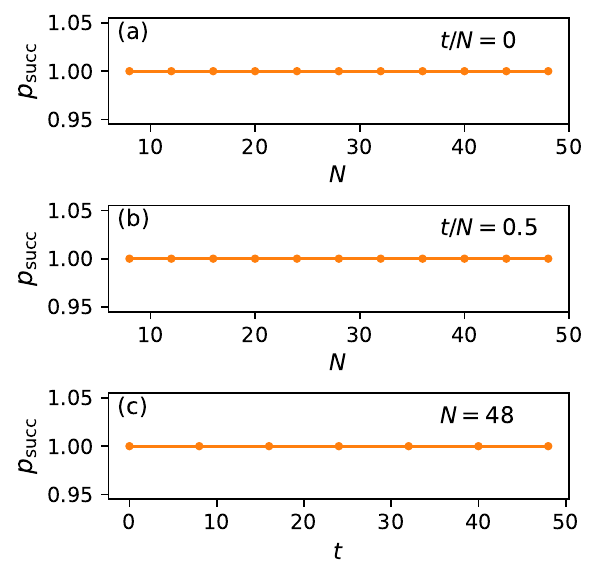}
\caption{Success probabilities of the stabilizer group learning algorithm finding the correct stabilizer group for $U_{\mathcal{C}}\ket{N,t}$ with (a) $t/N=0$, (b) $t/N=0.5$ and (c) $N=48$. Each data point averages over 1,000 random realizations of $U_{\mathcal{C}}\ket{N,t}$.}
\label{fig:learnstab}
\end{figure}

As a benchmark, we test this algorithm on MPS for states prepared as $U_{\mathcal{C}}\ket{N,t}$, where $U_{\mathcal{C}}$ is an 1D brick-wall circuit with random 2-qubit Clifford gates, and $\ket{N,t}=\ket{+}^{\otimes N-t}\ket{m}^{\otimes t}$, with $\ket{+}=(\ket{0}+\ket{1})/\sqrt{2}$ and $\ket{m}$ being random single-qubit non-stabilizer states. Fig.~\ref{fig:learnstab} shows 100\% success rate over 1,000 random realizations of $U_{\mathcal{C}}\ket{N,t}$ for various values of $N$ and $t$. 

In fact, the Clifford disentangling algorithm OBD is also applicable to the learning of stabilizer group for MPS as well as the calculation of stabilizer nullity. However, since it is a heuristic optimization algorithm, it might suffer from the issues of local minima or barren plateaus. The improved algorithm in this section is deterministic and more efficient for stabilizer group learning.

As have been noted, the purpose of reserving only the generator set, rather than all the Pauli strings as done in the original algorithm, is to reduce redundancy. The generator set is sufficient to describe the entire stabilizer group, which contains an exponential number of elements. Specifically, suppose $\{\bm{\sigma}^1,\bm{\sigma}^2,\cdots,\bm{\sigma}^{|G|}\}$ are the stabilizer generators for the MPS $\ket{\psi}$. Their substrings, up to the $n$-th qubit, $\{\bm{\sigma}_{1:n}^1,\bm{\sigma}_{1:n}^2,\cdots,\bm{\sigma}_{1:n}^{|G|}\}$, will be included in the Pauli string set that satisfies the threshold condition up to the $n$-th qubit. The group generated from them, which is of size at most $2^{|G|}$, will also be included. Since these Pauli strings will always satisfy the threshold condition in the remaining sampling steps, we can avoid this exponential overhead by keeping track of the generator set until the end. However, it might occur that two Pauli substrings $\Tilde{\bm{\sigma}}_{1:n}^1$, $\Tilde{\bm{\sigma}}_{1:n}^2$, which individually are not stabilizer Pauli substrings, can combine to form a stabilizer Pauli substring $\bm{\sigma}_{1:n}=\Tilde{\bm{\sigma}}_{1:n}^1\Tilde{\bm{\sigma}}_{1:n}^2$, and get selected into the reserved generator set, i.e., $\{\Tilde{\bm{\sigma}}_{1:n}^1,\Tilde{\bm{\sigma}}_{1:n}^2,\cdots,\bm{\sigma}_{1:n}^{|G|}\}$. This would result in a final generator set that misses two correct stabilizer generators. Fortunately, based on the benchmarking results (Fig.~\ref{fig:learnstab}), the improved algorithm seems to be able to rule out this possibility.  

As a result, the stabilizer group learning algorithm for MPS is improved in the following aspects: (1) Only a single iteration over the qubits is required to learn all the stabilizer generators. (2) No additional Clifford circuit layers are needed to modify the MPS for sampling the undiscovered stabilizers as done in the original algorithm. (3) Since the maximum number of generators for the $n$-qubit Pauli substrings is $2n$, the maximal number of reserved Pauli substrings during the autoregressive sampling do not need to be a heuristically tuned parameter for large system sizes and now is determinable as $2N$.

\section{Alternative View on the $t$-doped Clifford Circuits Simulation}

In this section, we provide an alternative view on the classical simulation of $t$-doped Clifford circuits with CAMPS. We start with the empirical observations before putting forward the criterion for the classical simulability based on the principle of OFD.  

\begin{figure*}
\centering
\includegraphics[width=1.0\textwidth]{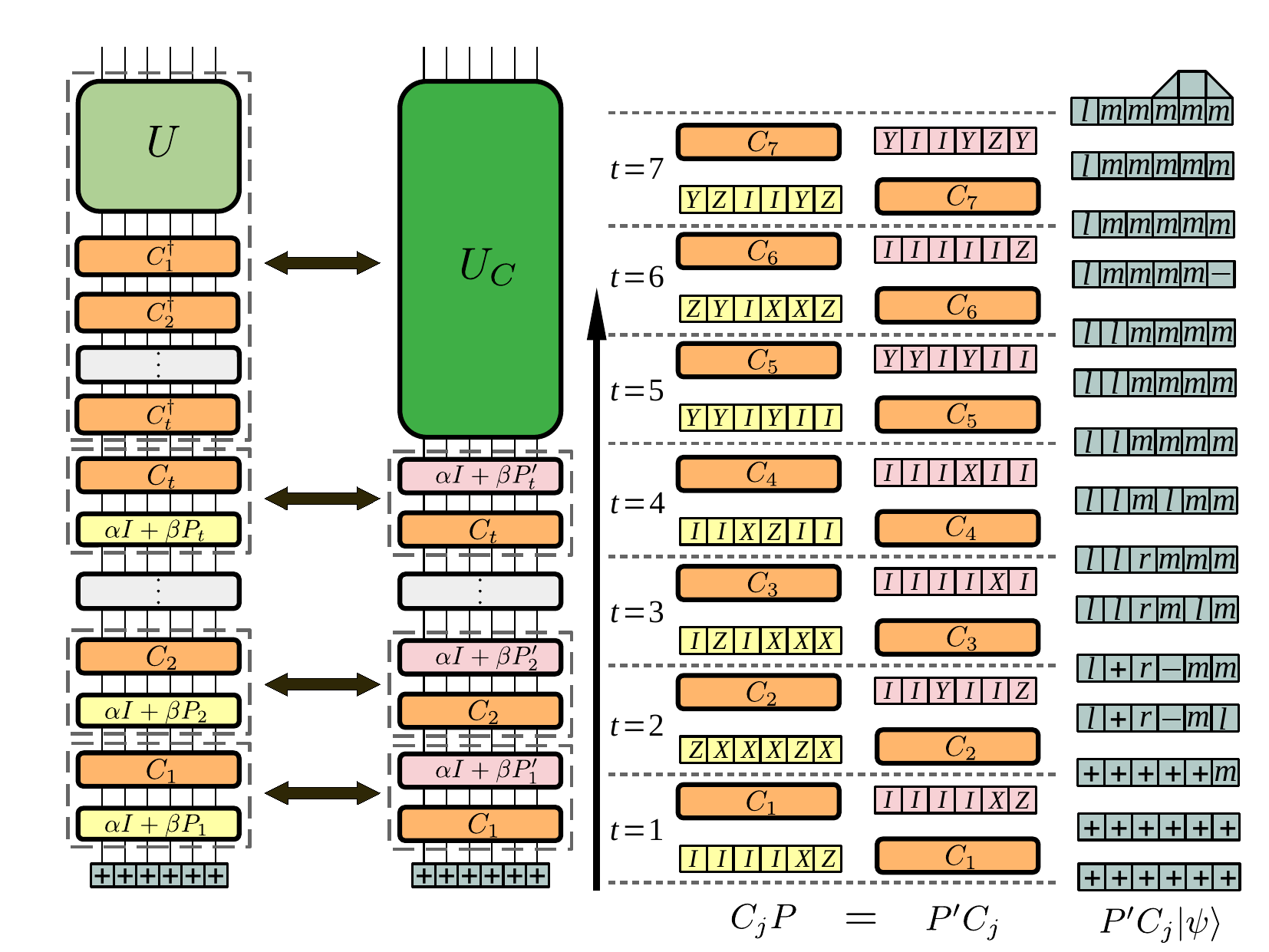}
\caption{Visualization of the simulation of $t$-doped Clifford circuits with CAMPS.}
\label{fig:diagram_cp}
\end{figure*}

In Fig.~\ref{fig:diagram_cp}, we present a schematic diagram illustrating the evolution of CAMPS for a real circuit example ($N=6$, $N_T=1$).
Initially, $T$-gates are injected into the Clifford circuit $U$. During the simulation, the Clifford gates are absorbed into the stabilizer tableau, while the non-Clifford $T$-gates are commuted through the Clifford gates, resulting in twisted operators adjacent to $\ket{\psi}$. Specifically, as shown on the left of Fig.~\ref{fig:diagram_cp}, each block labeled $\alpha I+\beta P_j$ ($j=1,2,...,t$)~\footnote{for convenience, the superscript of $P^{[j]}$ from the main text is moved to subscript in this appendix section.} represents the twisted operator from the $j$-th $T$-gate:
\begin{equation}
T^{[j]}U\prod_{k=1}^{j-1}C_k^\dagger = U\left(\prod_{k=1}^{j-1}C_k^\dagger\right)\left(\alpha I+\beta P_j\right)    
\label{eq:TtoP}
\end{equation}
where $C_k$ is the disentangling Clifford circuit from the Clifford disentangling algorithm for $k$-th $T$-gate. $C_j$ is inserted into the circuit as $C_j^\dagger C_j$, and $C_j^\dagger$ is absorbed into the stabilizer tableau for the Clifford circuit, denoted as $U_C$ in Fig.~\ref{fig:diagram_cp}, while $C_j$ and $\alpha I+\beta P_j$ are absorbed into $\ket{\psi}$ together to reduce its entanglement.

Now, alternatively, after identifying $C_j$, instead of allowing the twisted operator $\alpha I+\beta P_j$ to act on the MPS $\ket{\psi}$ first, we can commute it backwards. Specifically, we apply Algorithm~\ref{alg:CommutePauli} to find $P'_j=C_jP_jC_j^\dagger$ such that $C_j(\alpha I+\beta P_j)\ket{\psi}=(\alpha I+\beta P'_j)C_j\ket{\psi}$, as illustrated from the left half side of Fig.~\ref{fig:diagram_cp}. By doing so, $C_j$ acts on $\ket{\psi}$ first, and we then examine how the new twisted operator $\alpha I+\beta P'_j$ modifies the resulting $C_j\ket{\psi}$. 

On the right side of Fig.~\ref{fig:diagram_cp}, we provide an visualization of this simulation process. The yellow blocks in the left column represent $P_j$, the pink blocks in the middle column correspond to $P'_j$, and the blue blocks in the right column show the states $\ket{\psi}$ after the actions of the operators beneath them. For instance, the third blue block from the bottom, $\ket{+++++m}$ is equivalent to $(\alpha I+\beta IIIIXZ)C_1\ket{++++++}$, and the fourth block, $\ket{l+r-ml}$, corresponds to $C_2\ket{+++++m}$. Here the eigenstates for the Pauli operators $\{X,Y,Z\}$ with eigenvalues $\pm 1$ are denoted as $+/-$, $l/r$ and $0/1$, respectively, while the magic states are represented as $m$. The height of the blocks reflects the bond dimension within the MPS. For example, the last blue block has its fifth component with height 2 on both sides, indicating the matrix for the fifth qubit is $2\times2$, meaning this qubit is entangled with its two neighbors. In contrast, all other blocks have a height of 1 for each component, indicating that they are product states.  

Based on the visualization of the simulation process, we heuristically conclude that two conditions must be met simultaneously to prevent an increase in entanglement within the MPS:
\begin{enumerate}
\item [(C1):] When the disentangling Clifford circuit $C_j$ for the $j$-th $T$-gate transforms the MPS $\ket{\psi_j}$ into $\ket{\psi_j'}=C_j\ket{\psi_j}$, it preserves the bond dimension of the MPS. In other words, if $\ket{\psi_j}$ is a product state, $\ket{\psi_j'}$ remains a product state;   
\item [(C2):] The Pauli string $P_j'$ may contain multiple non-identity operators, but its effect on $\ket{\psi_j'}$ must be equivalently to that of another Pauli string $\widetilde{P}_j'$, which has only one non-identity operator acting on $\ket{\psi_j'}$, i.e., $\widetilde{P}_j'\ket{\psi_j'}=P_j'\ket{\psi_j'}$. If the qubit affected by the non-identity operator is an eigenstate of a Pauli operator, it becomes a magic qubit after the action of $\alpha I+\beta P_j'$; if it is already a magic qubit, the entanglement of the MPS remains unchanged, and the number of magic qubits does not increase.
\end{enumerate}
These conditions (C1) and (C2) can be verified from the example in Fig.~\ref{fig:diagram_cp}. Notably, whenever either of these conditions is violated, entanglement increases. For instance, at the final step ($t=7$) in Fig.~\ref{fig:diagram_cp}, the effect of $P'_7$ cannot be reduced to a Pauli string with only one non-identity Pauli operator, causing the fifth qubit to become entangled with its neighbors.  

\section{Conjectures for an Ideal Clifford disentangling Algorithm}
\label{append:conject_globalMagicDistill}
Throughout this work, we introduced two Clifford disentangling algorithms for simulating $t$-doped Clifford circuits: the first algorithm finds an optimal two-qubit Clifford gate to minimize entanglement (OBD), while the second utilizes control gates to reduce discrepancies (i.e., the number of qubits in differing states) between the combined quantum states (OFD). Both methods belong to \emph{local} Clifford disentangling algorithms, leveraging only local quantum information and two-qubit Clifford operations. We have also identified cases where these local techniques fail to reduce MPS entanglement. However, it remains as an open question whether we can find a \emph{global} Clifford disentangling algorithm that utilizes the global information of states and the full $N$-qubit Clifford group.   

In this section, instead of developing a global algorithm, we propose conjectures about its potential properties, should such a method be devised for simulating $t$-doped Clifford circuits.

We formalize the simulation process slightly different from the main text, though effectively they are equivalent. For a given $t$-doped Clifford circuit, we first commute all $T$-gates toward the MPS, which is in the state $\ket{00\dots0}$ initially. This yields a set of Pauli strings ${P_j}_{j=1}^t$, allowing us to represent the state to be simulated as
\begin{equation}
\ket{\Psi}=U\prod_{j=1}^t\left(\alpha I+\beta P_j\right)\ket{00\dots0}    
\end{equation}
For a global Clifford disentangling algorithm, we anticipate finding an $N$-qubit Clifford circuit $C$ that retains the MPS as a product state. Specifically, we would have:
\begin{equation}
\begin{split}    
\ket{\Psi}&=U\prod_{j=1}^t\left(\alpha I+\beta P_j\right) C^\dagger C\ket{00\dots0}\\
&=UC^\dagger\prod_{j=1}^t\left(\alpha I+\beta P'_j\right) C\ket{00\dots0}
\end{split}
\end{equation}
where $P'_j=CP_jC^\dagger$. The first desirable property for such an algorithm is: 
\begin{enumerate}
\item[(P1):] {$C\ket{00\dots0}$ is a product state.}
\end{enumerate}
Thus, we can write as $C\ket{00\dots0}=\otimes_{k=1}^N\ket{s_k}$, where $\ket{s_k}$ is an eigenstate of Pauli operator. Furthermore, as we will show later, we can transform $\ket{s_k}$ back to $\ket{0}$ without altering the results. 

The second desired property concerns the commutation structure of the transformed set $\{P'_j\}$, which retains the commutation relations of $\{P_j\}$. We define a commutation table $\mathbb{S}$, where each element $\mathbb{S}_{jk}\in\{0,1\}$ encodes the commutation relation between $P_j$ and $P_k$: $P_jP_k=(-1)^{\mathbb{S}_{jk}}P_kP_j$. Consequently, $\mathbb{S}$ is Hermitian and $\mathbb{S}_{jj}=0$. Additionally, we define a table $\mathbb{P}$ for the transformed operators $\{P'_j\}$, with entries $\mathbb{P}_{jk}$ denoting the Pauli operator of $P'_j$ on the $k$-th qubit. To ensure that $\ket{\Psi}$ retains a product state structure in the MPS portion, we conjecture the second property as: 
\begin{enumerate}
\item[(P2):] {Given $C\ket{00\dots0}=\ket{00\dots0}$, $C$ transforms $\{P_j\}$ to $\{P'_j\}$ so that $\mathbb{P}_{jj}=Z$ and $\mathbb{P}_{jk}=X^{\mathbb{S}_{jk}}$ for $j\neq k$.}
\end{enumerate}
This construction preserves the commutation relations of $\{P_j\}$. For instance, at $N=4$, $t=3$, given 
\begin{equation}
\mathbb{S}=\begin{bmatrix}
0 & 0 & 1 \\
0 & 0 & 1 \\
1 & 1 & 0
\end{bmatrix},    
\end{equation}
according to (P2), we obtain
\begin{equation}
\mathbb{P}=\begin{bmatrix}
Z & I & X & I \\
I & Z & X & I \\
I & I & Z & I 
\end{bmatrix}    
\label{eq:PPexample}
\end{equation}
This setup allows the operators  $\prod_{j=1}^t\left(\alpha I+\beta P'_j\right)$ acting on $C\ket{00\dots0}=\ket{00\dots0}$ to produce at most one magic qubit. The example above, for instance, yields $\ket{00m0}$. However, this construction ceases to work if $t>N$, at which point the MPS begins to develop entanglement.

Lastly, we present an alternative form of (P2) for cases where $C$ maps $\ket{0}$ to another eigenstate ($\ket{s_k}$) of Pauli operator $P[s_k]$:  
\begin{enumerate}
\item[(P2'):] {Given $C\ket{00\dots0}=\otimes_{k=1}^N\ket{s_k}$, $C$ transforms $\{P_j\}$ to $\{P'_j\}$, so that $\mathbb{P}_{jj}=P[s_j]$ and $\mathbb{P}_{jk}=\bar{P}[s_k]^{\mathbb{S}_{jk}}$, $j\neq k$.}
\end{enumerate}
where $\bar{P}[s_k]$ is a Pauli operator other than $P[s_k]$. The two properties (P2) and (P2') are equivalent. This equivalence is confirmed by noting the existence of a single-qubit Clifford gate $U(s_k)$ that transforms $\ket{s_k}$ back to $\ket{0}$, i.e., $U(s_k)\ket{s_k}=\ket{0}$. Meanwhile, $U(s_k)P[s_k]U(s_k)^\dagger=Z$, and $U(s_k)P[s_k]U(s_k)^\dagger=\bar{Z}$, where $\bar{Z}$ is a Pauli operator other than $Z$. Therefore, if we let
\begin{equation}
C\gets \left(\prod_{k=1}^N U(s_k)\right)C\ket{00\dots0}     
\end{equation}
we will get 
\begin{equation}
P'_j \gets \left(\prod_{k=1}^N U(s_k)\right)P'_j\left(\prod_{k=1}^N U(s_k)\right)^\dagger      
\end{equation}
which is exactly the construction from (P2).

\section{Numerical Results on Pauli Obsevables}
\label{sec:pauliexpect}
\begin{figure*}
\centering
\includegraphics[width=\textwidth]{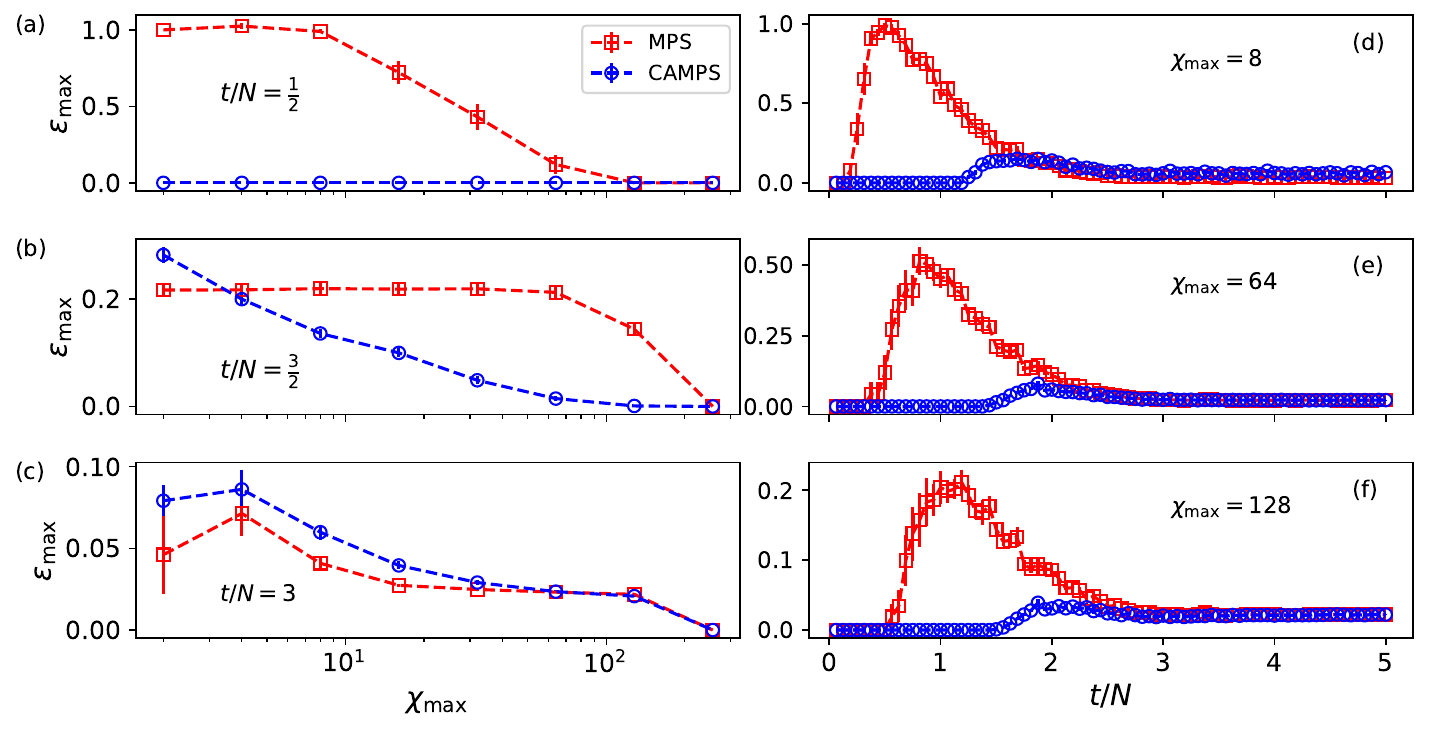}
\caption{Maximum error $\varepsilon_\mathrm{max}$ of the expectation from the sampled Pauli string set ($N=16$, $N_T=1$) with varying maximal MPS bond dimension $\chi_\mathrm{max}$. Each data is averaged over 16 circuit
realizations and error bars are present.}
\label{fig:pauliobservable}
\end{figure*}

Besides the observables from the main text, we also compare the observable expectation from Clifford-MPS with those from MPS. In Fig.~\ref{fig:pauliobservable}, we show $\varepsilon_\mathrm{max}$ versus MPS bond dimension $\chi_\mathrm{max}$ and $T$ gate number $t$. $\varepsilon_\mathrm{max}$ is evaluated as follows: we use the exact wave function (MPS with full bond dimension $\chi_\mathrm{max}=2^{N/2}$) to sample a set of Pauli strings $\{P_j\}_{j=1}^J$ and the corresponding exact expectations $\{p_j\}_{j=1}^J$ (total number $J=1000$ here); next we use Clifford-MPS and MPS with varying $\chi_\mathrm{max}$ to evaluate the expectation $\{p^c_j\}_{j=1}^J$, $c\in\{\mathrm{CAMPS},\mathrm{MPS}\}$ respectively; at last, we use 
\begin{equation}
\varepsilon^c_\mathrm{max}=\max_{j}|p^c_j-p_j|
\end{equation}
to evaluate the accuracy of Clifford-MPS or MPS. Note that when circuits get deeper, the density matrix of the states will distribute the weights over more Pauli strings, and have smaller expectation values for Pauli strings, thus we observe smaller absolute errors at large $t$ from Fig.~\ref{fig:pauliobservable}. The later rise of errors from CAMPS suggests more accurate representation ability from CAMPS than MPS.
\end{document}